\RequirePackage{fix-cm}
\documentclass[natbib,twocolumn]{svjour3}          
\smartqed  
\usepackage{graphicx}
\usepackage[bookmarks = true, bookmarksnumbered = true, pdfpagemode =None, pdfstartview = FitH, pdfpagelayout = SinglePage, colorlinks = true, urlcolor = blue, citecolor = blue]{hyperref}
\usepackage{lineno}
\usepackage{xcolor}
 \journalname{Space Science Reviews}
\begin{document}

\title{Ice Giant Circulation Patterns: Implications for Atmospheric Probes 
}
\subtitle{}

\titlerunning{Ice Giant Circulation}        

\author{Leigh N. Fletcher         \and
        Imke de Pater \and
        Glenn S. Orton \and
        Mark D. Hofstadter \and
        Patrick G.J. Irwin \and
        Michael T. Roman \and
        Daniel Toledo}


\institute{L.N. Fletcher and M.T. Roman \at
              Department of Physics and Astronomy, University of Leicester, University Road, Leicester, LE1 7RH, United Kingdom \\
              Tel.: +44-116-252-3585\\
              \email{leigh.fletcher@le.ac.uk}  \\
 \and M.D. Hofstadter and G.S. Orton \at
 Jet Propulsion Laboratory, 4800 Oak Grove Drive, Pasadena, CA 91109, USA \\
 \and I. de Pater \at
 Department of Astronomy, 501 Campbell Hall, University of California, Berkeley, CA 94720, USA
 \and P.G.J. Irwin and D. Toledo \at
 Atmpspheric, Oceanic and Planetary Physics, University of Oxford, Parks Road, Oxford, OX1 3PU, UK.
            }

\date{Received: 04-Jul-2019 / Accepted: 14-Feb-2020}

\maketitle

\tableofcontents


\begin{abstract}
Atmospheric circulation patterns derived from multi-spectral remote sensing can serve as a guide for choosing a suitable entry location for a future \textit{in situ} probe mission to the Ice Giants.  Since the Voyager-2 flybys in the 1980s, three decades of observations from ground- and space-based observatories have generated a picture of Ice Giant circulation that is complex, perplexing, and altogether unlike that seen on the Gas Giants.  This review seeks to reconcile the various competing circulation patterns from an observational perspective, accounting for spatially-resolved measurements of: zonal albedo contrasts and banded appearances; cloud-tracked zonal winds; temperature and para-H$_2$ measurements above the condensate clouds; and equator-to-pole contrasts in condensable volatiles (methane, ammonia, and hydrogen sulphide) in the deeper troposphere.  These observations identify three distinct latitude domains:  an equatorial domain of deep upwelling and upper-tropospheric subsidence, potentially bounded by peaks in the retrograde zonal jet and analogous to Jovian cyclonic belts; a mid-latitude transitional domain of upper-tropospheric upwelling, vigorous cloud activity, analogous to Jovian anticyclonic zones; and a polar domain of strong subsidence, volatile depletion, and small-scale (and potentially seasonally-variable) convective activity.  Taken together, the multi-wavelength observations suggest a tiered structure of stacked circulation cells (at least two in the troposphere and one in the stratosphere), potentially separated in the vertical by (i) strong molecular weight gradients associated with cloud condensation, and by (ii) transitions from a thermally-direct circulation regime at depth to a wave- and radiative-driven circulation regime at high altitude.  The inferred circulation can be tested in the coming decade by 3D numerical simulations of the atmosphere, and by observations from future world-class facilities.  The carrier spacecraft for any probe entry mission must ultimately carry a suite of remote-sensing instruments capable of fully constraining the atmospheric motions at the probe descent location.

\end{abstract}

\section{Introduction}
\label{intro}

Although three decades have passed since the Voyager 2 spacecraft encountered Uranus and Neptune, our understanding of Ice Giant meteorology and atmospheric circulation remains in its infancy, largely due to the tremendous challenge of observing these distant worlds.  The scientific case for an \textit{in situ} entry probe for one or both of these worlds \citep{18mousis} rests on its ability to uniquely determine atmospheric composition and structure during its descent, providing access to chemical species and altitude domains that are inaccessible to remote sensing.  However, our only previous experience with Giant Planet entry probes (the descent of the Galileo probe into Jupiter in 1995) demonstrated that the interpretation of measurements of condensable species required a good understanding of the local meteorology, which is in turn determined by larger-scale atmospheric circulation patterns \citep{98orton}.  The experience with Galileo argues in favour of multi-probe missions, but it is clear that any future atmospheric probes must be carefully targeted to maximise the scientific return - such as the deliberate targeting of strong regions of upwelling, or the desire to sample more `representative` regions of an Ice Giant.  This article reviews our current understanding of Ice Giant circulation patterns as determined from remote sensing, revealing that different altitude domains (the stratosphere, upper troposphere, and deeper troposphere below the clouds) may exhibit different patterns, leading to a stacked tier of different - but connected - circulation cells.  

\textit{How do we explore Ice Giant circulation patterns?} Remote sensing is required to diagnose these circulation regimes, to serve as a guide for the scientifically optimal locations for probe entry.  These observations can be subdivided into three categories:  (1) observations of reflected sunlight in the ultraviolet, visible, and near-infrared, probing in and out of strong methane absorption bands to sense the aerosol distribution (condensate clouds and photochemical hazes) as a function of altitude; (2) observations of thermal emission in the mid-infrared, far-infrared, sub-millimetre and radio, sensing atmospheric temperatures and gaseous composition from the stratosphere to the deep troposphere; and (3) observations of thermospheric emission from H$_2$ and H$_3^+$ to determine the circulation patterns in the upper atmosphere.  This review will focus on the first two categories. Diagnosing atmospheric circulation requires high spatial resolution, which for reflected light and thermal infrared observations either demands the use of 8-to-10-m diameter observatories on the ground (e.g., Gemini, Keck, Subaru and the Very Large Telescope, VLT), or the stable conditions of the Hubble Space Telescope (HST).  Furthermore, amateur observers have recently begun tracking prominent atmospheric storms and bands on both Uranus and Neptune \citep{17hueso}.  Although these are not yet of sufficient quality to aid in the exploration of Ice Giant circulation patterns, we might reasonably expect significant improvements in the decade before any Ice Giant mission.

As spatial resolution decreases with increasing wavelength, the majority of the literature deals with imaging and spectroscopy at visible and near-infrared wavelengths, providing insights into the distribution of clouds and hazes as a function of time, and revealing the banded structure of the Ice Giant clouds \citep[see the recent review by][and references therein]{18sanchez_jets}.  Thermal-infrared observations from Voyager 2 \citep{98conrath} and ground-based facilities \citep{07orton, 14fletcher_nep, 15orton, 19roman} have revealed structures on large scales, but not at the same resolution as the cloud banding.  Finally, centimetre and millimetre-wavelength arrays like the Karl G. Jansky Very Large Array (VLA) and Atacama Large Millimeter Array (ALMA) \citep{88depater, 03hofstadter, 14depater} provide maps of gaseous contrasts at greater depths, beneath the levels sampled by images in reflected sunlight.  In particular in the past few years, after a sensitivity upgrade to the VLA and commissioning of most ALMA antennas, such maps rival the reflected sunlight images in terms of spatial resolution \citep{19tollefson, 18depater}.  This review will seek to piece together these different strands of observational evidence to generate a picture of Ice Giant circulations that can be tested by future observing campaigns.

\textit{What drives atmospheric circulation?} Planetary atmospheres respond to differences in energy inputs as a function of altitude, location, and time, resulting in a circulation that is a delicate balance between solar inputs from above (with the axial tilt generating seasonally-dependent hemispheric contrasts) and spatially-variable heating from internal sources  (e.g., residual energy from planetary formation, rain-out, or other ongoing gravitational settling driving convective motions, atmospheric instabilities, etc.).  This circulation is largely axisymmetric as a result of the planetary rotation, and the latitudinal extent of the circulation cells depends on the rotation period \citep[e.g.,][]{80held}.  Our terrestrial troposphere features a thermally-direct Hadley circulation cell in the tropics (i.e., air rises near the equator where it is warm, and sinks where it is cold), which is prevented from extending all the way to the poles by the angular momentum of the rotating Earth.  In addition, a weaker thermally-indirect eddy-driven Ferrel circulation cell exists in the extra-tropics, which exhibits rising air at its polar-boundary, equatorward transport at altitude, and subsidence at the edge of the thermally-direct Hadley cell \citep[e.g.,][]{13showman}.  For the latter, the generation of eddies on small scales provides a `stirring' mechanism to generate larger-scale Rossby waves \citep{06vallis}, which are able to propagate energy latitudinally away from their generation region, which leads to the flux of momentum back into their generation region to accelerate the extra-tropical jets.  Note that the overall circulation at Earth's mid-latitudes is still thermally direct (e.g., heat is transported polewards), because the heat transport by the Ferrel circulation is overwhelmed by the energy transport associated with the eddies.  A similar ``Ferrel-like'' process may be responsible for the formation of the multiple belts and zones (and their associated zonal jets) on the Gas Giants, Jupiter and Saturn \citep[see recent reviews by][]{18sanchez_jets, 18showman, 19fletcher_beltzone}.  The Gas Giants rotate much faster than the Earth ($\sim10$-hour rotation periods), resulting in finer-scale banding than the terrestrial case.  These belts and zones are associated with overturning circulations (upwelling and subsidence) at the scales of the individual bands.  Eddy momentum fluxes into the prograde zonal jets have been observed on both Jupiter \citep[e.g.,][]{06salyk} and Saturn \citep{12delgenio}, and are hypothesised to produce super-rotating equatorial jets on hot Jupiters \citep{11showman}.  The different `flavours' of the bands (cyclonic or anticyclonic) also exhibit characteristic meteorology on smaller scales.

As Uranus and Neptune have similar 16-to-17-hour rotation periods that shape their global dynamics, one might expect their atmospheric circulation regimes to be intermediate between the Gas Giant and terrestrial cases.  However, our knowledge of the formation of the Ice Giant bands, and the relation with the observed wind field and storms, is nowhere near the maturity of our understanding of the Gas Giants from the Voyager, Galileo, Juno, and Cassini missions.  Given the appearance of the planets in reflected sunlight, it is not immediately clear even how to define an Ice Giant belt or zone.  As on Saturn, where the relationship between the temperature/wind fields and the visible banding is unclear \citep{18showman}, we will define Ice Giant bands as cyclonic belts (where the circulation between the peripheral jets is in the same sense as the planetary rotation) and anticyclonic zones (where the circulation opposes the planetary rotation), based on their measured temperature contrasts and zonal winds. Axisymmetric albedo contrasts within these cyclonic and anticyclonic regions will be discussed below.  Furthermore, we might expect stark differences between the circulation patterns on Uranus and Neptune.  Uranus' atmosphere is unique in the Solar System, receiving negligible heat flux from the deep interior and experiencing extreme seasonal forcing due to the $98^\circ$ obliquity. Neptune, with its powerful self-luminosity \citep[$2.61\pm0.28\times$ the solar input,][]{91pearl} and strong meteorological activity (dark and drifting ovals, rapidly-evolving bright clouds), provides an important counter-example of a convectively-active Ice Giant weather layer.  Taken together, these two worlds may be end-members of a whole category of astrophysical object \citep[the Neptunes and sub-Neptunes,][]{18fulton}, and provide an extreme test of our understanding of atmospheric circulation.

\textit{What are Ice Giants made of?} Before exploring Ice Giant circulation, we first briefly review their composition and cloud structure.  While the Gas Giants Jupiter and Saturn have a composition that is, broadly speaking, similar to that of our Sun, the Ice Giants Uranus and Neptune are much more enriched in ``heavy" elements (more massive than He). In all Giant Planet atmospheres, we expect the elements O, N, S, and C to be in the form of H$_2$O, NH$_3$, H$_2$S, and CH$_4$. At the higher altitudes, where the temperature is lower, these gases will condense when their partial pressure exceeds their saturated vapour pressure curve. Moreover, some NH$_3$ and H$_2$S will be dissolved in a deep water (solution) cloud, and at higher altitudes we expect NH$_3$ and H$_2$S to combine to form a solid NH$_4$SH cloud layer. For near-solar composition atmospheres (or atmospheres in which heavy elements are enhanced uniformly over solar values, as on Jupiter) we thus expect a deep water or solution cloud to form, topped off by water ice, an NH$_4$SH cloud, and above that an NH$_3$-ice cloud.  If it gets cold enough, like on the Ice Giants, the topmost layer will be CH$_4$-ice \citep{69lewis, 73weidenschilling, 85atreya}.  Microwave observations suggested that H$_2$S on Uranus and Neptune was much more enriched than NH$_3$, when compared to solar composition \citep{78gulkis, 91depater}, an observation that was confirmed both by measuring the microwave opacity of H$_2$S in the lab \citep{96deboer} and by detecting H$_2$S directly in the near-infrared \citep{18irwin_h2s, 19irwin_h2s}.  Throughout this review we will therefore assume that H$_2$S gas (and not NH$_3$) is present above the NH$_4$SH cloud layer, and that it will form an H$_2$S-ice cloud near the 3-6 bar level.

The structure of this review is as follows.  Section \ref{uptrop} explores circulation hypotheses related to reflected sunlight observations of planetary banding, cloud-tracking of zonal winds, and temperature measurements in the thermal-infrared, which all broadly sound the upper troposphere above the condensate clouds (methane and H$_2$S ice).  Section \ref{mitrop} then explores observations of equator-to-pole contrasts in the primary gaseous volatiles (methane, ammonia, and hydrogen sulphide) inferred from both near-infrared and microwave spectra, and Section \ref{stackedcells} shows how the upper-tropospheric and mid-tropospheric circulations might be reconciled.  Section \ref{stratosphere} looks at the large-scale overturning circulation in the stratosphere, and how it might be coupled to the circulation patterns and winds in the troposphere.  Finally, Section \ref{conclusions} attempts to draw these hypotheses together to understand Ice Giant circulation as a function of depth, concluding with an assessment of where a planetary entry probe might be best targeted.  

\section{Upper Tropospheric Circulation: Winds, Temperatures, and Reflectivity}
\label{uptrop}

\begin{figure*}
\begin{centering}
\centerline{\includegraphics[angle=0,scale=.4]{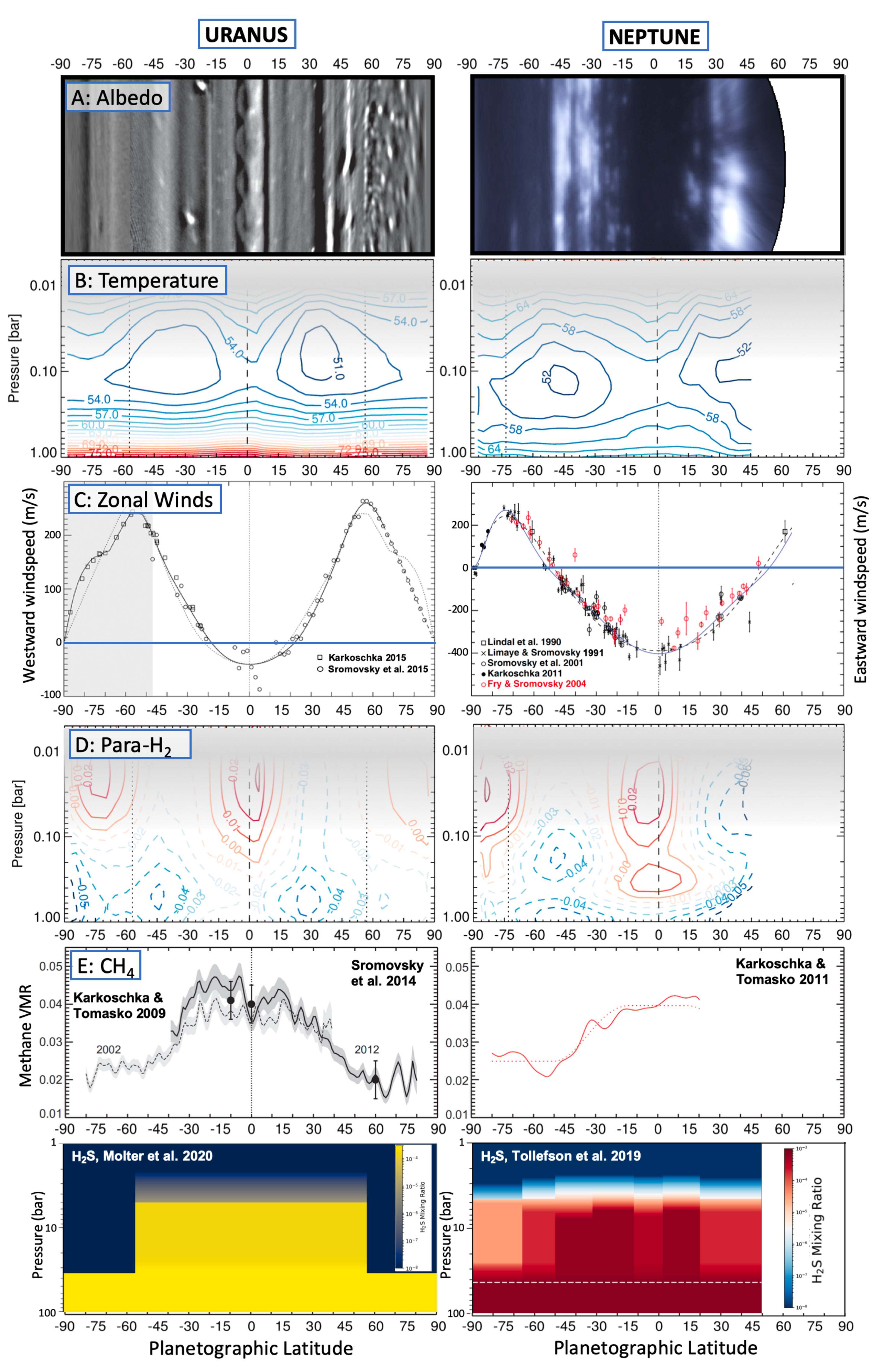}}
\caption{Collation of the key observables for the atmospheric circulation of Uranus (left) and Neptune (right).  Row A shows examples of the albedo structures.  For Uranus, we use a combination of Voyager-2 imagery poleward of $60^\circ$S \citep{15karkoschka} and Keck H-band imagery in 2012 northward of $60^\circ$S \citep{15sromovsky}.  For Neptune, we use Keck H-band imagery from October 2003 \citep{14depater}.  Temperatures in row B were derived from Voyager/IRIS observations \citep{15orton, 14fletcher_nep}.  Zonal winds in row C were derived from a variety of sources \citep[only Voyager 2 in the grey region,][]{15karkoschka}, and this row has been modified from \citep{18sanchez_jets}.  The extent of disequilibrium for para-H$_2$ in row D (sub-equilibrium in dotted lines indicating upwelling, super-equilibrium in solid lines indicating subsidence) were derived from Voyager/IRIS observations \citep{15orton, 14fletcher_nep}. Grey shading in B and D indicates that extrapolations of temperature and para-H$_2$ for $p<70$ mbar (i.e., above the tropopause) are not constrained by the Voyager/IRIS data data. The latitudinal distribution of methane in row E is from \citet{09karkoschka, 14sromovsky, 11karkoschka_ch4} - the dotted line for Neptune is an idealised form of the CH$_4$ distribution.  The deep distribution of H$_2$S in row F is based upon a combination of VLA and ALMA observations of Uranus \citep{20molter} and Neptune \citet{19tollefson}.  Figures have been modified from their original sources for ease of comparison.}
\label{data}
\end{centering}
\end{figure*}

Figure \ref{data} presents a comparison of the latitudinal distributions of albedo (Section \ref{albedo}), zonal winds (Section \ref{winds}), temperatures and para-H$_2$ (Section \ref{temp}), methane (Section \ref{methane}), and microwave absorbers like H$_2$S and NH$_3$ (Section \ref{h2s}), which will be referred to repeatedly through the following sections.  We note that temporal variations in albedo have been observed on both worlds, related to weather phenomena, seasonal changes, and solar-driven variability.  We mention them briefly in the sections that follow, but given that we have only been able to resolve atmospheric features for about 1.5 Uranian seasons (each season lasting 21 years), and have yet to see in detail even a complete 41-year Neptunian season, our understanding of these variations is necessarily limited.

\subsection{Aerosols and Albedo}
\label{albedo}

\subsubsection{Ice Giant Banding}
Jupiter provides the archetype for understanding planetary banding, with a clear distinction between reflective zones and darker belts, and a close correlation between albedo and the latitudinal variations of temperatures and winds.  However, we should not let our historical familiarity with Jupiter's bands bias our understanding of atmospheric circulation on the Ice Giants.  The atmospheres of Uranus \citep{86smith, 87lindal, 86tyler} and Neptune \citep{89smith, 89conrath, 89broadfoot, 89tyler, 92lindal_nep} were first explored by the Voyager-2 spacecraft in 1986 and 1989, respectively, and revealed worlds that were dramatically different from Jupiter.  Planetary banding on Neptune appeared to be much more subtle than Jupiter's, and the bland appearance of Uranus suggested a wholly different type of circulation regime to the other giants \citep[e.g., see the comprehensive review by][]{18sanchez_jets}.  However, modern image processing techniques, and improvements in Earth-based observations, are beginning to shift this view.

Voyager 2 encountered Uranus near its southern summer solstice, with the northern hemisphere hidden in winter darkness.  Visible-light maps of the southern hemisphere reprocessed by \citet{15karkoschka} showed that Uranus was not so bland after all:  although somewhat subjective, these albedo maps showed narrow, reflective bands near $25-30^\circ$S, $52^\circ$S, $68^\circ$S and $77^\circ$S, with finer-scale dark banding observed equatorward of $45^\circ$S and between $68-77^\circ$S.  The darkest and blandest band was observed between $78-83^\circ$S, and can be seen in Fig. \ref{data}A.   Uranus' zonal banding was found to be probably due to variations in aerosol optical depth, although one of the bands could have been caused by variations in aerosol absorption, suggestive of different materials in some bands \citep{15karkoschka}.  

As Uranus passed through northern spring equinox in 2007, the northern hemisphere came into view and could be captured via advanced ground-based imaging techniques, providing the unprecedented views of fine-scale banding shown in Fig. \ref{data}A.  Bright storm clouds reaching high above the surrounding clouds (presumably of methane ice) became increasingly visible in the years following equinox, and have been tracked with high-resolution ground-based imaging \citep{15depater, 17irwin, 15sromovsky}.   Bright features can be readily seen in methane bands and at red wavelengths \citep{95sromovsky}.  Long exposures in the near-infrared (H band, 1.4-1.8 $\mu$m) using instruments on Keck and Gemini in 2012-14 were `derotated` using a knowledge of the zonal wind field in order to enhance the visibility of features in Fig. \ref{data}A \citep{12fry, 15sromovsky}.  Images at these wavelengths reveal Uranus' banded structure, albeit interrupted by bright clouds, vortices, and ephemeral storms.  Zonal medians of the near-infrared reflectivity maps \citep{15sromovsky} indicate that these zonal contrasts are not static, but change from observation to observation, potentially due to obscuration of the banded structure by discrete features.  Some persist - brighter bands in Fig. \ref{data}A could be seen between $40-50^\circ$S, $10-20^\circ$S, $0-8^\circ$N, $10-12^\circ$N, $18-31^\circ$N, $38-42^\circ$N, and $48-52^\circ$N, along with fainter zonal albedo contrasts on a $\sim5^\circ$-latitude scale.  \citet{15sromovsky} also found an equatorial wave feature with diffuse bright features every $30-40^\circ$ longitude. In the north polar region, brightness minima occurred near $53-54^\circ$N, $60-61^\circ$N, $70-71^\circ$N, and $76-80^\circ$N, and \citet{15sromovsky} showed that the springtime polar region beyond $55^\circ$N was characterised by small, bright cloud features with 600-800 km diameters, potentially of convective origin.  This also corresponded to a region of solid body rotation, as explained in Section \ref{winds}.  Taken together, the contrast-enhanced views of Uranus' southern \citep{15karkoschka} and northern \citep{15sromovsky} hemispheres revealed banded albedo patterns on a much finer scale than the temperature and wind fields, which is totally unlike Jupiter and Saturn.  

Neptune's banded pattern was evident in Voyager-2 flyby images \citep{89smith, 91limaye, 01sromovsky}, although the overall morphology changed considerably between 1989 and 2001. In particular, the bright bands with fine-scale zonal structures at southern and northern mid-latitudes became very prominent only after the Voyager flyby \citep{02sromovsky, 11karkoschka, 12martin}.  Given Neptune's long orbital period, we have only ever been able to observe the southern hemisphere (summer solstice was in 2005).  Neptune in Fig. \ref{data}A exhibits bright and variable cloud activity at mid-latitudes, with a number of narrow, bright bands near $25-50^\circ$S and $25-45^\circ$N, but bands of lower reflectivity equatorward of $\pm25^\circ$ and poleward of $50^\circ$S \citep{14depater}.  Neptune's equator is relatively quiescent, lacking the bright cloud activity compared to the stormy mid-latitudes, where cloud activity seems to peak in latitude bands centred near $25^\circ$S and $30^\circ$N.  Large anticyclones on Uranus and Neptune can form and dissipate on the timescales of years \citep{95hammel}.  A small dark spot was observed on Uranus in 2006 at $28^\circ$N \citep{09hammel}, and since that time several more have been observed \citep[e.g.,][]{15sromovsky}.  Before disappearing by 1994, the Great Dark Spot dominated the Voyager-2 flyby of Neptune \citep{89smith, 94baines, 93sromovsky, 98lebeau}, and drifted equatorward before disappearing.  Dark ovals on both Ice Giants are usually accompanied by bright companion clouds due to air being forced upwards over the underlying vortex \citep[orographic clouds,][]{01stratman}.  Since the Voyager flyby, several new dark spots have been observed on Neptune by Hubble, allowing a comprehensive exploration of their drift rates and lifetimes \citep{01sromovsky, 18wong, 19simon}.  Since the dark spots on Uranus are much smaller than those on Neptune, they are harder to observe, and may be more rare on this planet.  The latitudinal drifting of these dark anticyclones sets the Ice Giants apart from the Gas Giants, where vortices remain in their latitude bands due to the strong shears associated with the zonal winds.  This again hints at atmospheric circulation patterns that differ between the Gas and Ice Giants.  

\subsubsection{Temporal Variability in Reflectivity}
\begin{figure*}
\begin{centering}
\centerline{\includegraphics[angle=0,scale=.4]{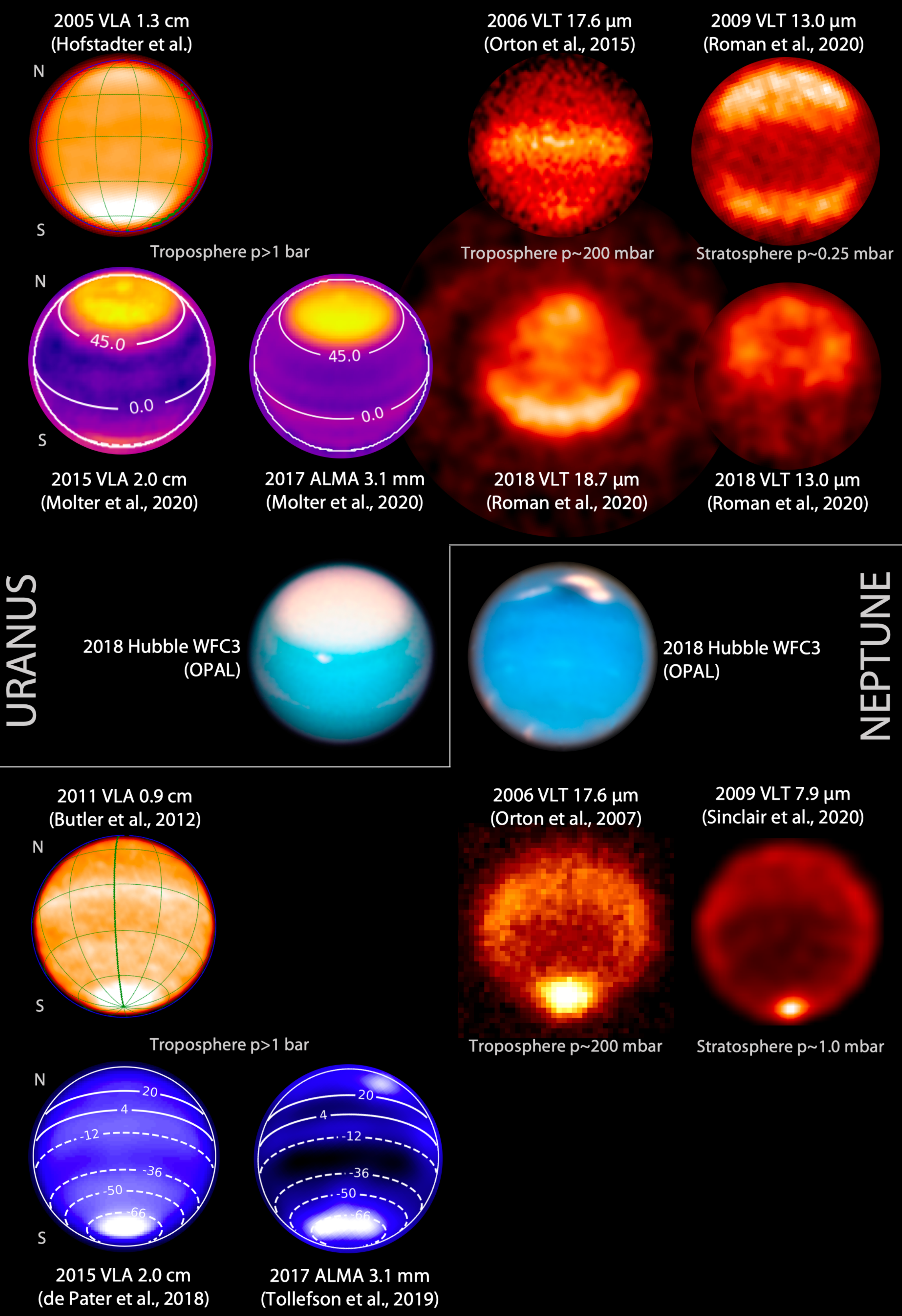}}
\caption{Montage of thermal emission observations of Uranus (top) and Neptune (bottom) that help characterise ice giant circulation.  On the left, centimetre-wave observations from the VLA \citep{12butler_dps, 18depater, 20molter} and millimetre-wave observations from ALMA \citep{19tollefson, 20molter} sense opacity variations in the deep troposphere.  Fainted banded structure is visible on both planets, although maps were constructed from many hours of data, smearing features in longitude.  On the right, 17-18 $\mu$m observations sense upper tropospheric temperatures \citep{07orton, 15orton}, whereas 7.9 and 13.0 $\mu$m sense stratospheric temperatures via methane and acetylene emission, respectively \citep{19roman, 20sinclair}.  Two sets of Uranus data are shown, one near equinox (top row) when both poles were visible, and one in 2015-18 (second row) when the north pole was in view.  Hubble/WFC3 images of Uranus and Neptune in 2018 are shown in the centre for context, courtesy of the OPAL programme (\url{https://archive.stsci.edu/prepds/opal/}).  All images have been oriented so that the north pole is at the top.   On both planets, the dominant features at centimetre/millimetre wavelengths are the very bright poles, interpreted as regions of dry, subsiding air parcels at pressures greater than $\sim1$ bar.  Note that Uranus' large polar region extends to $\sim\pm45^\circ$, while Neptune's extends only to $\sim65^\circ$S.}

\label{images}
\end{centering}
\end{figure*}

Observations of Uranus now span 1.5 Uranian seasons (Uranus' year is 84 Earth years long), which has revealed seasonal changes in its albedo.  High southern latitudes (poleward of $45^\circ$S) were found to be more reflective than mid-latitudes during the Voyager encounter \citep[potentially due to the increased optical depth of the methane cloud near 1.2-1.3 bar,][]{91rages}, with an absence of any small-scale convective structures \citep{12sromovsky} suggesting suppressed convection at the summer pole under a `south polar cap', or convective features being hidden by this cap of high-altitude aerosols.  This albedo pattern changed over the subsequent years:  Hubble 700-1000 nm images between 1994-2002 \citep{04rages} revealed a darkening of the south pole, the formation of a bright ring near $70^\circ$S, and a south polar collar at $45^\circ$S.  As the northern hemisphere came into view after 2007, the south polar collar diminished in brightness \citep{10irwin, 18roman}, and a north polar collar became steadily brighter at $45^\circ$N \citep{12irwin}, with a bright `north polar cap' observed after 2014 \citep{15depater, 15sromovsky, 18toledo}, as shown in the centre of Fig. \ref{images}.  Unlike the southern polar region, the north polar region exhibited numerous small-scale features in 2012-14, suggestive of convective clouds and an asymmetry between the southern summer and northern spring poles \citep{12sromovsky}.  These cloud features are still visible at the time of writing (2019).

These changes in Uranus' albedo may be due to both a high-latitude depletion of methane (Section \ref{methane}), coupled with seasonally-changing aerosols.  Stratospheric aerosols were revealed by Voyager 2 \citep{87pollack, 91rages}, most likely related to photochemistry and the resulting condensation of hydrocarbon ices.  But the rapid decline of the south polar cap and the development of a cap in the north is much faster than would be expected from radiative timescales \citep{90conrath}, photochemical timescales \citep{18moses}, and aerosol microphysical timescales \citep{19toledo_ura}.  Furthermore, stratospheric hazes formed from the condensation of photochemically-produced hydrocarbons would need to be transported meridionally from upwelling regions to subsiding regions in the upper troposphere to influence the reflectivity of the $p\sim1$-bar cloud layer (there is insufficient UV-penetration to the 1-2 bar region to drive photochemical haze production in situ). Accumulation, sedimentation and coagulation are relatively slow processes that would produce a substantial seasonal lag, again inconsistent with the rapid changes observed on Uranus \citep{19toledo_ura}.  This all hints at meridional circulation patterns in the stratosphere and upper troposphere, to which we will return in the coming sections.  Finally, we note that Uranus is not alone in displaying albedo variations - long-term monitoring of Neptune's disc-averaged visible magnitude shows a consistent brightening over many decades \citep{06lockwood, 07hammel, 19lockwood} and a modulation by solar ultraviolet and galactic cosmic rays that drive aerosol nucleation via the production of ions \citep{92moses, 16aplin}.  Furthermore, \citet{11karkoschka} used fourteen years of HST observations to identify intriguing periodicities of $\sim5$ years in Neptune's discrete cloud activity, suggesting that the Ice Giants might exhibit atmospheric cycles on timescales much shorter than a season.

\subsection{Zonal Winds}
\label{winds}
The measurement of Ice Giant winds via cloud tracking probably accounts for the largest body of literature for these two worlds.  Winds have been measured from Voyager 2 \citep{86smith, 89smith, 89hammel, 91limaye, 93sromovsky, 15karkoschka}, the Hubble Space Telescope \citep{95sromovsky, 97hammel, 01sromovsky, 98karkoschka, 01hammel}, and ground-based facilities like Keck and Gemini \citep{04fry_dps, 05sromovsky, 05hammel, 09sromovsky, 12martin, 15sromovsky, 18tollefson}.  In contrast to Uranus' stable zonal wind patterns \citep{15sromovsky}, there is a good deal of dispersion in the motions of the cloud features on Neptune: it is not always clear that the features are genuinely tracking the underlying winds, and the different features are also likely representative of different altitudes \citep{14fitzpatrick, 18tollefson}.   Nevertheless, these studies have shown that both worlds feature retrograde jets at their equators and a single prograde jet at high latitudes in each hemisphere (Fig. \ref{data}C).  Neptune's prograde jet peaks between $70-80^\circ$S (a symmetric jet is expected, but not yet observed, in the north), whereas Uranus' jets peak at $50-60^\circ$.  These wind patterns are very different from the multi-jet circulation patterns on Jupiter and Saturn \citep[see the comprehensive review by][]{18sanchez_jets}, leading to the suggestion that some latitudinal gradients of the zonal winds might not have been fully resolved by the measurements to date (see Section \ref{unres_winds}).  Nevertheless, the winds on the two Ice Giants are surprisingly similar considering their different spin axes and internal heat fluxes.

Most of the tracked cloud features on Uranus are near the 1.2-bar methane condensation level or in the deeper 2-3 bar main clouds of H$_2$S ice \citep{18irwin_h2s}\footnote{Note that infrared observations are most sensitive to the cloud tops, whereas the base pressure could be at higher pressures.}, but some can reach the 250-600 mbar region in the upper troposphere \citep{07sromovsky, 12sromovsky, 18roman}.  A long-lived feature known as the `Berg` had bright features near 550-750 mbar, but with the main parts of the structure near 1.7-3.5 bars \citep{11depater}.  On Neptune, clouds are seen in the stratosphere at 20-60 mbar, at altitudes much higher than those on Uranus \citep{03gibbard, 16irwin_nep}, making them visible in the K' filter (sensing strong methane absorption, and therefore high altitudes, between 2.0-2.4 $\mu$m), tracking features in the 20-300 mbar range \citep{01sromovsky}, well above the methane condensation level near 1.4 bar.  At these altitudes, windshear could add several tens of m/s dispersion to the results in Fig. \ref{data}C.  Adding to this complexity, Neptune's clouds evolve over very short timescales \citep{12martin, 16simon, 16stauffer}, displaying both episodic and continuous cloud activity \citep{94baines, 95sromovsky, 18molter} that can even be observed from the ground by amateur astronomers \citep{17hueso}.  Neptune's rapidly evolving convective clouds seem to prevail at mid-latitudes, and clouds in the main storm band at $20-40^\circ$S may have become increasingly vigorous since the Voyager flyby \citep{07hammel_var}.  Similar rapid evolution is observed in Uranus' small-scale clouds \citep{17irwin}, making it challenging to track them for long periods, with the highest-intensity storms occurring near $30-40^\circ$N \citep{15depater}. The canonical zonal wind profiles in Fig. \ref{data}C are therefore subject to considerable uncertainty, but the jets do demarcate a polar domain, mid-latitude domain, and equatorial domain on both worlds. 

A crucial open question is how the strength of the winds varies with altitude.  Section \ref{temp} shows how the zonal jets should decay with altitude through the upper troposphere ($p<1$ bar).  Analyses of Uranus' and Neptune's gravity fields out to the fourth-order harmonic \citep{91hubbard, 13kaspi} suggest that the zonal wind patterns are restricted to the outermost 1000 km of the planet's radii, so the winds must ultimately be decaying with depth too, although the decay function is poorly constrained \citep{13kaspi}.  Exploring the interface between these two domains is vitally important for an understanding of the processes driving and maintaining the zonal winds.  \citet{18tollefson} detected vertical wind shear at Neptune's equator by tracking bright cloud features in the H- (1.4-1.8 $\mu$m) and K' (2.0-2.4 $\mu$m) bands with Keck in 2013-14.  The higher-altitude features at K' (sensing features at  $\sim10$ mbar) showed stronger retrograde velocities than deeper  features seen in the H-band ($p>1$ bar), suggesting that the winds increase in strength with height, opposite to that inferred from the thermal field (Section \ref{temp}).  The deep features have a less negative retrograde velocity (by $\sim100$ m/s) than the higher-altitude features, suggesting a significant windshear, at least at Neptune's low latitudes (the equatorial region, between $\pm25^\circ$ latitude).  A potential reconciliation of the windshear and temperature fields will be presented in Section \ref{temp}.  

A final complication is that the zonal wind field is potentially variable with the seasons.  On Uranus, the only Ice Giant where winds have been observable for more than a single season (and even here it is only ~35 years or 1.5 seasons), an asymmetry between northern spring in 2012-14 \citep[where a broad region of solid-body rotation from 62-83$^\circ$N was identified,][]{15sromovsky} and southern summer in 1986 \citep[with a much smaller region of solid-body rotation and a large gradient in drift rates,][]{15karkoschka} might potentially reverse as Uranus approaches northern summer solstice in 2028. The zonal motions, as well as the contrasting distributions of small-scale convective clouds, might be intricately linked with the seasonal insolation and thus, with any large-scale hemispheric circulation.

\subsection{Atmospheric Temperatures and Ortho/Para-H$_2$}
\label{temp}

Cloud-tracked zonal winds are typically relevant to a narrow altitude range, with the precise cloud-top and base pressures subject to large uncertainties.  Thermal-infrared remote sensing, on the other hand, can provide vertical profiles of atmospheric temperature as a function of position, extending the two-dimensional windfields into three dimensions.  The upper tropospheric temperatures derived from Voyager 2 in Fig. \ref{data}B show cool mid-latitudes in the 80-800 mbar range, contrasted with the warmer equator and poles, as revealed by both Voyager 2  \citep{87flasar, 91conrath, 98conrath} and subsequent ground-based mid-infrared imaging in Fig. \ref{images} \citep{06hammel, 07orton_nep, 14fletcher_nep, 15orton}.  In the years after the Voyager-2 encounter, a warm summertime vortex developed over Neptune's southern pole in the upper troposphere and stratosphere \citep{07orton, 14fletcher_nep}.  Note that, like the zonal winds, the temperature bands also demarcate polar, mid-latitude, and equatorial domains, but do not display the fine-scale banding observed in the albedo in Fig. \ref{data}A.  In the absence of any appreciable latitudinal differences in heating, horizontal temperature contrasts can be used as an indirect measure of vertical motions.  For example, if we assume latitudinal uniformity in radiative heating (in gases and aerosol layers), exothermic chemical reactions, latent heat of volatile condensation and/or para-hydrogen equilibriation, then cool temperatures suggest rising motion with adiabatic cooling at mid-latitudes, accompanied by subsidence and adiabatic warming at the equator and poles \citep{87flasar, 91conrath, 91bezard, 98conrath}.  

Tropospheric temperatures were derived from the 17-50 $\mu$m region of the Ice Giant spectra, which is dominated by a collision-induced continuum of hydrogen and helium.  The shape of this continuum is also governed by the ratio of the two spin isomers of H$_2$ \citep{82massie, 98conrath}:  the S(1) absorption near 17 $\mu$m is formed from transitions within ortho-H$_2$ (the odd spin state of H$_2$ with parallel spins), whereas S(0) near 28 $\mu$m is formed from transitions within para-H$_2$ (the even spin state of H$_2$ with anti-parallel spins). Populating or depopulating the S(0) states therefore affects the shape and gradient of the far-IR continuum, with cooler regions near the tropopause having a higher fraction of para-H$_2$.  This allows us to identify regions of sub-equilibrium conditions, where the para-H$_2$ fraction is lower than would be expected from thermal equilibrium due to upwelling of low-para-H$_2$ air from the deeper troposphere.  Conversely, super-equilibrium conditions are consistent with the subsidence of high-para-H$_2$ air from the tropopause region.  Fig. \ref{data}D shows the disequilibrium of para-H$_2$, derived from Voyager/IRIS measurements using modern opacities for the collision-induced absorption \citep{14fletcher_nep, 15orton, 18fletcher_cia}, and confirming the presence of mid-latitude upwelling and equatorial/polar subsidence \citep{87flasar, 98conrath}.  However, we caution that the IRIS data were only sensitive to the 100-800 mbar range \citep[i.e., at depths below the tropopause,][]{98conrath}, with speculative results at higher altitudes in Fig. \ref{data}D resulting from smooth relaxation of retrieved profiles to their priors.  At present, para-H$_2$ does not provide any constraint on lower stratospheric circulation in the text that follows.

\begin{figure*}
\begin{centering}
\centerline{\includegraphics[angle=0,scale=.6]{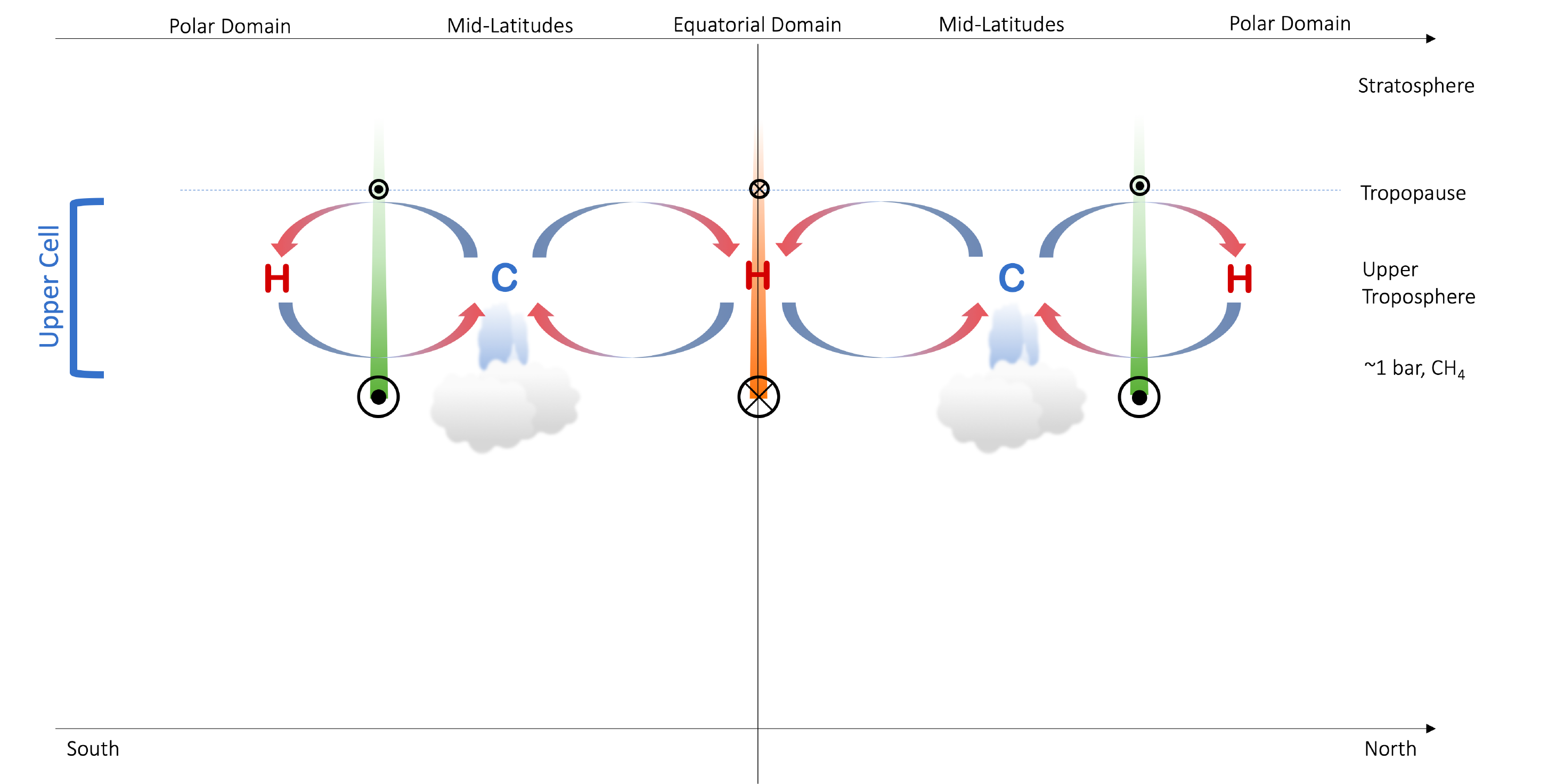}}
\caption{Schematic depicting the meridional circulation in the upper tropospheres of Uranus and Neptune based on (i) tropospheric temperatures, denoted by `C' and `H`' for cold and hot, respectively; (ii) the deviation of para-H$_2$ from equilibrium; (iii) simplistic inferences of enhanced cloud activity at mid-latitudes; and (iv) the inferred decay of the winds with altitude.  Retrograde winds are indicated by orange bars and circles with crosses; prograde winds are indicated by green bars with circles with dots.  This circulation pattern is suggested to be present between the tropopause at $\sim0.1$ bar and the CH$_4$ condensation level at $p>1$ bar.}  
\label{uppertrop}
\end{centering}
\end{figure*}

The Ice Giant circulations inferred from the temperature, para-H$_2$ (below the tropopause), and winds are shown in schematic form in Fig. \ref{uppertrop}, representing the 0.1-1.0 bar range, approximately. Geostrophy implies that temperatures and winds are in balance with one another via the thermal windshear equation \citep{04holton}, and \citet{87flasar} proposed a model where the acceleration of the zonal flow due to conservation of angular momentum was balanced by frictional damping (i.e., vertical shear), potentially due to the breaking of vertically-propagating waves in the upper troposphere.  The upper tropospheric temperature gradients in the 80-800 mbar range imply maximum positive windshears near $\pm(15-30)^\circ$ latitude (i.e., on the flanks of the equatorial retrograde jet) and maximum negative windshear near $\pm(60-75)^\circ$ (i.e., near the prograde jets at high latitudes).  This is shown by coloured bars of decreasing size and contrast in Fig. \ref{uppertrop}.  The windshear is minimal (i.e., close to barotropic, with no wind variability with height) in the $\pm30-50^\circ$ latitude range associated with the mid-latitude temperature minima.  It is in this mid-latitude domain, with the coolest temperatures and sub-equilibrium para-H$_2$, that the most frequent and notable storm activity occurs, and where the seasonal polar collars emerge on Uranus.  The upper tropospheric circulation in Fig. \ref{uppertrop} might also suggest a meridional transport of aerosols from regions of mid-latitude upwelling to a region of equatorial subsidence.  This is consistent with a reflective band seen near the equator in Hubble and VLT imaging \citep{18toledo}, suggesting an accumulation of haze at the equator.

It is interesting to consider whether the symmetric circulation depicted in Fig. \ref{uppertrop} could be seasonally variable.  For example, averaged over a Uranian year, Uranus receives more heat at the poles than at the equator, so that energy must be transported equatorward.  Following earlier work by \citet{87friedson} and \citet{87flasar},  \citet{90conrath} used radiative-dynamical balance equations to explore the relationship between temperatures and winds on Uranus and Neptune as a function of time.  Their model features atmospheric motions forced by radiative heating in the stratospheres (by weak methane absorption bands), or by the mechanical forcing from zonal winds imposed at the lower boundaries, without any momentum convergence to accelerate the flow, and with a simplified frictional drag to decay the winds with altitude.  For Uranus, the radiative time constant ($\sim130$ years) was longer than the orbital period, such that the atmospheric temperatures remained close to the annual-average radiative equilibrium values, despite the large amplitude of the radiative forcing, consistent with the lack of observed temperature variations over the full season between solstice (1986), equinox (2007), and the present day \citep{15orton, 19roman}.  Polar regions were only slightly warmer than the equator, depending only on the integrated solar irradiance at the top of the atmosphere, which is some $\sim30$\% larger at the poles.  This result would be changed if a latitude-dependent absorbing aerosol were present (see Section \ref{albedo}), but the small magnitude of the temperature differences due to Uranus' small haze opacity should result in a weak annual-mean meridional circulation, with a low-latitude cell of rising motion between $10-30^\circ$ in both hemispheres and subsidence at the equator.  \citet{18li} calculated updated estimates of the radiative heating/cooling rates, using modern estimates of temperature and hydrocarbon profiles.  Uranus exhibits a longer radiative time constant than at Neptune because it has the coldest atmosphere and the lowest  methane abundance (and resulting photochemical products).  As a result, cooling of Uranus is primarily due to the collision-induced opacity of H$_2$.  The calculated cooling rates are larger than the heating rates on both worlds, creating the stratospheric energy crisis (i.e., the stratospheres are much warmer than would be calculated from a pure radiative model).  \citet{18li} noted that their radiative time constants were much shorter than the older results of \citet{90conrath}, indicating that seasonal effects might be more important than previously ascertained - however, this is inconsistent with the absence of seasonal change on Uranus noted by \citet{15orton} and \citet{19roman}, perhaps suggesting much longer radiative time constants or efficient transport at these atmospheric depths.

\begin{figure*}
\begin{centering}
\centerline{\includegraphics[angle=0,scale=.6]{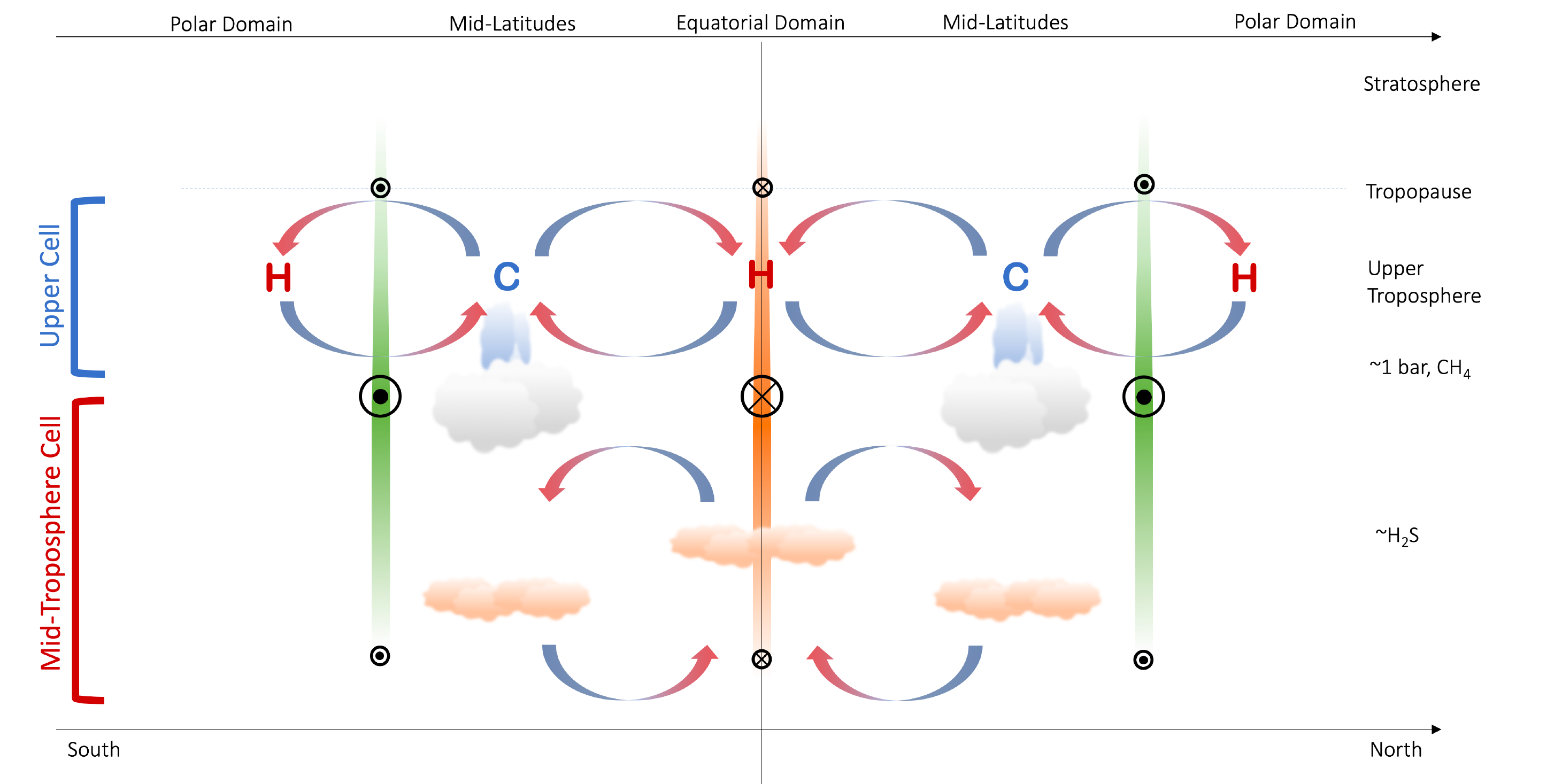}}
\caption{Modified schematic of the meridional circulation in the upper tropospheres of Uranus and Neptune (Fig. \ref{uppertrop}) to account for the observation that Neptune's equatorial jet becomes more retrograde with altitude, potentially requiring the equator to be cooler than mid-latitudes (see main text).  We have arbitrarily hypothesised the same trend in Uranus' equatorial retrograde jet, and in the polar prograde jets on both planets, neither of which have been proven or dis-proven by the available data.   Retrograde winds are indicated by orange bars and circles with crosses; prograde winds are indicated by green bars with circles with dots.}  
\label{uppertropwinds}
\end{centering}
\end{figure*}

We now return to the conundrum in Section \ref{winds}, where \citet{18tollefson} had identified a strengthening of the retrograde flow with increasing altitude, counter to the sense of the windshear in Fig. \ref{uppertrop}.  Using a modified thermal wind equation, \citet{18tollefson} showed that the required windshear could be produced either by (i) having a cool equator compared to mid-latitudes at $p>1$ bar (i.e., in the opposite sense to that in Fig. \ref{data}B and Fig. \ref{images}); (ii) by having methane strongly enriched at the equator compared to latitudes $>\pm40^\circ$; or (iii) some combination of the two.  They showed that this could be made consistent with the Voyager/IRIS measurements, provided this `reversed pattern' of a cool equator and warm mid-latitudes were restricted to pressures greater than 1 bar.  The implied circulation is shown schematically in Fig. \ref{uppertropwinds}, showing the retrograde jet weakening into the deeper atmosphere, and a cool equator (upwelling) at $p>1$ bar sitting beneath the warm subsiding region at $p<1$ bar.  This deeper circulation is more reminiscent of the cool, upwelling equators of Jupiter and Saturn.  Stacked tiers of opposing circulation regimes have been proposed for the gas giants too \citep[see review by][]{19fletcher_beltzone}, to resolve the discrepancies between the meridional motions at the cloud tops associated with eddy momentum convergence on the prograde jets, and meridional motions observed in the upper troposphere \citep{00ingersoll, 05showman, 11fletcher_vims}.  We shall return to this interpretation in Section \ref{mitrop}, but first we explore further observational constraints on atmospheric circulation in the troposphere and stratosphere.

\section{Mid-Tropospheric Circulation: Equator-to-Pole Contrasts}
\label{mitrop}

\subsection{Atmospheric Composition}

The visible and near-infrared reflectivity explored in Section \ref{albedo} depends on the vertical distribution of aerosols in the tropospheres of Uranus and Neptune, which in turn depend on the distributions of condensable volatile species.  Based on cosmochemical abundances, and observations of Jupiter and Saturn, the most abundant condensable species in Ice Giant tropospheres are expected to be CH$_4$, H$_2$O, NH$_3$, and H$_2$S.  Of these, only CH$_4$ has been directly measured \citep{11karkoschka_ch4, 09karkoschka}, and is found to be enriched relative to H$_2$ by factors of 10 to 100 times over solar abundance ratios.  The bulk mixing ratios of most other condensable species are thought to be similarly enriched, while H$_2$O is expected to be even higher, accounting for ~60\% of the planet's total mass \citep[e.g.,][]{95guillot}.  One must remember, however, that the bulk abundances in the planet may not be reflected in abundances within the troposphere.  For example, radio observations found, surprisingly, that NH$_3$ is strongly depleted in at least the upper $\sim30$ bars of Uranus' atmosphere, and perhaps deeper \citep{78gulkis}.  Whether this reflects a misunderstanding of the overall composition of planets \citep[e.g., the lack of nitrogen in Uranus and Neptune could be caused by the inefficient trapping of N$_2$ by clathration in water ice,][]{04hersant}, or is due to chemical trapping of NH$_3$ in the interior, is not yet clear.

Based on these chemical abundances, the topmost clouds are expected to be comprised of a thin cloud of CH$_4$ ice with a base near 1.3 bar, where latent heat released by condensation was observed to modify the temperature lapse rate observed in Voyager radio occultations \citep{87lindal, 92lindal_nep}.  Below, at the 3-6 bar level, lie clouds of H$_2$S ice \citep{91depater}, whereas H$_2$S vapour has been detected above these cloud decks \citep{18irwin_h2s, 19irwin_h2s}.  Clouds of NH$_4$SH may exist in the 30-to-40-bar region \citep{73weidenschilling, 05atreya}, and are expected to lock away any remaining tropospheric NH$_3$ that is not trapped at much deeper levels \citep{91depater, 03hofstadter}.  Dry or wet adiabatic extrapolation of the temperature profiles from 2 bars \citep[where the temperature was determined by radio occultations,][]{87lindal, 92lindal_nep} to deeper pressure levels are used to estimate where the water cloud base forms.  If the atmospheric mixing ratio of H$_2$O were solar, it would occur near 50 bars.  Both planetary interior models \citep{95guillot} and recent measurements of CO \citep{13luszczcook, 17cavalie} suggest that the H$_2$O abundance is much higher, potentially several hundred times solar, putting the water cloud base at pressures exceeding 200 bars.  This condensation region may be stable against convection, and could separate the troposphere above it from the much deeper atmosphere and the suspected deep, superionic, watery ocean with its own interior circulation patterns at great depths (see Section \ref{deep}).  If the condensables are sufficiently abundant, then density stratification (i.e., the mean molecular weight gradient) could have a stabilising influence on convective motions \citep{87gierasch, 95guillot}, producing distinct vertical boundaries to heat transport that act like an insulating layer and trap energy. The spatial distribution of volatiles, particularly CH$_4$ (Section \ref{methane}) and H$_2$S (Section \ref{h2s}), are therefore crucial to our assessment of atmospheric circulation. 

We briefly note that measurements of disequilibrium tropospheric species like CO and PH$_3$ could also provide an indirect way of measuring atmospheric circulation and constraining the deep oxygen abundances \citep{05visscher, 17cavalie}.  PH$_3$ is used as a tracer of vertical mixing on Jupiter and Saturn, but only upper limits are available from millimetre observations on Uranus and Neptune \citep{09moreno_dps, 19teanby}.  Maybe the large water abundance is responsible for converting all the PH$_3$ into other compounds in the 1000-bar region \citep{85fegley, 05visscher}.  Tropospheric CO has been detected on both planets, but Neptune has an extremely high abundance \citep{93marten, 05marten, 05lellouch, 10lellouch, 10fletcher_akari}, implying that its deep atmosphere is probably fully convective and well-mixed. \citet{13luszczcook} used the far wings of CO lines to show that a brightness temperature increase from Neptune's southern mid-latitudes to the south pole could be explained by a drop in CO opacity.  However, for both PH$_3$ and CO, we have only a very limited sense of their spatial distributions with which to derive atmospheric circulation - they will not be considered again in this review.  Instead, we turn to what can be learned from the spatial distributions of CH$_4$ and H$_2$S.

\subsection{Methane}
\label{methane}

Fig. \ref{data}E shows the similarities in the spatial distributions of tropospheric methane on both Ice Giants, where there is a significant decline from the equatorial domain to the polar domain.  Rather than being a gradual decline, there are hints of a step-like structure, with a transition in the $\pm30-40^\circ$ latitude region on both worlds.  Derivations of the methane abundance are usually entangled with assumptions about the scattering properties and spatial distributions of aerosols, but at some near-infrared wavelengths the collision-induced absorption of H$_2$ (near 825 and 1080 nm) can be used to separate these variables, allowing a direct measurement of the methane abundance.  Narrow-band observations at these H$_2$-dominated wavelengths, in conjunction with data at 1.29 $\mu$m probing similar depths, have been exploited using data from the Hubble Space Telescope and ground-based near-IR spectroscopy \citep{14sromovsky, 18sromovsky}.

Uranus' south polar depletion of methane was first detected via Hubble 300-1000 nm (STIS) spectroscopy in 2002 \citep{09karkoschka}, as the southern hemisphere was approaching autumn conditions.  The suggested polar subsidence was thought to be responsible for the notable absence of any small-scale convective structures near the south pole \citep{12sromovsky}.  However, further Hubble observations in 2012 \citep{14sromovsky} and 2015 \citep{18sromovsky}, combined with ground-based data in the near-IR, revealed that the north polar region, now emerging into spring, was similarly depleted in methane and yet did display small-scale convective clouds.  The symmetric polar depletions might therefore be separated from the seasonal differences in the appearance of small-scale clouds in the polar domain.  These STIS observations, depicted in Fig. \ref{data}E,  also show that the methane drop from $\sim4$\% near the equator to $\sim2$\% at high latitudes is relatively stable over time.  The ongoing brightening of the north polar region during spring (Section \ref{albedo}, and shown in the centre of Fig. \ref{images}) must therefore be related to changes in aerosol scattering properties, rather than changes in the methane depletion.  Uranus' polar depletion was also observed from ground-based facilities in the near-infrared, including the IRTF \citep{13tice}, Keck \citep{15dekleer}, Palomar \citep{18roman}, the VLT and Gemini in the H (1.4-1.8 $\mu$m) band \citep{18toledo, 19irwin_h2s}.

Models reproducing the spectroscopy suggest that Uranus' methane variability is restricted to the upper troposphere (i.e., relatively shallow), with a latitudinally variable decrease in abundance from $\sim3-5$ bar to the condensation altitude near 1.1 bar \citep{14sromovsky, 18sromovsky}.   The latitudinally-uniform deep abundance for $p>5$ bar depends on the assumed aerosol scattering properties, but ranges from $\sim3.5$\% to $\sim2.7$\% for Uranus.  Although the equator-to-pole structure is statistically significant, the methane abundances derived in these studies also show smaller-scale latitudinal structure at a lower significance level.  For example, the STIS observations in 2012 and 2015 suggest a local minimum in CH$_4$ abundance right at the equator, flanked by (weak) local maxima at $\pm5-20^\circ$ latitude \citep[Figs. 24-25 of][]{18sromovsky}, then a decrease by a factor of 2 to $45^\circ$N and a factor of 3 by $60^\circ$N, but with abundances that are ultimately sensitive to the aerosol properties.  A principal component analysis of Gemini H-band spectra by \citet{19irwin_h2s} reproduced the general equator-to-pole contrast, and also showed tentative hints that CH$_4$ is not maximal right at the equator (their Fig. 13).  The suggested circulation pattern consists of low-latitude upwelling of moist, CH$_4$-rich air; poleward motion as the air dries by condensation, precipitation and sedimentation; and high-latitude subsidence of CH$_4$-depleted air (and modified by more complex equatorial circulation).  As will be described below, this requires alterations to the $p>1$-bar schematic in Fig. \ref{uppertropwinds}.

Neptune's methane distribution, at least for the southern hemisphere visible from Earth, is shown in Fig. \ref{data}E, as derived from Hubble STIS 300-1000 nm spectroscopy \citep{11karkoschka_ch4}.  These results are consistent with a deep mixing ratio of $\sim4$\% at $p>3$ bar, but a decrease by a factor of three from the equator to mid-latitudes in the $p<3$ bar region.  Fig. 14 of \citet{11karkoschka_ch4} shows a transition between elevated CH$_4$ equatorward of $\pm20^\circ$ latitude to depleted CH$_4$ poleward of $45^\circ$S, again depicting three distinct regions:  a well-mixed equatorial region, a region of decline at mid-latitudes, then a well-mixed polar region with depressed abundances.  As on Uranus, smaller-scale and low-significance methane variations are indicated in Fig. \ref{data}E, with suggestions of depletion at $25^\circ$S and $45-55^\circ$S and enhancements near $35^\circ$S. Neptune's $45-55^\circ$S region is one of the most active regions of cloud formation \citep{11karkoschka}, but subsidence would be inferred from the methane depletion results there \citep{11karkoschka_ch4}.  Using millimetre observations sensing the 1-10 bar range, \citet{19tollefson} find elevated methane between $12-32^\circ$S and $2-20^\circ$N, but depressed abundances poleward of $66^\circ$S and near the equator at $12^\circ$S-$2^\circ$N - i.e., a local equatorial minimum flanked by weak off-equatorial maxima, similar to that seen on Uranus.  However, \citet{19tollefson} stress that their results are not entirely consistent with those of \citet{11karkoschka_ch4}, potentially due to the different modelling approaches in the near-infrared and microwave.  A recent principal-component analysis (PCA) of VLT MUSE visible-light spectroscopy of Neptune confirms the  equator-to-pole gradient in Neptune's methane observed by \citet{11karkoschka_ch4}, alongside a slight lowering of methane at the equator consistent with the millimetre observations \citep{19irwin_muse}.  It is clear that methane is tracking both a global circulation pattern, alongside more local meteorology on a finer zonal scale.

The connection between the equator-to-pole gradient in methane gas, and the observed distributions of clouds potentially associated with methane ice, is perplexing.  Strong subsidence in the polar domain would tend to inhibit convection and CH$_4$ cloud formation \citep{14sromovsky}, and yet we see discrete clouds at the poles, and a seasonally-changing opacity of aerosols in Uranus' 1-2 bar region (Section \ref{albedo}).  Furthermore, the CH$_4$ distributions do not seem to be tracing the upper tropospheric ($p<1$ bar) mid-latitude upwelling responsible for the cool temperatures, sub-equilibrium para-H$_2$, and polar collars of aerosols in Uranus' $40-50^\circ$ domain.   Strong equatorial upwelling might promote cloud and haze formation around the equator, coupled with cool temperatures from adiabatic expansion - neither of which are observed.  The hints of small-scale methane variations near the equator might suggest that the upper tropospheric equatorial subsidence (and warming $p<1$ bar) is modulating the $p>1$-bar methane distribution, but the two circulation regimes do seem somewhat separate. The near-IR spectroscopy also suggests that the methane variability is restricted to $p<3-5$ bar.  To further investigate this deeper circulation, we now look at the motions inferred from millimetre- and centimetre-wave observations.

\subsection{H$_2$S}
\label{h2s}
Millimetre- and centimetre-wave observations (i.e., microwave) of the Ice Giants \citep{91depater} sense thermal emission modulated by the pressure-broadened wings of NH$_3$ and H$_2$S (along with other contributions from CH$_4$, CO, H$_2$O, potentially PH$_3$, and the hydrogen-helium continuum).  Uniquely distinguishing between these contributions is a challenge, and spatial variations of temperatures in the deep troposphere (including the lapse-rate changes associated with cloud condensation) could also contribute to the observed emissions in Fig. \ref{images}. Nevertheless, spectral fitting appears consistent with H$_2$S as the dominant absorber in the upper $\sim20$ bars of the troposphere, with NH$_3$ removed via the formation of the NH$_4$SH cloud just below this level and (probably) dissolution in the aqueous layers at higher pressures.  This was recently supported by the detection of H$_2$S absorption features in the H-band near 1.57-1.58 $\mu$m on both Uranus \citep{18irwin_h2s} and Neptune \citep{19irwin_h2s} using Gemini-North/NIFS - this suggests that the atmospheric S/N ratio is greater than unity, implying that the main cloud deck visible on both worlds comprise H$_2$S ices.  As millimetre and centimetre wavelengths probe beneath the clouds down to 50-100 bars, recent advances in spatially-resolved microwave observations (Fig. \ref{images}) have provided an invaluable tool for tracing deep atmospheric circulation.

VLA centimetre observations of Uranus between 1982 and 1994 \citep{88depater, 91depater_radio, 03hofstadter}, probing down to 50 bar, showed an equator-to-pole gradient, with the south pole considerably brighter than the equator, and a boundary somewhere near $45^\circ$S.  VLA observations since 2003 (Fig. \ref{images}) also indicated that the north polar region was similarly bright \citep{04hofstadter_dps, 18depater}.  ALMA observations in Fig. \ref{images} taken in 2015-2018 at 1-3 mm, probing the 1-10 bar range, also show a bright north polar region \citep{20molter}.  The latitudinal distribution of H$_2$S, as derived from a combined analysis of ALMA and VLA data, is shown in Fig. 1F \citep{20molter}. The low opacity over Uranus' poles is explained as a complete absence of H$_2$S down to $p\sim35$ bar, although a small amount of NH$_3$ gas (few times $10^{-7}$) appear to be required \citep{20molter}.  The NH$_3$ abundance is essentially zero at mid-latitudes and the equator.  The images in Fig. \ref{images} further show relatively bright bands at latitudes of $\sim20^\circ$N and $\sim20^\circ$S at 1-3 mm and 1-2 cm, while a third band is visible at the ALMA millimetre wavelengths at the equator. These observations are indicative of a lower opacity at these latitudes at 0.5-5 bar, likely due to variations in the H$_2$S relative humidity and/or CH$_4$ abundance.

These observations show a morphological similarity to the distribution of CH$_4$ in the 1-5 bar range \citep{14sromovsky}, suggesting that the polar subsidence (volatile depletion) and equatorial upwelling (volatile enrichment) might extend over great depths.  However, the latitudinal distribution may be subtly different at the altitudes of the H$_2$S ice cloud (2-4 bars): a PCA analysis of Gemini H-band observations by \citet{19irwin_h2s} suggested that Uranus' H$_2$S column abundance above the 2-to-4 bar cloud is largest at mid-latitudes, and displays local minima at the equator and north pole (the south pole was not in view).  At first glance, this might be more consistent with modulation by the upper tropospheric circulation pattern shown in Fig. \ref{uppertrop}, with time-variable mid-latitude upwelling being responsible for Uranus' polar collars.   

VLA 1.3-6 cm maps of Neptune (lower part of Fig. \ref{images}) probe the 10-50 bar range \citep{14depater} and showed strong depletion of volatiles at the south pole \citep{12butler_dps, 13luszczcook, 14depater} in a region coinciding with a warm summertime polar vortex in the upper troposphere and stratosphere at $p<1$ bar \citep{07hammel, 07orton_nep, 14fletcher_nep}.  The warm south polar region extends to approximately $65^\circ$S with a low abundance of H$_2$S down to $\sim40$ bar - this is a smaller region of volatile depletion than at Uranus' poles, but consistent with Neptune's prograde jet being at a higher latitude than Uranus' prograde jet.  The centimetre maps from \citet{14depater} do not show an equatorial brightening, but 0.9-3.0 cm observations from the upgraded VLA shown in Fig. \ref{images} \citep{12butler_dps, 18depater} support a picture of equatorial subsidence to the 5-10 bar level, modulating the general equator-to-pole gradient in H$_2$S.  Recent ALMA millimetre observations of Neptune \citep{19tollefson}, probing the 1-10 bar range, have identified distinct bands of warmer brightness temperatures, explained by H$_2$S abundance variations at $p<10$ bar, with a severe depletion at the south pole.  The circulation inferred from these millimetre observations suggests air rising at mid-latitudes ($12-32^\circ$S) and north of the equator ($2-10^\circ$N), and sinking in the $2^\circ$N-$12^\circ$S region and poleward of $66^\circ$S.  As on Uranus, this suggests equator-to-pole transport in the mid-troposphere, coupled with a complicated pattern of equatorial subsidence and near-equatorial upwelling being modulated by the upper tropospheric circulation in Fig. \ref{uppertrop}.  However, the recent detection of H$_2$S in the near-IR \citep{19irwin_h2s} has complicated this picture, showing an enhanced H$_2$S relative humidity in the south polar region above the clouds, potentially associated with the warmer temperatures of the polar vortex, with local microphysical effects at the cloud tops, or with an absence of aerosols permitting longer path lengths through the polar atmosphere.  A reconciliation of this cloud-top polar enhancement with the stark polar depletion observed in the microwave has yet to be performed.

\section{Stacked Circulation Cells}
\label{stackedcells}

\begin{figure*}
\begin{centering}
\centerline{\includegraphics[angle=0,scale=.6]{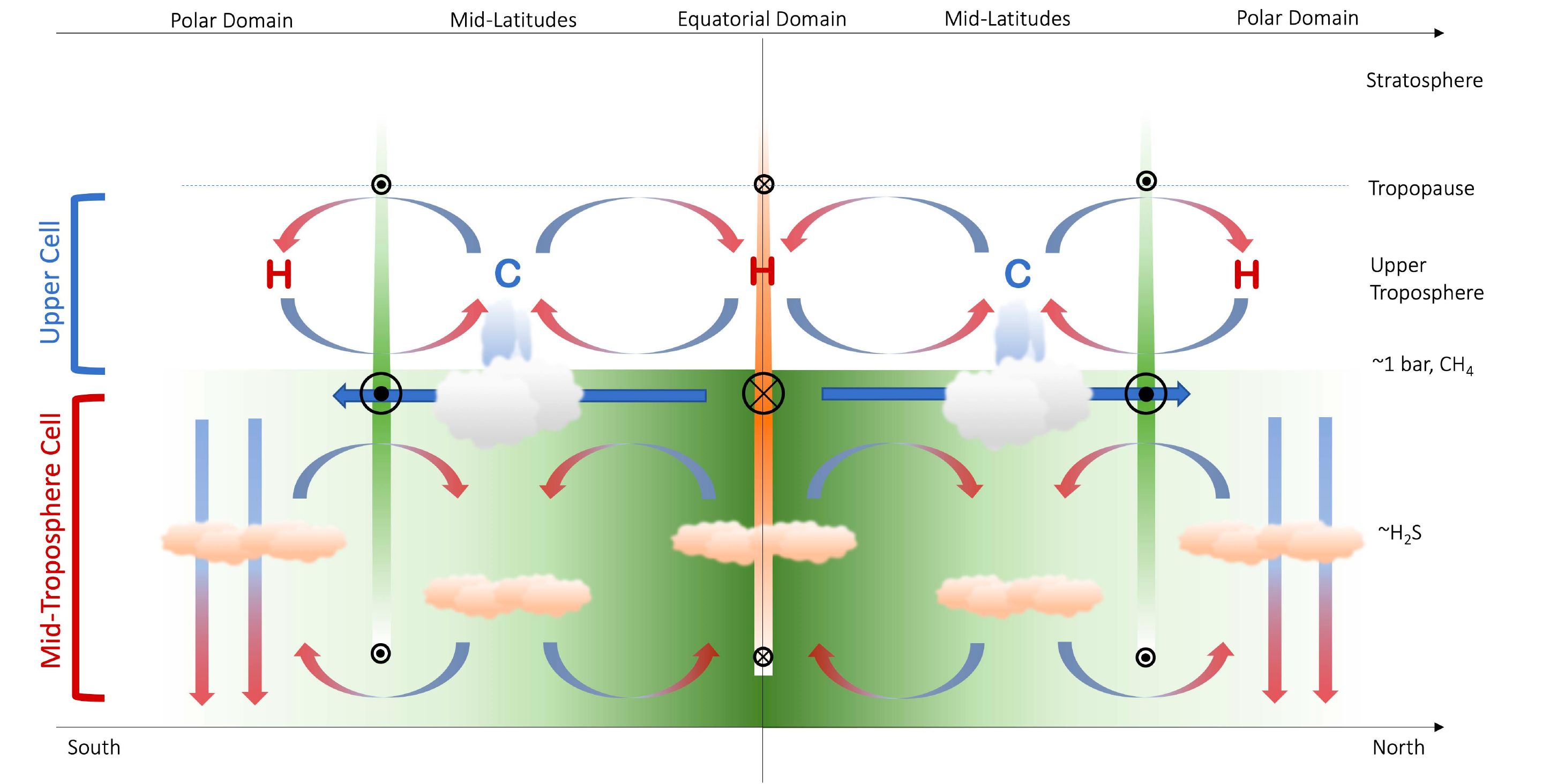}}
\caption{Modified schematic of the meridional circulation, now extending from the upper troposphere into the mid-troposphere.  Large-scale equator-to-pole transport, with rising motions at low latitudes (i.e., within 20-$30^\circ$ of the equator, consistent with the wind patterns inferred in Fig. \ref{uppertropwinds}) and strong polar subsidence, has been included to account for the latitudinal distributions of CH$_4$ and H$_2$S - the green equator-to-pole shading represents this gradient.  Small-scale rising at high latitudes may explain the existence of polar clouds \citep{14sromovsky} and excess H$_2$S humidity at Neptune's south pole \citep{19irwin_h2s}, but this has to exist within a region of net subsidence to explain microwave observations.  Here we see a tier of two stacked cells, potentially separated in the $p\sim1$-bar region.  The sense of the mid-tropospheric circulation near 1 bar, equatorward across the prograde jets, would be consistent with an (unproven) eddy-driven prograde jet, as is found on Jupiter and Saturn.  The closure of the circulation at high pressures is arbitrary, but microwave observations suggest the polar subsidence persists to at least $p\sim50$ bar.  }  
\label{midtrop}
\end{centering}
\end{figure*}

\subsection{Tropospheric Circulation}

Fig. \ref{midtrop} updates our previous schematics of the upper-tropospheric circulation (Figs. \ref{uppertrop}-\ref{uppertropwinds}) to incorporate the findings from the CH$_4$ and H$_2$S distributions.  This follows similar attempts to construct two-dimensional tropospheric circulations by previous authors \citep{03hofstadter, 14depater, 14sromovsky, 19tollefson}. The equatorial upwelling and $p\sim1$ bar cooling required to produce the windshear measured by \citet{18tollefson} is included\footnote{We note that the situation is more complex than this, as a cold equator and warm mid-latitudes are only needed if methane (and hence atmospheric density) are meridionally uniform.  However, the results of \citet{18tollefson} are somewhat degenerate, as the inclusion of an equator-to-pole depletion in methane by a factor of four allows for the warm equator and cool mid-latitudes that are actually observed, consistent with Neptune's ``upper cell'' of air rising at mid-latitudes and sinking over the equator.  This study makes it clear that both the temperature gradients and density gradients should be accounted for when trying to interpret vertical wind shear.}, but new horizontal arrows depict the general equator-to-pole transport of both methane and H$_2$S, and the strong polar depletions are depicted by downward arrows to the bottom of our domain (50-100 bars).  While not indicated in the figure, it should be kept in mind that the polar region of depletion is much larger on Uranus, extending down to $\pm45^\circ$ in both hemispheres, than it is on Neptune, where it extends to $\sim65^\circ$S (the only hemisphere currently observed).  The $p>1$-bar equatorial upwelling meets the $p<1$-bar equatorial subsidence somewhere near the methane-ice cloud tops, where a meridional divergence towards the poles would transport material to higher, off-equatorial latitudes.  This could be responsible for the local minima in CH$_4$ and H$_2$S at the equator \citep{18sromovsky, 19irwin_h2s, 19tollefson}, flanked by equatorial maxima where the upper tropospheric subsidence is weaker.  

At mid-latitudes we retain the upwelling inferred from the temperature and para-H$_2$ distributions for $p<1$ bar, which may also be responsible for mid-latitude peaks in Neptune's H$_2$S abundance\footnote{We caution the reader that these observations measure column abundances above the clouds, so if the cloud tops are lower then the column abundance will increase without any need to increase the H$_2$S abundance.} at these altitudes \citep{19irwin_h2s} and the formation of the polar collars on Uranus.  The equator-to-pole gradients in methane and H$_2$S mixing ratios could have a significant effect on the zonal windshear through density changes \citep{18tollefson}, which is quite unlike anything found on the Gas Giants, and the decrease of density with latitude should supplement the observed thermal wind shear, opposing the observed prograde jets as shown in Fig. \ref{midtrop}.  Such shears should persist to great depths, following the contrasts in volatile absorbers shown in Fig. \ref{data}F.  

The mid-latitude subsidence at $p>1$ bar is more speculative, but could be responsible for the mid-latitude transitions (from strongly enriched to strongly depleted) observed in both the CH$_4$ and H$_2$S distributions in Fig. \ref{data}E-F.  Furthermore, the $p>1$ bar circulation cell across the prograde jets in Fig. \ref{midtrop} are highly speculative, suggesting equatorward flow across the jet near 1 bar, balanced by poleward flow at some uncertain depth.  This structure has been hypothesised by analogy to Jupiter's prograde jets, which are forced by eddy-momentum fluxes converging and transporting energy into the jet streams \citep{04ingersoll, 06salyk}.  This convergence is balanced by a meridional circulation from the cyclonic belts to anticyclonic zones at the cloud-tops, and from the zones to the belts at depth \citep[see the review by][]{19fletcher_beltzone}.  In this picture, the Ice Giant poles are analogous to jovian belts, the mid-latitudes to jovian zones, but we stress that there is currently no observational evidence for eddies forcing the prograde jets on the Ice Giants.  This scenario is also qualitatively consistent with zones (cold and volatile-enriched) and belts (warm and volatile-depleted) in Jupiter's upper troposphere at $p<1$ bar.  Nevertheless, there are several problems with this conceptual picture.  Firstly, the equatorward flow near $p\sim1$ bar is in the opposite direction to the strong CH$_4$ and H$_2$S gradients. Secondly, it is unclear how any polar upwelling at depth (associated with the cross-jet flow) is mixed with the strong volatile depletions shown in the microwave observations.  And thirdly, it remains to be seen how the circulation patterns modulate the fine-scale albedo bands described in Section \ref{albedo}.  It is possible to envision a scenario where all of these processes are at work, but that one dominates the equator-to-pole gradients in composition, and the others contribute to local meteorological and cloud phenomena. Furthermore, this circulation pattern may ultimately be unnecessary if the equator-to-pole contrasts are shown to be driven by local microphysical and chemical processes, rather than large-scale transport. 

By treating the polar domains as analogous to jovian belts, we might gain insights into a further conundrum:  large-scale polar subsidence should be inhibiting cloud formation and small-scale convection, but we see evidence for both of these on Uranus.  Jupiter's cyclonic belts exhibit frequent small-scale moist convective activity over small areas, despite being locations of net subsidence in the upper troposphere.  Indeed, maps of the distribution of lightning show a higher occurrence in cyclonic domains, and cyclonic features in particular \citep{99little, 00gierasch}.  Although lightning has been detected on both Uranus \citep{86zarka} and Neptune \citep{90gurnett}, requiring the presence of polarisable, mixed-phase materials like water, the spatial distribution of Ice Giant lighting is unknown.  Jupiter's belts are strongly depleted in ammonia \citep{06achterberg, 16fletcher_texes, 16depater, 17li}, in the same way as Ice Giant poles are strongly depleted in H$_2$S. 

But the exact mechanisms for moist convection in these domains of net subsidence and depleted volatiles remains unclear without some supply of condensable volatiles (and their latent heat) to the cloud base, possibly via the deep, poleward transport depicted in Fig. \ref{midtrop}.  The need for a deeper circulation cell, transporting H$_2$S to high latitudes at the cloud level, was also noted by \citet{14sromovsky} as part of their multi-tiered structure for Uranus, and may be responsible for the enhanced H$_2$S relative humidity at the cloud tops of Neptune observed by \citet{19irwin_h2s}.  Maybe the strength of the polar convection changes with season, due to a destabilising effect as the troposphere warms in the spring, resulting in more convective transport of volatile-laden air at $p<1$ bar to generate the bright polar caps described in Section \ref{albedo}.

However, these multi-tiered structures in Fig. \ref{midtrop} are seemingly at odds with the single deep cells proposed for Neptune by \citet{14depater} and \citet{19tollefson}, and for Uranus by \citet{20molter}, extending from the stratosphere down into the deep atmosphere.  Maybe strong molecular weight gradients at the cloud condensation altitudes \citep{95guillot} could serve to keep circulation cells separated in layers that have yet to be properly resolved by any of the remote-sensing investigations presented so far.  Nevertheless, the multi-tiered stack of circulation cells provides a hypothesis to be tested by future measurements and numerical simulations.  

\subsection{Connection to the Interior}
\label{deep}

It is reasonable to ask how far down the atmospheric circulation in Fig. \ref{midtrop} extends - we currently assume that the bottom of this figure is at the 50-100 bar level, below the NH$_4$SH clouds but still above the H$_2$O cloud base.  We have seen that these condensation zones for CH$_4$ (and, to a lesser extent, H$_2$S) may be accompanied by significant molecular weight gradients and alterations to the temperature lapse rate \citep{87lindal, 95guillot}, which may serve to separate different layers or strata of circulation and convection \citep{87gierasch} and act as insulators for the escape of internal heat \citep[e.g.,][]{11fortney}.  In particular, a stable layer at the depth of the water clouds could effectively separate the water-rich interior from the dry exterior of an Ice Giant.  Episodic erosion of such a stable layer might be responsible for intermittent outbursts of storm activity or release of internal energy \citep{87gierasch, 15li, 17friedson}.  And the potential stable layers within Uranus might be preventing it from following the expected thermal evolution, partially explaining its apparent absence of internal heat \citep{91pearl}. 

The deep water-ice clouds may be at the top of a massive aqueous water cloud extending to thousands of bars. These values are highly uncertain, given the lack of knowledge of the temperature lapse rate, particularly in the zone of water condensation where there might be significant departures from adiabatic behaviour \citep{95guillot, 17leconte}. The bulk water abundance, inferred from the planet's bulk density and using thermochemical modelling with disequilibrium trace species, may be enhanced by hundreds of times the solar abundance \citep{94lodders, 13luszczcook, 17cavalie}.  Molecular dynamics calculations suggest that water may form a superionic icy ocean at great depths \citep{05goldman, 19millot}, separated from the molecular envelope of H$_2$ and removing (via dissolution) ammonia from the upper levels of the troposphere \citep{91depater}.  This fluid ionic “watery” ocean is likely the region that generates each planet’s internal dynamo \citep{86ness, 89ness}.  Intriguingly, dynamo simulations predict large circulation cells in the deep interior with upwelling (and a peak in the radial heat flux) near the equator and, depending on the thickness of the convecting layer, polar meridional cells may also be present \citep{13soderlund}. The connection between interior circulation and atmospheric circulation would be an intriguing subject for future work.

\section{Stratospheric Circulation: Chemical Tracers}
\label{stratosphere}

Having explored the potentially multi-tiered circulation cells in the upper troposphere and mid-troposphere, we now turn our attention to atmospheric motions above the tropopause. On the Gas Giants, the zonal winds and banded structures persist high into the stratosphere, and are superimposed onto larger interhemispheric circulations, both non-seasonal on Jupiter \citep{07nixon, 13zhang, 18melin} and seasonal on Saturn \citep[see review by][]{18fletcher_book}.  The temperature and composition of the Gas Giant stratospheres is also modulated by the presence of polar vortices and wave-driven equatorial oscillations.  In contrast, our understanding of spatial contrasts in the Ice Giant stratospheres is in its infancy due to their great distance from ground- and space-based telescopes, and the lack of a long-lived orbiter observing them from close range.

Stratospheric composition is driven by the UV photolysis of methane, resulting in a complicated mix of hydrocarbons \citep{83atreya, 05moses, 10dobrijevic, 18moses} that can be investigated via mid-infrared remote sensing \citep{87orton, 97feuchtgruber, 98encrenaz, 10fletcher_akari, 11greathouse, 14orton, 14fletcher_nep, 19roman} and UV occultations \citep{87herbert, 90bishop}.  In addition, oxygen species like CO, CO$_2$ and H$_2$O are also present in the upper stratosphere, originating from cometary impacts, satellite debris, and ablation of interplanetary dust and ring particles.  These compounds potentially play an important role in the photochemical pathways  \citep{97feuchtgruber, 05lellouch, 10lellouch, 13luszczcook, 14cavalie, 14orton, 16poppe, 17moses}. The abundances are sensitive to the strength of atmospheric mixing, so Uranus' sluggish motions produces a lower methane homopause than on Neptune \citep{87herbert, 90bishop}, ensuring that photochemistry on Uranus occurs in a different physical regime (higher pressures) than on any other giant planet, suppressing photochemical networks \citep{91atreya}.  The effects of this are apparent from our observations:  Uranus' hydrocarbons are confined to altitudes below the 0.1-mbar level, and the ratio of ethane to acetylene is very different on Uranus compared to all the other giants \citep{14orton}.  Furthermore, photolysis of CO and CO$_2$ can lead to secondary peaks of hydrocarbon production at higher altitudes \citep{18moses}.  The spatial distribution of these stratospheric species control the local radiative balance \citep[ethane and acetylene are excellent coolants, but their efficiency leads to a stratospheric energy crisis, e.g.,][]{18li} and the condensation of thin stratospheric haze layers in the 0.1-30 mbar range \citep{91rages, 93romani, 17moses, 18toledo}.  Aerosol layers of water, benzene, CO$_2$, acetylene, ethane and propane are just some of the various condensed layers that might be expected at these low temperatures.  As they sediment downwards into the troposphere, they can also modify the optical properties (or serve as nucleation sites for) the tropospheric aerosols (see Section \ref{albedo}).  Thus any redistribution of the hydrocarbons via atmospheric circulation would have important implications for the energetics and hazes of the stratosphere.

The terrestrial stratosphere exhibits a wave-driven `Brewer-Dobson' circulation \citep[BDC,][]{87andrews}, transporting air (and ozone) from the equator to the pole.  The circulation, with air rising at low latitudes and descending at mid- to high-latitudes, is driven primarily by Rossby (planetary) waves from the troposphere.  At even higher altitudes, an upward motion in the summer hemisphere and downward motion in the winter hemisphere is known as the solsticial mesospheric circulation, and is primarily driven by gravity waves.  Wave propagation and breaking on the Ice Giants may be a key mechanism for energy transport to partially resolve the stratospheric and thermospheric energy crises, where solar heating alone is insufficient to explain the high temperatures \citep{87herbert, 93stevens, 18li, 19melin}.  So wave-driven circulations might be at work in giant planet stratospheres, in addition to thermally-driven Hadley-like circulations such as those modelled by \citet{90conrath}, who predicted mid-latitude rising and equatorial subsidence on Uranus.  For example, \citet{12friedson} provided models for a seasonally reversing circulation in Saturn's stratosphere with rising motion in the summer hemisphere and sinking motion in the winter hemisphere, and \citet{09guerlet} interpreted a local maximum in Saturn's hydrocarbons at $25^\circ$N as evidence for the descending branch of a Hadley-like circulation reaching into the stratosphere.  

Assessing these circulations requires spatially-resolved observations of stratospheric temperatures, composition, and hazes, but these are extremely challenging.  The stratospheric distribution of methane is particularly uncertain - it could be uniformly mixed; it could be reaching the stratosphere via convective overshooting from storm systems at mid-latitudes; it could be enhanced at the equator due to the tropospheric circulation in Section \ref{methane}; or it could be leaking through the warm polar vortices \citep[where the cold-trap is less efficient,][]{07orton_nep}.  Distinguishing between these possibilities remains a considerable challenge.  Ground-based observations in the thermal-infrared have revealed relatively uniform stratospheric temperatures and composition on Neptune \citep{91bezard, 11greathouse, 14fletcher_nep}, but with suggestions of a mid-latitude minimum and a rise in emission over the warm summertime polar vortex \citep{06hammel, 07orton_nep, 12orton, 14depater, 14fletcher_nep} suggestive of polar subsidence within $\sim30^\circ$ of the south pole.  

\begin{figure*}
\begin{centering}
\centerline{\includegraphics[angle=0,scale=.2]{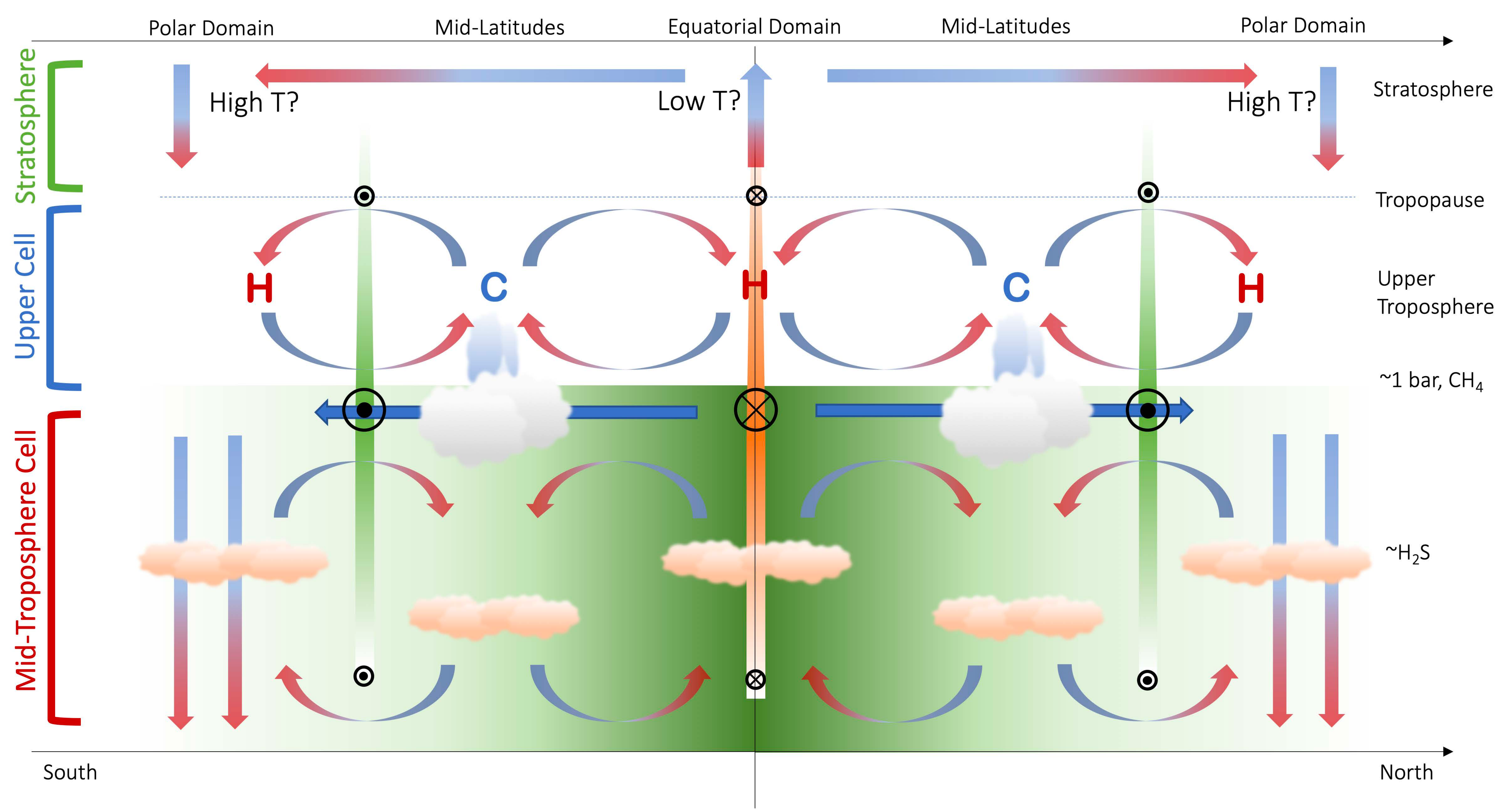}}
\caption{Modified schematic of the tropospheric circulation shown in Fig. \ref{midtrop}, adding the large-scale equator-to-pole motions inferred in the stratospheres of Uranus and Neptune if we consider the observed brightness to be a result of latitudinal temperature variations (an alternative is shown in Fig. \ref{strat_comp}).  Note that the suggested latitudes of stratospheric subsidence (poleward of $\sim\pm25^\circ$ on Uranus, poleward of $\sim70^\circ$S on Neptune) are different between the two worlds. At pressures exceeding 5 bar, the region of polar subsidence on Uranus is smaller than the stratospheric region, extending down to $\pm45^\circ$, whereas on Neptune the stratospheric and deep tropospheric areas of subsidence poleward of $65^\circ$S cover similar spatial areas.} 
\label{strat_temp}
\end{centering}
\end{figure*}

The colder temperatures of Uranus have made similar measurements of the Uranian stratosphere more challenging, but imaging data near 13 $\mu$m (sensitive to stratospheric acetylene) have revealed a contrasting pattern of emission with a distinct equatorial minimum in emission that rises sharply poleward of $\sim25^\circ$ in both hemispheres \citep{18orton_cospar, 19roman}.  The observed gradient is inconsistent with the smooth trend predicted by radiative and photochemical modelling \citep{18moses}, and is the complete opposite of the warm equator observed in the upper troposphere (Section \ref{temp}).  If the emission contrast is due a sharp latitudinal gradient in stratospheric temperatures, this implies an additional equator-to-pole circulation above the tropopause, with low-latitude upwelling and high-latitude subsidence, as depicted in Fig. \ref{strat_temp}.  The strong mid-latitude temperature gradient would also have implications for the thermal wind balance (see Section \ref{unres_winds}). However, if Uranus' stratospheric warming poleward of $\sim25^\circ$ is caused by subsidence, it implies an extremely broad area of polar downwelling compared to that observed on Neptune.  Furthermore, \citet{18orton_cospar} reported an absence of major meridional structure in the stratospheric temperatures unambiguously probed using the H$_2$ quadrupole lines, albeit sensing slightly deeper than the acetylene emission.  We also note that the stratospheric circulation in Fig. \ref{strat_temp} might appear, at first glance, to be inconsistent with the distribution of para-H$_2$ disequilibrium in Fig. \ref{data}D.  However, Voyager IRIS spectroscopy only constrained tropospheric para-H$_2$, with lower-stratospheric distributions being the result of a speculative relaxation to uniform priors.

\begin{figure*}
\begin{centering}
\centerline{\includegraphics[angle=0,scale=.2]{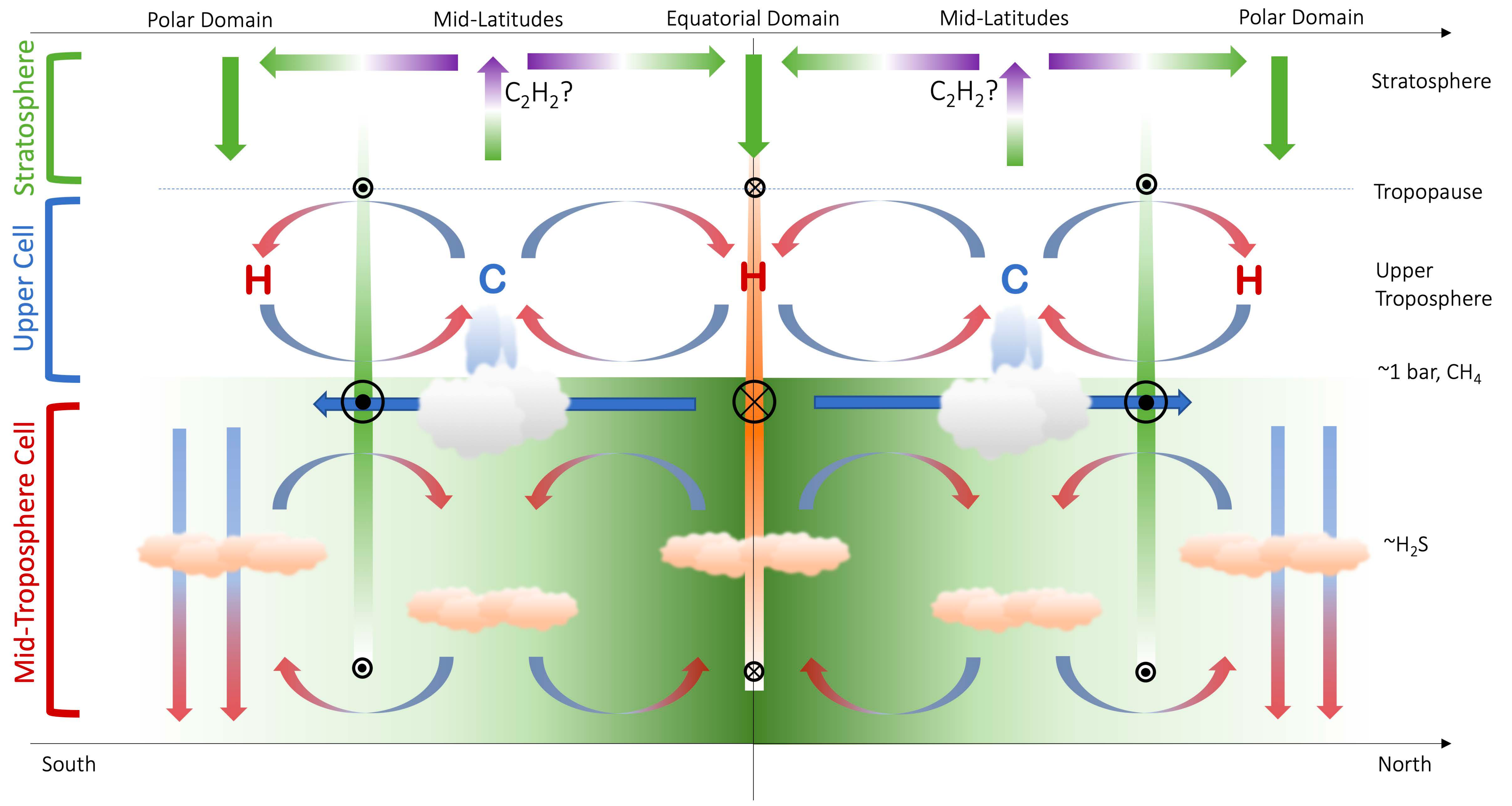}}
\caption{An alternative schematic for the stratospheric circulation following \citet{19roman}, if we assume the the observed brightness of Uranus is the result of latitudinal hydrocarbon variations rather than temperatures.  In this case, the mid-latitude upwelling overshoots from the troposphere into the stratosphere, carrying CH$_4$ aloft to be photolysed to C$_2$H$_2$, which is subsequently transported towards the equator and pole.} 
\label{strat_comp}
\end{centering}
\end{figure*}

An alternative explanation was proposed by \citet{19roman} and is depicted as a schematic in Fig. \ref{strat_comp}.  They showed that the observed stratospheric emission may, in part, be due to a mid-latitude enhancement of acetylene rather than temperatures. Noting that the structure of zonally averaged stratospheric emission over most latitudes appeared to be the inverse of what is observed in the upper troposphere (i.e., the stratosphere was brightest where the troposphere was coldest), \citet{19roman} argued that the resulting stratospheric structure was likely an extension of the upper-tropospheric circulation.  In this scenario, upwelling at mid-latitudes transports methane from the troposphere, through the cold trap, and into the stratosphere for subsequent photolysis to acetylene. A similar explanation had been previously proposed to explain inferences of enhanced stratospheric hydrocarbons found in UVS data by \citet{89yelle} and \citet{92mcmillan}. The weaker emission at the equator compared to the pole (which tentatively appears bright in both the stratosphere and troposphere) would imply that either very little acetylene survives the journey to the equator or that there is an asymmetry in the meridional transport that leads to relatively greater acetylene abundances and/or temperatures at the poles.  Data are still sparse and details of this model remain to be worked out, but more stringent constraints on the stratospheric circulation of both Ice Giants are expected from the James Webb Space Telescope \citep{18moses}.

\section{Unresolved Equatorial Winds}
\label{unres_winds}

Before discussing the different options for probe entry locations, we make one final speculative modification to the schematics presented in Figs. \ref{uppertrop}, \ref{midtrop}, \ref{strat_temp} and \ref{strat_comp}.  On Jupiter and Saturn the relationship between the atmospheric temperatures and winds, via the thermal wind equation, is well established:  prograde zonal jets occur poleward of cool anticyclonic zones and equatorward of warm, cyclonic belts; retrograde jets occur poleward of warm, cyclonic belts and equatorward of cool anticyclonic zones.  In all cases, the maximum temperature gradient (and hence windshear) is co-located with the peak of a zonal jet, either eastward or westwards.  At high latitudes on Uranus and Neptune, this relationship also seems to be true, with the prograde jet separating the cyclonic polar domain from the anticyclonic mid-latitude domain.  But at low latitudes, both Voyager/IRIS (Fig. \ref{data}B) and ground-based observations of the tropospheres \citep{87flasar, 98conrath, 14fletcher_nep, 15orton}, suggest a strong temperature gradient in the $15-30^\circ$ latitude range,  spatially coincident with the strong gradient in hydrocarbons or temperatures in the stratosphere \citep{18orton_cospar, 19roman}.  This is partially supported by \citet{87flasar}, who computed the vertical windshear on Uranus to show maximal shear in the $\pm15^\circ$ latitude region.

\begin{figure*}
\begin{centering}
\centerline{\includegraphics[angle=0,scale=.2]{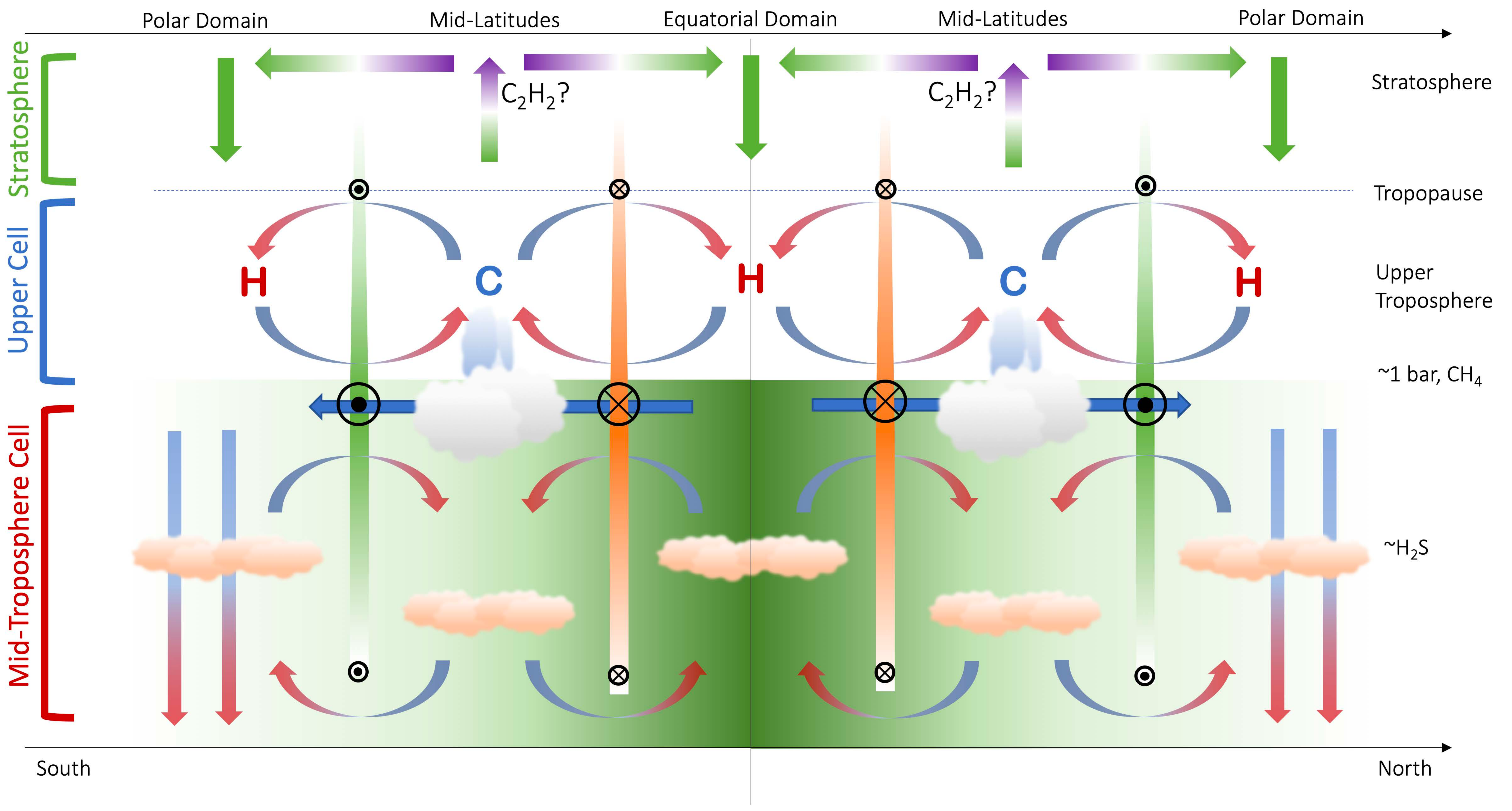}}
\caption{Modified meridional circulation pattern in the troposphere and stratosphere (Fig. \ref{strat_comp}) to split the equatorial retrograde jet, moving its peak winds to be coincident with the locations of maximal thermal windshear.  This is entirely hypothetical, but could explain contrasts in temperature, composition, and albedo observed by different authors.  Furthermore, small-scale structure in the low-latitude zonal winds could indeed be present, but not yet identified in Voyager or Earth-based cloud-tracking observations.}  
\label{splitjet}
\end{centering}
\end{figure*}

Could this be associated with peaks in the retrograde flow that have yet to be resolved in the zonal winds shown in Fig. \ref{data}C?  Could such an unresolved jet be separating the mid-latitude anticyclonic domain (upwelling) from the equatorial cyclonic domain (subsidence)?  We depict this in the schematic in Fig. \ref{splitjet}.   The quality of the available zonal wind data to date \citep[e.g.,][]{18tollefson} do not preclude this idea, and the albedo patterns do show a good deal of zonal banding at low latitudes that could be connected with unresolved zonal jets.  The jets need not be symmetric about the equator, given that the methane and H$_2$S distributions in Section \ref{midtrop} show some asymmetries in terms of upwelling and subsiding regions.  The existence of such off-equatorial zonal wind maxima would be analogous to those on Jupiter (where prograde winds peak at the edges of the Equatorial Zone at $\pm7^\circ$ latitude) and Saturn \citep[e.g., ][]{11garcia}.  Future observations of cloud-tracked winds will be needed to identify whether the equatorial retrograde jet shows a splitting in this manner.

\section{Conclusion:  Where to target a probe?}
\label{conclusions}

Our insights into Ice Giant circulation patterns have progressed slowly over the three decades since the Voyager observations, but enhanced techniques at visible wavelengths (including the Hubble Space Telescope) have tracked clouds and provided spatial distributions of methane; improvements in thermal-infrared observations allow us to characterise upper tropospheric and stratospheric temperatures; and the development of millimetre and centimetre capabilities now allow us to study the distribution of important volatiles (H$_2$S and NH$_3$) in the atmosphere well below the top-most clouds.  Taken together, the distributions of these species present a puzzling, and sometimes conflicting, picture of meridional circulations on Uranus and Neptune.  Indeed, many of our inferences may be biased by the more comprehensive studies of Jupiter and Saturn.  An orbital mission in the coming decades, capable of multi-spectral remote sensing, will be essential to further reveal these circulation patterns.  In addition, an \textit{in situ} atmospheric probe falling to the 10-bar level (or deeper) would provide essential `ground-truth' for the remote measurements, providing vertical profiles of temperature, density, gaseous abundances, para-hydrogen, aerosol properties, and Doppler-tracked horizontal winds \citep{18mousis}.  As we described in Section \ref{intro}, knowledge of the local circulation and meteorology of a probe entry location will be essential for the interpretation of the compositional and atmospheric structure measurements.

The selection of a probe entry location will depend upon a multitude of factors, primarily orbital mechanics and the requirement to have direct communications between a probe and its carrier spacecraft \citep[e.g.,][]{18mousis}.  Nevertheless, the circulation patterns inferred in the previous sections can tell us about the expected conditions in different atmospheric domains (Fig. \ref{strat_comp}-\ref{splitjet}).  Note that in this section, we are discussing how probe data can advance our knowledge of condensable species and atmospheric circulation.  Measurements of noble gases and their isotopes, which are critical to testing planetary formation models, can be made at any latitude without regard to atmospheric patterns and weather due to the non-reactive nature of those species.

\begin{itemize}
    
    \item \textbf{Equatorial Domain from $0^\circ$ to $15-30^\circ$N/S:}  A deep atmosphere enriched in methane and H$_2$S (and possibly other volatiles such as NH$_3$ and H$_2$O) by rising motion from the hundred-bar level or deeper.  This rising motion meets a region of atmospheric subsidence in the upper troposphere, characterised by warm temperatures, super-equilibrium para-H$_2$ fractions, and small-scale modulations of the CH$_4$, H$_2$S, and aerosol albedo.  A region of vertical convergence must exist somewhere in the CH$_4$- and H$_2$S cloud forming regions (1-4 bar), with air moving meridionally away from the equator towards mid-latitudes.  The meridional temperature gradient for $p>1$ bar provides a windshear that makes retrograde winds stronger with altitude, whereas the temperature gradient for $p<1$ bar makes the retrograde winds decay with height in the upper troposphere.  Above the tropopause we may transition into a region dominated by wave-driven circulation, where recent work finds suppressed stratospheric emission associated with either low-latitude upwelling and adiabatic cooling, or with an absence of stratospheric hydrocarbons \citep{19roman}.  This multi-tier structure, with vertical convergence in the cloud-forming region leading to meridional divergence and poleward flow, might be analogous to cyclonic jovian belts: regions of net subsidence in the upper troposphere, but net upwelling (and associated lightning from moist convection) in the deeper troposphere. The equatorial domain might be bounded by off-equatorial maxima in the retrograde zonal jet, which are currently unresolved in the available cloud tracking measurements.  A probe descending right at the equator might encounter local minima in volatile species in the upper troposphere, but will ultimately reach the deeper cell where the abundances of CH$_4$ and H$_2$S are at their maximum.  Local meteorological features, such as Uranus' equatorial wave phenomena \citep{15sromovsky}, should be avoided to mitigate the risks of encountering strong downdrafts like those encountered by the Galileo probe \citep{98orton}.  
    
    \item \textbf{Mid-Latitude Domain from $15-30^\circ$N/S to $60-75^\circ$N/S:}  A transitional domain where the abundance of CH$_4$ and H$_2$S declines dramatically with increasing latitude due to increasingly strong deep-atmosphere subsidence.  This domain is situated between the low-latitude retrograde jet and the high-latitude prograde jet, both of which decay with increasing depth below the clouds, and decay with increasing altitude above the clouds.  Cold temperatures, sub-equilibrium para-H$_2$ fractions, sporadic convective storm activity, and `polar collars` of aerosols all imply upwelling motions in the upper troposphere.  This suggests vertical divergence somewhere near the cloud level (between upper tropospheric upwelling, and deep tropospheric subsidence), and thus meridional convergence of the air from the equatorial and polar domains.  In this regard, the Ice Giant mid-latitudes are analogous to anticyclonic jovian zones, which are regions of net rising in the upper troposphere.  Jovian zones also exhibit equatorward flow at cloud level across their bounding prograde jets (due to the meridional circulation balancing eddy momentum flux converging on the prograde jets), although such motions have not been observed on the Ice Giants.  Probes entering these mid-latitude domains might encounter localised storm activity in the upper-tropospheric upwelling, enhanced optical depths of clouds and hazes, but negligible vertical shear on the zonal winds.
    
    \item \textbf{Polar domain poleward of $60-75^\circ$N/S:}  The polar domains, bounded by the high-latitude prograde jets, are most depleted in volatiles CH$_4$, H$_2$S, NH$_3$, and perhaps H$_2$O due to strong atmospheric subsidence.  Thus a probe entering this location might only be capable of returning upper limits on key elemental abundances and isotopic ratios (we note that noble gas measurements, however, can be made in this region).  However, a polar probe would also sample a unique region where small-scale convective activity persists in a region of net subsidence, potentially leading to enhanced humidity of H$_2$S immediately above the clouds.  This small-scale convective activity is reminiscent of the plumes and lightning observed in Jupiter's cyclonic belts, and maybe the puffy clouds observed at high latitudes on Saturn.  It would also sample warm polar vortices evident in the troposphere and stratosphere (although more spatially confined to the pole on Neptune than on Uranus), potentially accessing unique chemical domains not found elsewhere on the planet.  
    
\end{itemize}

Given the primary goal of an entry probe would be the measurement of bulk chemical enrichment and isotopic ratios, targeting the peak abundances at low latitudes would be a sensible first step.  However, avoidance of low-latitude meteorological features is key to sampling a `representative' region of an Ice Giant, so an off-equatorial entry site might be optimal, before encountering the storm bands and strong upwelling of the mid-latitudes.  A secondary probe, if available, could target the polar domain to provide an extreme counterpoint for the low-latitude measurement.  Such a comparison would provide the much-needed ground truth for remote-sensing investigations, able to sample the vertical distributions of temperatures, clouds, aerosols and gaseous species with a far greater vertical resolution than could ever be achieved from orbit.  

Spatial variations in temperatures, clouds, and composition are extremely challenging to monitor from 20 or 30 AU away, even with future facilities like the 30-40-m class observatories or the James Webb Space Telescope.  Multi-spectral remote sensing of the probe entry site will be invaluable to diagnose the probe results, including (i) tracking of winds and cloud features; (ii) UV-visible-near-infrared spectroscopy to assess aerosol distributions with height; (iii) thermal emission from the infrared to the sub-millimetre to determine temperatures and upwelling/subsiding motions; and (iv) microwave observations to connect the cloud-top meteorology to circulation patterns at great depth.  Carefully planned, but ultimately serendipitous, ground-based observations were able to provide some of this information for the Galileo probe \citep{98orton}, but any probe mission should carefully consider having these capabilities on an associated orbiter.

In seeking to consider the observational evidence for atmospheric circulation on the Ice Giants, we have hypothesised a multi-tiered structure of stacked circulation cells, with motions that are potentially in opposition to one another.  Can reality really be this complex?  The case for stacked cells has been postulated on Jupiter and Saturn for some time \citep{00ingersoll, 05showman, 11fletcher_vims}, and for Uranus \citep{14sromovsky} and Neptune \citep{18tollefson}, and was reviewed by \citet{19fletcher_beltzone}.  The terrestrial atmosphere exhibits a transition from the troposphere (thermally-direct Hadley cells and mid-latitude heat transport by eddies) to the stratosphere (wave-driven thermally indirect circulation), with an associated reversal in the sense of the annual mean temperature gradient that might be considered as a multi-tiered circulation structure.  The enrichments in volatiles in the Ice Giant atmospheres may generate substantial density stratifications where they condense, leading to circulation regimes that are only weakly connected to one another and a natural transition point between the stacked cells.  However, there is an absence of numerical simulations of Ice Giant atmospheres (which must crucially include condensable species as active agents in the circulation) against which such inferences can be tested.  We hope that any numerical simulations developed in the coming decade take all the observations presented in Fig. \ref{data} into account as a test of their credibility, as well as future proposed observations from next-generation ground- and space-based observatories.  This may well lead to a comprehensive rejection of the qualitative and complex multi-tiered circulations in Fig. \ref{splitjet}, but would be invaluable in expanding and maturing our understanding of Ice Giant circulation patterns, and to guide our targeting of humankind's first atmospheric probe for these distant and enigmatic worlds.

\begin{acknowledgements}
Fletcher was supported by a Royal Society Research Fellowship at the University of Leicester.  Fletcher and Roman also received funding from a European Research Council Consolidator Grant (under the European Union's Horizon 2020 research and innovation programme, grant agreement No 723890).  A portion of this work was performed by Orton at the Jet Propulsion Laboratory, California Institute of Technology, under a contract with NASA.  de Pater was in part supported by the National Science Foundation, NSF grant AST-1615004 and NASA Grant NNX16AK14G through the Solar System Observations (SSO) program to the University of California, Berkeley.  Irwin and Toledo acknowledge support from the UK Science and Technologies Facilities Council (STFC).  We are grateful for support from the \textit{In Situ Exploration of the Ice Giants: Science and Technology} workshop held in February 2019 at the Laboratoire d'Astrophysique de Marseille for inspiring this review.

\end{acknowledgements}

\bibliographystyle{aps-nameyear}      
\bibliography{references.bib}                

\begin{thebibliography}{183}
\ifx \bisbn   \undefined \def \bisbn  #1{ISBN #1}\fi
\ifx \binits  \undefined \def \binits#1{#1} \fi
\ifx \bauthor  \undefined \def \bauthor#1{#1} \fi
\ifx \bjtitle  \undefined \def \bjtitle#1{\textrm{#1}}\fi
\ifx \batitle  \undefined \def \batitle#1{#1} \fi
\ifx \bctitle  \undefined \def \bctitle#1{#1} \fi
\ifx \bvolume  \undefined \def \bvolume#1{\textbf{#1}}\fi
\ifx \byear  \undefined \def \byear#1{#1} \fi
\ifx \bissue  \undefined \def \bissue#1{#1} \fi
\ifx \bfpage  \undefined \def \bfpage#1{#1} \fi
\ifx \blpage  \undefined \def \blpage #1{#1} \fi
\ifx \burl  \undefined \def \burl#1{#1} \fi
\ifx \doiurl  \undefined \def \doiurl#1{#1} \fi
\ifx \betal  \undefined \def \betal{et al.} \fi
\ifx \binstitute  \undefined \def \binstitute#1{#1} \fi
\ifx \beditor  \undefined \def \beditor#1{#1} \fi
\ifx \bpublisher  \undefined \def \bpublisher#1{#1} \fi
\ifx \bbtitle  \undefined \def \bbtitle#1{\textit{#1}} \fi
\ifx \bedition  \undefined \def \bedition#1{#1} \fi
\ifx \bseriesno  \undefined \def \bseriesno#1{#1} \fi
\ifx \blocation  \undefined \def \blocation#1{#1} \fi
\ifx \bsertitle  \undefined \def \bsertitle#1{#1} \fi
\ifx \bsnm \undefined \def \bsnm#1{#1} \fi
\ifx \bsuffix \undefined \def \bsuffix#1{#1} \fi
\ifx \bparticle \undefined \def \bparticle#1{#1} \fi
\ifx \barticle \undefined \def \barticle#1{#1} \fi
\ifx \botherref \undefined \def \botherref #1{#1} \fi
\ifx \url \undefined \def \url#1{#1} \fi
\ifx \bchapter \undefined \def \bchapter#1{#1} \fi
\ifx \bbook \undefined \def \bbook#1{#1} \fi
\ifx \bcomment \undefined \def \bcomment#1{#1} \fi
\ifx \oauthor \undefined \def \oauthor#1{#1} \fi
\ifx \citeauthoryear \undefined \def \citeauthoryear#1{#1} \fi
\ifx \texttildelow  \undefined \def \texttildelow{\symbol{126}} \fi
\def \endbibitem {}
\ifx \bconflocation  \undefined \def \bconflocation#1{#1} \fi

\bibitem[\protect\citeauthoryear{{Achterberg} et~al.}{2006}]{06achterberg}
\begin{barticle}
\bauthor{\binits{R.K.} \bsnm{{Achterberg}}},
\bauthor{\binits{B.J.} \bsnm{{Conrath}}},
\bauthor{\binits{P.J.} \bsnm{{Gierasch}}},
\batitle{{Cassini CIRS retrievals of Ammonia in Jupiter's Upper Troposphere}}.
\bjtitle{Icarus}
\bvolume{182},
\bfpage{169}--\blpage{180}
(\byear{2006}).
doi:\doiurl{10.1016/j.icarus.2005.12.020}
\end{barticle}
\endbibitem

\bibitem[\protect\citeauthoryear{{Andrews} et~al.}{1987}]{87andrews}
\begin{bbook}
\bauthor{\binits{D.G.} \bsnm{{Andrews}}},
\bauthor{\binits{J.R.} \bsnm{{Holton}}},
\bauthor{\binits{C.B.} \bsnm{{Leovy}}},
\bbtitle{{Middle atmosphere dynamics}}
(\bpublisher{Academic Press, New York}, \blocation{???}, \byear{1987})
\end{bbook}
\endbibitem

\bibitem[\protect\citeauthoryear{{Aplin} and {Harrison}}{2016}]{16aplin}
\begin{barticle}
\bauthor{\binits{K.L.} \bsnm{{Aplin}}},
\bauthor{\binits{R.G.} \bsnm{{Harrison}}},
\batitle{{Determining solar effects in Neptune's atmosphere}}.
\bjtitle{Nature Communications}
\bvolume{7},
\bfpage{11976}
(\byear{2016}).
doi:\doiurl{10.1038/ncomms11976}
\end{barticle}
\endbibitem

\bibitem[\protect\citeauthoryear{{Atreya} and {Ponthieu}}{1983}]{83atreya}
\begin{barticle}
\bauthor{\binits{S.K.} \bsnm{{Atreya}}},
\bauthor{\binits{J.J.} \bsnm{{Ponthieu}}},
\batitle{{Photolysis of methane and the ionosphere of Uranus}}.
\bjtitle{Plan. \& Space Sci.}
\bvolume{31},
\bfpage{939}--\blpage{944}
(\byear{1983}).
doi:\doiurl{10.1016/0032-0633(83)90149-6}
\end{barticle}
\endbibitem

\bibitem[\protect\citeauthoryear{{Atreya} and {Romani}}{1985}]{85atreya}
\begin{botherref}
\oauthor{\binits{S.K.} \bsnm{{Atreya}}},
\oauthor{\binits{P.N.} \bsnm{{Romani}}},
{Photochemistry and clouds of Jupiter, Saturn and Uranus.},
ed. by G.E. {Hunt}
1985,
pp. 17--68
\end{botherref}
\endbibitem

\bibitem[\protect\citeauthoryear{{Atreya} and {Wong}}{2005}]{05atreya}
\begin{barticle}
\bauthor{\binits{S.K.} \bsnm{{Atreya}}},
\bauthor{\binits{A.-S.} \bsnm{{Wong}}},
\batitle{{Coupled Clouds and Chemistry of the Giant Planets -- A Case for
  Multiprobes}}.
\bjtitle{Space Sci. Rev.}
\bvolume{116},
\bfpage{121}--\blpage{136}
(\byear{2005}).
doi:\doiurl{10.1007/s11214-005-1951-5}
\end{barticle}
\endbibitem

\bibitem[\protect\citeauthoryear{{Atreya} et~al.}{1991}]{91atreya}
\begin{botherref}
\oauthor{\binits{S.K.} \bsnm{{Atreya}}},
\oauthor{\binits{B.R.} \bsnm{{Sandel}}},
\oauthor{\binits{P.N.} \bsnm{{Romani}}},
{Photochemistry and vertical mixing},
ed. by J.T. {Bergstralh}, E.D. {Miner}, M.S. {Matthews}
(Arizona Press, 1991),
pp. 110--146
\end{botherref}
\endbibitem

\bibitem[\protect\citeauthoryear{{Baines} and {Hammel}}{1994}]{94baines}
\begin{barticle}
\bauthor{\binits{K.H.} \bsnm{{Baines}}},
\bauthor{\binits{H.B.} \bsnm{{Hammel}}},
\batitle{{Clouds, hazes, and the stratospheric methane abundance in Neptune}}.
\bjtitle{Icarus}
\bvolume{109},
\bfpage{20}--\blpage{39}
(\byear{1994}).
doi:\doiurl{10.1006/icar.1994.1075}
\end{barticle}
\endbibitem

\bibitem[\protect\citeauthoryear{{B{\'e}zard} et~al.}{1991}]{91bezard}
\begin{barticle}
\bauthor{\binits{B.} \bsnm{{B{\'e}zard}}},
\bauthor{\binits{P.N.} \bsnm{{Romani}}},
\bauthor{\binits{B.J.} \bsnm{{Conrath}}},
\bauthor{\binits{W.C.} \bsnm{{Maguire}}},
\batitle{{Hydrocarbons in Neptune's stratosphere from Voyager infrared
  observations}}.
\bjtitle{J. Geophys. Res.}
\bvolume{96},
\bfpage{18961}
(\byear{1991})
\end{barticle}
\endbibitem

\bibitem[\protect\citeauthoryear{{Bishop} et~al.}{1990}]{90bishop}
\begin{barticle}
\bauthor{\binits{J.} \bsnm{{Bishop}}},
\bauthor{\binits{S.K.} \bsnm{{Atreya}}},
\bauthor{\binits{F.} \bsnm{{Herbert}}},
\bauthor{\binits{P.} \bsnm{{Romani}}},
\batitle{{Reanalysis of Voyager 2 UVS occultations at Uranus - Hydrocarbon
  mixing ratios in the equatorial stratosphere}}.
\bjtitle{Icarus}
\bvolume{88},
\bfpage{448}--\blpage{464}
(\byear{1990}).
doi:\doiurl{10.1016/0019-1035(90)90094-P}
\end{barticle}
\endbibitem

\bibitem[\protect\citeauthoryear{{Broadfoot} et~al.}{1989}]{89broadfoot}
\begin{barticle}
\bauthor{\binits{A.L.} \bsnm{{Broadfoot}}},
\bauthor{\binits{S.K.} \bsnm{{Atreya}}},
\bauthor{\binits{J.L.} \bsnm{{Bertaux}}},
\bauthor{\binits{J.E.} \bsnm{{Blamont}}},
\bauthor{\binits{A.J.} \bsnm{{Dessler}}},
\bauthor{\binits{T.M.} \bsnm{{Donahue}}},
\bauthor{\binits{W.T.} \bsnm{{Forrester}}},
\bauthor{\binits{D.T.} \bsnm{{Hall}}},
\bauthor{\binits{F.} \bsnm{{Herbert}}},
\bauthor{\binits{J.B.} \bsnm{{Holberg}}},
\batitle{{Ultraviolet Spectrometer Observations of Neptune and Triton}}.
\bjtitle{Science}
\bvolume{246}(\bissue{4936}),
\bfpage{1459}--\blpage{1466}
(\byear{1989}).
doi:\doiurl{10.1126/science.246.4936.1459}
\end{barticle}
\endbibitem

\bibitem[\protect\citeauthoryear{{Butler} et~al.}{2012}]{12butler_dps}
\begin{bchapter}
\bauthor{\binits{B.J.} \bsnm{{Butler}}},
\bauthor{\binits{M.} \bsnm{{Hofstadter}}},
\bauthor{\binits{M.} \bsnm{{Gurwell}}},
\bauthor{\binits{G.} \bsnm{{Orton}}},
\bauthor{\binits{J.} \bsnm{{Norwood}}},
\bctitle{{The Deep Atmosphere of Neptune From EVLA Observations}},
in \bbtitle{AAS/Division for Planetary Sciences Meeting Abstracts \#44}.
\bsertitle{AAS/Division for Planetary Sciences Meeting Abstracts},
\byear{2012},
pp. \bfpage{504}--\blpage{06}
\end{bchapter}
\endbibitem

\bibitem[\protect\citeauthoryear{{Cavali{\'e}} et~al.}{2014}]{14cavalie}
\begin{barticle}
\bauthor{\binits{T.} \bsnm{{Cavali{\'e}}}},
\bauthor{\binits{R.} \bsnm{{Moreno}}},
\bauthor{\binits{E.} \bsnm{{Lellouch}}},
\bauthor{\binits{P.} \bsnm{{Hartogh}}},
\bauthor{\binits{O.} \bsnm{{Venot}}},
\bauthor{\binits{G.S.} \bsnm{{Orton}}},
\bauthor{\binits{C.} \bsnm{{Jarchow}}},
\bauthor{\binits{T.} \bsnm{{Encrenaz}}},
\bauthor{\binits{F.} \bsnm{{Selsis}}},
\bauthor{\binits{F.} \bsnm{{Hersant}}},
\bauthor{\binits{L.N.} \bsnm{{Fletcher}}},
\batitle{{The first submillimeter observation of CO in the stratosphere of
  Uranus}}.
\bjtitle{Astron. Astrophys}
\bvolume{562},
\bfpage{33}
(\byear{2014}).
doi:\doiurl{10.1051/0004-6361/201322297}
\end{barticle}
\endbibitem

\bibitem[\protect\citeauthoryear{{Cavali{\'e}} et~al.}{2017}]{17cavalie}
\begin{barticle}
\bauthor{\binits{T.} \bsnm{{Cavali{\'e}}}},
\bauthor{\binits{O.} \bsnm{{Venot}}},
\bauthor{\binits{F.} \bsnm{{Selsis}}},
\bauthor{\binits{F.} \bsnm{{Hersant}}},
\bauthor{\binits{P.} \bsnm{{Hartogh}}},
\bauthor{\binits{J.} \bsnm{{Leconte}}},
\batitle{{Thermochemistry and vertical mixing in the tropospheres of Uranus and
  Neptune: How convection inhibition can affect the derivation of deep oxygen
  abundances}}.
\bjtitle{Icarus}
\bvolume{291},
\bfpage{1}--\blpage{16}
(\byear{2017}).
doi:\doiurl{10.1016/j.icarus.2017.03.015}
\end{barticle}
\endbibitem

\bibitem[\protect\citeauthoryear{{Conrath} et~al.}{1991}]{91conrath}
\begin{barticle}
\bauthor{\binits{B.J.} \bsnm{{Conrath}}},
\bauthor{\binits{F.M.} \bsnm{{Flasar}}},
\bauthor{\binits{P.J.} \bsnm{{Gierasch}}},
\batitle{{Thermal structure and dynamics of Neptune's atmosphere from Voyager
  measurements}}.
\bjtitle{J. Geophys. Res.}
\bvolume{96},
\bfpage{18931}
(\byear{1991})
\end{barticle}
\endbibitem

\bibitem[\protect\citeauthoryear{{Conrath} et~al.}{1990}]{90conrath}
\begin{barticle}
\bauthor{\binits{B.J.} \bsnm{{Conrath}}},
\bauthor{\binits{P.J.} \bsnm{{Gierasch}}},
\bauthor{\binits{S.S.} \bsnm{{Leroy}}},
\batitle{{Temperature and circulation in the stratosphere of the outer
  planets}}.
\bjtitle{Icarus}
\bvolume{83},
\bfpage{255}--\blpage{281}
(\byear{1990}).
doi:\doiurl{10.1016/0019-1035(90)90068-K}
\end{barticle}
\endbibitem

\bibitem[\protect\citeauthoryear{{Conrath} et~al.}{1998}]{98conrath}
\begin{barticle}
\bauthor{\binits{B.J.} \bsnm{{Conrath}}},
\bauthor{\binits{P.J.} \bsnm{{Gierasch}}},
\bauthor{\binits{E.A.} \bsnm{{Ustinov}}},
\batitle{{Thermal Structure and Para Hydrogen Fraction on the Outer Planets
  from Voyager IRIS Measurements}}.
\bjtitle{Icarus}
\bvolume{135},
\bfpage{501}--\blpage{517}
(\byear{1998}).
doi:\doiurl{10.1006/icar.1998.6000}
\end{barticle}
\endbibitem

\bibitem[\protect\citeauthoryear{{Conrath} et~al.}{1989}]{89conrath}
\begin{barticle}
\bauthor{\binits{B.} \bsnm{{Conrath}}},
\bauthor{\binits{F.M.} \bsnm{{Flasar}}},
\bauthor{\binits{R.} \bsnm{{Hanel}}},
\bauthor{\binits{V.} \bsnm{{Kunde}}},
\bauthor{\binits{W.} \bsnm{{Maguire}}},
\bauthor{\binits{J.} \bsnm{{Pearl}}},
\bauthor{\binits{J.} \bsnm{{Pirraglia}}},
\bauthor{\binits{R.} \bsnm{{Samuelson}}},
\bauthor{\binits{D.} \bsnm{{Cruikshank}}},
\bauthor{\binits{L.} \bsnm{{Horn}}},
\batitle{{Infrared observations of the Neptunian system}}.
\bjtitle{Science}
\bvolume{246},
\bfpage{1454}--\blpage{1459}
(\byear{1989}).
doi:\doiurl{10.1126/science.246.4936.1454}
\end{barticle}
\endbibitem

\bibitem[\protect\citeauthoryear{{de Kleer} et~al.}{2015}]{15dekleer}
\begin{barticle}
\bauthor{\binits{K.} \bsnm{{de Kleer}}},
\bauthor{\binits{S.} \bsnm{{Luszcz-Cook}}},
\bauthor{\binits{I.} \bsnm{{de Pater}}},
\bauthor{\binits{M.} \bsnm{{{\'A}d{\'a}mkovics}}},
\bauthor{\binits{H.B.} \bsnm{{Hammel}}},
\batitle{{Clouds and aerosols on Uranus: Radiative transfer modeling of
  spatially-resolved near-infrared Keck spectra}}.
\bjtitle{Icarus}
\bvolume{256},
\bfpage{120}--\blpage{137}
(\byear{2015}).
doi:\doiurl{10.1016/j.icarus.2015.04.021}
\end{barticle}
\endbibitem

\bibitem[\protect\citeauthoryear{{de Pater}}{1991}]{91depater_radio}
\begin{botherref}
\oauthor{\binits{I.} \bsnm{{de Pater}}},
{The Significance of Radio Observations for Planets}.
{Physics Reports},
1--50
(1991)
\end{botherref}
\endbibitem

\bibitem[\protect\citeauthoryear{{de Pater} and {Gulkis}}{1988}]{88depater}
\begin{barticle}
\bauthor{\binits{I.} \bsnm{{de Pater}}},
\bauthor{\binits{S.} \bsnm{{Gulkis}}},
\batitle{{VLA observations of Uranus at 1.3-20 CM}}.
\bjtitle{Icarus}
\bvolume{75},
\bfpage{306}--\blpage{323}
(\byear{1988}).
doi:\doiurl{10.1016/0019-1035(88)90007-3}
\end{barticle}
\endbibitem

\bibitem[\protect\citeauthoryear{{de Pater} et~al.}{1991}]{91depater}
\begin{barticle}
\bauthor{\binits{I.} \bsnm{{de Pater}}},
\bauthor{\binits{P.N.} \bsnm{{Romani}}},
\bauthor{\binits{S.K.} \bsnm{{Atreya}}},
\batitle{{Possible microwave absorption by H2S gas in Uranus' and Neptune's
  atmospheres}}.
\bjtitle{Icarus}
\bvolume{91},
\bfpage{220}--\blpage{233}
(\byear{1991}).
doi:\doiurl{10.1016/0019-1035(91)90020-T}
\end{barticle}
\endbibitem

\bibitem[\protect\citeauthoryear{{de Pater} et~al.}{2011}]{11depater}
\begin{barticle}
\bauthor{\binits{I.} \bsnm{{de Pater}}},
\bauthor{\binits{L.A.} \bsnm{{Sromovsky}}},
\bauthor{\binits{H.B.} \bsnm{{Hammel}}},
\bauthor{\binits{P.M.} \bsnm{{Fry}}},
\bauthor{\binits{R.P.} \bsnm{{LeBeau}}},
\bauthor{\binits{K.} \bsnm{{Rages}}},
\bauthor{\binits{M.} \bsnm{{Showalter}}},
\bauthor{\binits{K.} \bsnm{{Matthews}}},
\batitle{{Post-equinox observations of Uranus: Berg's evolution, vertical
  structure, and track towards the equator}}.
\bjtitle{Icarus}
\bvolume{215},
\bfpage{332}--\blpage{345}
(\byear{2011}).
doi:\doiurl{10.1016/j.icarus.2011.06.022}
\end{barticle}
\endbibitem

\bibitem[\protect\citeauthoryear{{de Pater} et~al.}{2014}]{14depater}
\begin{barticle}
\bauthor{\binits{I.} \bsnm{{de Pater}}},
\bauthor{\binits{L.N.} \bsnm{{Fletcher}}},
\bauthor{\binits{S.} \bsnm{{Luszcz-Cook}}},
\bauthor{\binits{D.} \bsnm{{DeBoer}}},
\bauthor{\binits{B.} \bsnm{{Butler}}},
\bauthor{\binits{H.B.} \bsnm{{Hammel}}},
\bauthor{\binits{M.L.} \bsnm{{Sitko}}},
\bauthor{\binits{G.} \bsnm{{Orton}}},
\bauthor{\binits{P.S.} \bsnm{{Marcus}}},
\batitle{{Neptune's global circulation deduced from multi-wavelength
  observations}}.
\bjtitle{Icarus}
\bvolume{237},
\bfpage{211}--\blpage{238}
(\byear{2014}).
doi:\doiurl{10.1016/j.icarus.2014.02.030}
\end{barticle}
\endbibitem

\bibitem[\protect\citeauthoryear{{de Pater} et~al.}{2015}]{15depater}
\begin{barticle}
\bauthor{\binits{I.} \bsnm{{de Pater}}},
\bauthor{\binits{L.A.} \bsnm{{Sromovsky}}},
\bauthor{\binits{P.M.} \bsnm{{Fry}}},
\bauthor{\binits{H.B.} \bsnm{{Hammel}}},
\bauthor{\binits{C.} \bsnm{{Baranec}}},
\bauthor{\binits{K.M.} \bsnm{{Sayanagi}}},
\batitle{{Record-breaking storm activity on Uranus in 2014}}.
\bjtitle{Icarus}
\bvolume{252},
\bfpage{121}--\blpage{128}
(\byear{2015}).
doi:\doiurl{10.1016/j.icarus.2014.12.037}
\end{barticle}
\endbibitem

\bibitem[\protect\citeauthoryear{{de Pater} et~al.}{2016}]{16depater}
\begin{barticle}
\bauthor{\binits{I.} \bsnm{{de Pater}}},
\bauthor{\binits{R.J.} \bsnm{{Sault}}},
\bauthor{\binits{B.} \bsnm{{Butler}}},
\bauthor{\binits{D.} \bsnm{{DeBoer}}},
\bauthor{\binits{M.H.} \bsnm{{Wong}}},
\batitle{{Peering through Jupiter's clouds with radio spectral imaging}}.
\bjtitle{Science}
\bvolume{352},
\bfpage{1198}--\blpage{1201}
(\byear{2016}).
doi:\doiurl{10.1126/science.aaf2210}
\end{barticle}
\endbibitem

\bibitem[\protect\citeauthoryear{{de Pater} et~al.}{2018}]{18depater}
\begin{barticle}
\bauthor{\binits{I.} \bsnm{{de Pater}}},
\bauthor{\binits{B.} \bsnm{{Butler}}},
\bauthor{\binits{R.J.} \bsnm{{Sault}}},
\bauthor{\binits{A.} \bsnm{{Moullet}}},
\bauthor{\binits{C.} \bsnm{{Moeckel}}},
\bauthor{\binits{J.} \bsnm{{Tollefson}}},
\bauthor{\binits{K.} \bsnm{{de Kleer}}},
\bauthor{\binits{M.} \bsnm{{Gurwell}}},
\bauthor{\binits{S.} \bsnm{{Milam}}},
\batitle{{Potential for Solar System Science with the ngVLA}}.
\bjtitle{Science with a Next Generation Very Large Array}
\bvolume{517},
\bfpage{49}
(\byear{2018})
\end{barticle}
\endbibitem

\bibitem[\protect\citeauthoryear{{DeBoer} and {Steffes}}{1996}]{96deboer}
\begin{barticle}
\bauthor{\binits{D.R.} \bsnm{{DeBoer}}},
\bauthor{\binits{P.G.} \bsnm{{Steffes}}},
\batitle{{Estimates of the Tropospheric Vertical Structure of Neptune Based on
  Microwave Radiative Transfer Studies}}.
\bjtitle{Icarus}
\bvolume{123}(\bissue{2}),
\bfpage{324}--\blpage{335}
(\byear{1996}).
doi:\doiurl{10.1006/icar.1996.0161}
\end{barticle}
\endbibitem

\bibitem[\protect\citeauthoryear{{Del Genio} and {Barbara}}{2012}]{12delgenio}
\begin{barticle}
\bauthor{\binits{A.D.} \bsnm{{Del Genio}}},
\bauthor{\binits{J.M.} \bsnm{{Barbara}}},
\batitle{{Constraints on Saturn's tropospheric general circulation from Cassini
  ISS images}}.
\bjtitle{Icarus}
\bvolume{219},
\bfpage{689}--\blpage{700}
(\byear{2012}).
doi:\doiurl{10.1016/j.icarus.2012.03.035}
\end{barticle}
\endbibitem

\bibitem[\protect\citeauthoryear{{Dobrijevic} et~al.}{2010}]{10dobrijevic}
\begin{barticle}
\bauthor{\binits{M.} \bsnm{{Dobrijevic}}},
\bauthor{\binits{T.} \bsnm{{Cavali{\'e}}}},
\bauthor{\binits{E.} \bsnm{{H{\'e}brard}}},
\bauthor{\binits{F.} \bsnm{{Billebaud}}},
\bauthor{\binits{F.} \bsnm{{Hersant}}},
\bauthor{\binits{F.} \bsnm{{Selsis}}},
\batitle{{Key reactions in the photochemistry of hydrocarbons in Neptune's
  stratosphere}}.
\bjtitle{Plan. \& Space Sci.}
\bvolume{58},
\bfpage{1555}--\blpage{1566}
(\byear{2010}).
doi:\doiurl{10.1016/j.pss.2010.07.024}
\end{barticle}
\endbibitem

\bibitem[\protect\citeauthoryear{{Encrenaz} et~al.}{1998}]{98encrenaz}
\begin{barticle}
\bauthor{\binits{T.} \bsnm{{Encrenaz}}},
\bauthor{\binits{H.} \bsnm{{Feuchtgruber}}},
\bauthor{\binits{S.K.} \bsnm{{Atreya}}},
\bauthor{\binits{B.} \bsnm{{Bezard}}},
\bauthor{\binits{E.} \bsnm{{Lellouch}}},
\bauthor{\binits{J.} \bsnm{{Bishop}}},
\bauthor{\binits{S.} \bsnm{{Edgington}}},
\bauthor{\binits{T.} \bsnm{{Degraauw}}},
\bauthor{\binits{M.} \bsnm{{Griffin}}},
\bauthor{\binits{M.F.} \bsnm{{Kessler}}},
\batitle{{ISO observations of Uranus: The stratospheric distribution of
  C\_2H\_2 and the eddy diffusion coefficient}}.
\bjtitle{Astron. Astrophys}
\bvolume{333},
\bfpage{43}--\blpage{46}
(\byear{1998})
\end{barticle}
\endbibitem

\bibitem[\protect\citeauthoryear{{Fegley} and {Prinn}}{1985}]{85fegley}
\begin{barticle}
\bauthor{\binits{B.} \bsnm{{Fegley}}},
\bauthor{\binits{R.G.} \bsnm{{Prinn}}},
\batitle{{Equilibrium and nonequilibrium chemistry of Saturn's atmosphere -
  Implications for the observability of PH$_3$, N$_2$, CO, and GeH$_4$}}.
\bjtitle{Astrophys. J.}
\bvolume{299},
\bfpage{1067}--\blpage{1078}
(\byear{1985}).
doi:\doiurl{10.1086/163775}
\end{barticle}
\endbibitem

\bibitem[\protect\citeauthoryear{{Feuchtgruber} et~al.}{1997}]{97feuchtgruber}
\begin{barticle}
\bauthor{\binits{H.} \bsnm{{Feuchtgruber}}},
\bauthor{\binits{E.} \bsnm{{Lellouch}}},
\bauthor{\binits{T.} \bsnm{{de Graauw}}},
\bauthor{\binits{B.} \bsnm{{Bezard}}},
\bauthor{\binits{T.} \bsnm{{Encrenaz}}},
\bauthor{\binits{M.} \bsnm{{Griffin}}},
\batitle{{External supply of oxygen to the atmospheres of giant planets}}.
\bjtitle{Nature}
\bvolume{389},
\bfpage{159}--\blpage{162}
(\byear{1997})
\end{barticle}
\endbibitem

\bibitem[\protect\citeauthoryear{{Fitzpatrick} et~al.}{2014}]{14fitzpatrick}
\begin{barticle}
\bauthor{\binits{P.J.} \bsnm{{Fitzpatrick}}},
\bauthor{\binits{I.} \bsnm{{de Pater}}},
\bauthor{\binits{S.} \bsnm{{Luszcz-Cook}}},
\bauthor{\binits{M.H.} \bsnm{{Wong}}},
\bauthor{\binits{H.B.} \bsnm{{Hammel}}},
\batitle{{Dispersion in Neptune's zonal wind velocities from NIR Keck AO
  observations in July 2009}}.
\bjtitle{\apss}
\bvolume{350}(\bissue{1}),
\bfpage{65}--\blpage{88}
(\byear{2014}).
doi:\doiurl{10.1007/s10509-013-1737-2}
\end{barticle}
\endbibitem

\bibitem[\protect\citeauthoryear{{Flasar} et~al.}{1987}]{87flasar}
\begin{barticle}
\bauthor{\binits{F.M.} \bsnm{{Flasar}}},
\bauthor{\binits{B.J.} \bsnm{{Conrath}}},
\bauthor{\binits{J.A.} \bsnm{{Pirraglia}}},
\bauthor{\binits{P.J.} \bsnm{{Gierasch}}},
\batitle{{Voyager infrared observations of Uranus' atmosphere - Thermal
  structure and dynamics}}.
\bjtitle{Journal of Geophysical Research}
\bvolume{92},
\bfpage{15011}--\blpage{15018}
(\byear{1987}).
doi:\doiurl{10.1029/JA092iA13p15011}
\end{barticle}
\endbibitem

\bibitem[\protect\citeauthoryear{{Fletcher} et~al.}{2018}]{18fletcher_cia}
\begin{barticle}
\bauthor{\binits{L.N.} \bsnm{{Fletcher}}},
\bauthor{\binits{M.} \bsnm{{Gustafsson}}},
\bauthor{\binits{G.S.} \bsnm{{Orton}}},
\batitle{{Hydrogen Dimers in Giant-planet Infrared Spectra}}.
\bjtitle{{Astrophys. J. Supplement}}
\bvolume{235},
\bfpage{24}
(\byear{2018}).
doi:\doiurl{10.3847/1538-4365/aaa07a}
\end{barticle}
\endbibitem

\bibitem[\protect\citeauthoryear{{Fletcher} et~al.}{2010}]{10fletcher_akari}
\begin{barticle}
\bauthor{\binits{L.N.} \bsnm{{Fletcher}}},
\bauthor{\binits{P.} \bsnm{{Drossart}}},
\bauthor{\binits{M.} \bsnm{{Burgdorf}}},
\bauthor{\binits{G.S.} \bsnm{{Orton}}},
\bauthor{\binits{T.} \bsnm{{Encrenaz}}},
\batitle{{Neptune's atmospheric composition from AKARI infrared spectroscopy}}.
\bjtitle{Astron. Astrophys.}
\bvolume{514},
\bfpage{17}
(\byear{2010}).
doi:\doiurl{10.1051/0004-6361/200913358}
\end{barticle}
\endbibitem

\bibitem[\protect\citeauthoryear{{Fletcher} et~al.}{2011}]{11fletcher_vims}
\begin{barticle}
\bauthor{\binits{L.N.} \bsnm{{Fletcher}}},
\bauthor{\binits{K.H.} \bsnm{{Baines}}},
\bauthor{\binits{T.W.} \bsnm{{Momary}}},
\bauthor{\binits{A.P.} \bsnm{{Showman}}},
\bauthor{\binits{P.G.J.} \bsnm{{Irwin}}},
\bauthor{\binits{G.S.} \bsnm{{Orton}}},
\bauthor{\binits{M.} \bsnm{{Roos-Serote}}},
\bauthor{\binits{C.} \bsnm{{Merlet}}},
\batitle{{Saturn's tropospheric composition and clouds from Cassini/VIMS
  4.6-5.1 {$\mu$}m nightside spectroscopy}}.
\bjtitle{Icarus}
\bvolume{214},
\bfpage{510}--\blpage{533}
(\byear{2011}).
doi:\doiurl{10.1016/j.icarus.2011.06.006}
\end{barticle}
\endbibitem

\bibitem[\protect\citeauthoryear{{Fletcher} et~al.}{2014}]{14fletcher_nep}
\begin{barticle}
\bauthor{\binits{L.N.} \bsnm{{Fletcher}}},
\bauthor{\binits{I.} \bsnm{{de Pater}}},
\bauthor{\binits{G.S.} \bsnm{{Orton}}},
\bauthor{\binits{H.B.} \bsnm{{Hammel}}},
\bauthor{\binits{M.L.} \bsnm{{Sitko}}},
\bauthor{\binits{P.G.J.} \bsnm{{Irwin}}},
\batitle{{Neptune at summer solstice: Zonal mean temperatures from ground-based
  observations, 2003-2007}}.
\bjtitle{Icarus}
\bvolume{231},
\bfpage{146}--\blpage{167}
(\byear{2014}).
doi:\doiurl{10.1016/j.icarus.2013.11.035}
\end{barticle}
\endbibitem

\bibitem[\protect\citeauthoryear{{Fletcher} et~al.}{2016}]{16fletcher_texes}
\begin{barticle}
\bauthor{\binits{L.N.} \bsnm{{Fletcher}}},
\bauthor{\binits{T.K.} \bsnm{{Greathouse}}},
\bauthor{\binits{G.S.} \bsnm{{Orton}}},
\bauthor{\binits{J.A.} \bsnm{{Sinclair}}},
\bauthor{\binits{R.S.} \bsnm{{Giles}}},
\bauthor{\binits{P.G.J.} \bsnm{{Irwin}}},
\bauthor{\binits{T.} \bsnm{{Encrenaz}}},
\batitle{{Mid-infrared mapping of Jupiter's temperatures, aerosol opacity and
  chemical distributions with IRTF/TEXES}}.
\bjtitle{Icarus}
\bvolume{278},
\bfpage{128}--\blpage{161}
(\byear{2016}).
doi:\doiurl{10.1016/j.icarus.2016.06.008}
\end{barticle}
\endbibitem

\bibitem[\protect\citeauthoryear{{Fletcher} et~al.}{2018}]{18fletcher_book}
\begin{bchapter}
\bauthor{\binits{L.N.} \bsnm{{Fletcher}}},
\bauthor{\binits{T.K.} \bsnm{{Greathouse}}},
\bauthor{\binits{J.I.} \bsnm{{Moses}}},
\bauthor{\binits{S.} \bsnm{{Guerlet}}},
\bauthor{\binits{R.A.} \bsnm{{West}}},
\bctitle{{Saturn's Seasonally Changing Atmosphere: Thermal Structure,
  Composition and Aerosols}},
ed. by \beditor{\binits{K.H.} \bsnm{{Baines}}},
\beditor{\binits{F.M.} \bsnm{{Flasar}}},
\beditor{\binits{N.} \bsnm{{Krupp}}},
\beditor{\binits{T.S.} \bsnm{{Stallard}}}
(\bpublisher{Cambridge Planetary Science. Cambridge Univ. Press, New York},
  \blocation{???}, \byear{2018}).
\bcomment{Chap. 10}
\end{bchapter}
\endbibitem

\bibitem[\protect\citeauthoryear{{Fletcher} et~al.}{2020}]{19fletcher_beltzone}
\begin{botherref}
\oauthor{\binits{L.N.} \bsnm{{Fletcher}}},
\oauthor{\binits{Y.} \bsnm{{Kaspi}}},
\oauthor{\binits{T.} \bsnm{{Guillot}}},
\oauthor{\binits{A.P.} \bsnm{{Showman}}},
{How well do we understand the belt/zone circulation of Giant Planet
  atmospheres?}
Space Science Reviews, accepted,
1907--01822
(2020)
\end{botherref}
\endbibitem

\bibitem[\protect\citeauthoryear{{Fortney} et~al.}{2011}]{11fortney}
\begin{barticle}
\bauthor{\binits{J.J.} \bsnm{{Fortney}}},
\bauthor{\binits{M.} \bsnm{{Ikoma}}},
\bauthor{\binits{N.} \bsnm{{Nettelmann}}},
\bauthor{\binits{T.} \bsnm{{Guillot}}},
\bauthor{\binits{M.S.} \bsnm{{Marley}}},
\batitle{{Self-consistent Model Atmospheres and the Cooling of the Solar
  System's Giant Planets}}.
\bjtitle{\apj}
\bvolume{729}(\bissue{1}),
\bfpage{32}
(\byear{2011}).
doi:\doiurl{10.1088/0004-637X/729/1/32}
\end{barticle}
\endbibitem

\bibitem[\protect\citeauthoryear{{Friedson} and {Moses}}{2012}]{12friedson}
\begin{barticle}
\bauthor{\binits{A.J.} \bsnm{{Friedson}}},
\bauthor{\binits{J.I.} \bsnm{{Moses}}},
\batitle{{General circulation and transport in Saturn's upper troposphere and
  stratosphere}}.
\bjtitle{Icarus}
\bvolume{218},
\bfpage{861}--\blpage{875}
(\byear{2012}).
doi:\doiurl{10.1016/j.icarus.2012.02.004}
\end{barticle}
\endbibitem

\bibitem[\protect\citeauthoryear{{Friedson} and {Gonzales}}{2017}]{17friedson}
\begin{barticle}
\bauthor{\binits{A.J.} \bsnm{{Friedson}}},
\bauthor{\binits{E.J.} \bsnm{{Gonzales}}},
\batitle{{Inhibition of ordinary and diffusive convection in the water
  condensation zone of the ice giants and implications for their thermal
  evolution}}.
\bjtitle{Icarus}
\bvolume{297},
\bfpage{160}--\blpage{178}
(\byear{2017}).
doi:\doiurl{10.1016/j.icarus.2017.06.029}
\end{barticle}
\endbibitem

\bibitem[\protect\citeauthoryear{{Friedson} and {Ingersoll}}{1987}]{87friedson}
\begin{barticle}
\bauthor{\binits{J.} \bsnm{{Friedson}}},
\bauthor{\binits{A.P.} \bsnm{{Ingersoll}}},
\batitle{{Seasonal meridional energy balance and thermal structure of the
  atmosphere of Uranus: A radiative-convective-dynamical model}}.
\bjtitle{Icarus}
\bvolume{69}(\bissue{1}),
\bfpage{135}--\blpage{156}
(\byear{1987}).
doi:\doiurl{10.1016/0019-1035(87)90010-8}
\end{barticle}
\endbibitem

\bibitem[\protect\citeauthoryear{{Fry} and {Sromovsky}}{2004}]{04fry_dps}
\begin{bchapter}
\bauthor{\binits{P.M.} \bsnm{{Fry}}},
\bauthor{\binits{L.A.} \bsnm{{Sromovsky}}},
\bctitle{{Keck 2 AO Observations of Neptune in 2003 and 2004.}},
in \bbtitle{AAS/Division for Planetary Sciences Meeting Abstracts \#36}.
\bsertitle{AAS/Division for Planetary Sciences Meeting Abstracts},
\byear{2004},
pp. \bfpage{18}--\blpage{02}
\end{bchapter}
\endbibitem

\bibitem[\protect\citeauthoryear{{Fry} et~al.}{2012}]{12fry}
\begin{barticle}
\bauthor{\binits{P.M.} \bsnm{{Fry}}},
\bauthor{\binits{L.A.} \bsnm{{Sromovsky}}},
\bauthor{\binits{I.} \bsnm{{de Pater}}},
\bauthor{\binits{H.B.} \bsnm{{Hammel}}},
\bauthor{\binits{K.A.} \bsnm{{Rages}}},
\batitle{{Detection and Tracking of Subtle Cloud Features on Uranus}}.
\bjtitle{Astronomical Journal}
\bvolume{143},
\bfpage{150}
(\byear{2012}).
doi:\doiurl{10.1088/0004-6256/143/6/150}
\end{barticle}
\endbibitem

\bibitem[\protect\citeauthoryear{{Fulton} and {Petigura}}{2018}]{18fulton}
\begin{barticle}
\bauthor{\binits{B.J.} \bsnm{{Fulton}}},
\bauthor{\binits{E.A.} \bsnm{{Petigura}}},
\batitle{{The California-Kepler Survey. VII. Precise Planet Radii Leveraging
  Gaia DR2 Reveal the Stellar Mass Dependence of the Planet Radius Gap}}.
\bjtitle{\aj}
\bvolume{156}(\bissue{6}),
\bfpage{264}
(\byear{2018}).
doi:\doiurl{10.3847/1538-3881/aae828}
\end{barticle}
\endbibitem

\bibitem[\protect\citeauthoryear{{Garc{\'{\i}}a-Melendo}
  et~al.}{2011}]{11garcia}
\begin{barticle}
\bauthor{\binits{E.} \bsnm{{Garc{\'{\i}}a-Melendo}}},
\bauthor{\binits{S.} \bsnm{{P{\'e}rez-Hoyos}}},
\bauthor{\binits{A.} \bsnm{{S{\'a}nchez-Lavega}}},
\bauthor{\binits{R.} \bsnm{{Hueso}}},
\batitle{{Saturn's zonal wind profile in 2004-2009 from Cassini ISS images and
  its long-term variability}}.
\bjtitle{Icarus}
\bvolume{215},
\bfpage{62}--\blpage{74}
(\byear{2011}).
doi:\doiurl{10.1016/j.icarus.2011.07.005}
\end{barticle}
\endbibitem

\bibitem[\protect\citeauthoryear{{Gibbard} et~al.}{2003}]{03gibbard}
\begin{barticle}
\bauthor{\binits{S.G.} \bsnm{{Gibbard}}},
\bauthor{\binits{I.} \bsnm{{de Pater}}},
\bauthor{\binits{H.G.} \bsnm{{Roe}}},
\bauthor{\binits{S.} \bsnm{{Martin}}},
\bauthor{\binits{B.A.} \bsnm{{Macintosh}}},
\bauthor{\binits{C.E.} \bsnm{{Max}}},
\batitle{{The altitude of Neptune cloud features from high-spatial-resolution
  near-infrared spectra}}.
\bjtitle{Icarus}
\bvolume{166},
\bfpage{359}--\blpage{374}
(\byear{2003}).
doi:\doiurl{10.1016/j.icarus.2003.07.006}
\end{barticle}
\endbibitem

\bibitem[\protect\citeauthoryear{{Gierasch} and {Conrath}}{1987}]{87gierasch}
\begin{barticle}
\bauthor{\binits{P.J.} \bsnm{{Gierasch}}},
\bauthor{\binits{B.J.} \bsnm{{Conrath}}},
\batitle{{Vertical temperature gradients on Uranus - Implications for layered
  convection}}.
\bjtitle{Journal of Geophysical Research}
\bvolume{92},
\bfpage{15019}--\blpage{15029}
(\byear{1987}).
doi:\doiurl{10.1029/JA092iA13p15019}
\end{barticle}
\endbibitem

\bibitem[\protect\citeauthoryear{{Gierasch} et~al.}{2000}]{00gierasch}
\begin{barticle}
\bauthor{\binits{P.J.} \bsnm{{Gierasch}}},
\bauthor{\binits{A.P.} \bsnm{{Ingersoll}}},
\bauthor{\binits{D.} \bsnm{{Banfield}}},
\bauthor{\binits{S.P.} \bsnm{{Ewald}}},
\bauthor{\binits{P.} \bsnm{{Helfenstein}}},
\bauthor{\binits{A.} \bsnm{{Simon-Miller}}},
\bauthor{\binits{A.} \bsnm{{Vasavada}}},
\bauthor{\binits{H.H.} \bsnm{{Breneman}}},
\bauthor{\binits{D.A.} \bsnm{{Senske}}},
\bauthor{\bsnm{{Galileo Imaging Team}}},
\batitle{{Observation of moist convection in Jupiter's atmosphere}}.
\bjtitle{Nature}
\bvolume{403},
\bfpage{628}--\blpage{630}
(\byear{2000}).
doi:\doiurl{10.1038/35001017}
\end{barticle}
\endbibitem

\bibitem[\protect\citeauthoryear{{Goldman}}{2005}]{05goldman}
\begin{bchapter}
\bauthor{\binits{N.} \bsnm{{Goldman}}},
\bctitle{{Simulations of water in giant planets: discovery of symmetric
  H-bonding in the superionic phase}},
in \bbtitle{APS Shock Compression of Condensed Matter Meeting Abstracts},
\byear{2005},
pp. \bfpage{5}--\blpage{002}
\end{bchapter}
\endbibitem

\bibitem[\protect\citeauthoryear{{Greathouse} et~al.}{2011}]{11greathouse}
\begin{barticle}
\bauthor{\binits{T.K.} \bsnm{{Greathouse}}},
\bauthor{\binits{M.} \bsnm{{Richter}}},
\bauthor{\binits{J.} \bsnm{{Lacy}}},
\bauthor{\binits{J.} \bsnm{{Moses}}},
\bauthor{\binits{G.} \bsnm{{Orton}}},
\bauthor{\binits{T.} \bsnm{{Encrenaz}}},
\bauthor{\binits{H.B.} \bsnm{{Hammel}}},
\bauthor{\binits{D.} \bsnm{{Jaffe}}},
\batitle{{A spatially resolved high spectral resolution study of Neptune's
  stratosphere}}.
\bjtitle{Icarus}
\bvolume{214},
\bfpage{606}--\blpage{621}
(\byear{2011}).
doi:\doiurl{10.1016/j.icarus.2011.05.028}
\end{barticle}
\endbibitem

\bibitem[\protect\citeauthoryear{{Guerlet} et~al.}{2009}]{09guerlet}
\begin{barticle}
\bauthor{\binits{S.} \bsnm{{Guerlet}}},
\bauthor{\binits{T.} \bsnm{{Fouchet}}},
\bauthor{\binits{B.} \bsnm{{B{\'e}zard}}},
\bauthor{\binits{A.A.} \bsnm{{Simon-Miller}}},
\bauthor{\binits{F.M.} \bsnm{{Flasar}}},
\batitle{{Vertical and meridional distribution of ethane, acetylene and propane
  in Saturn's stratosphere from CIRS/Cassini limb observations}}.
\bjtitle{Icarus}
\bvolume{203},
\bfpage{214}--\blpage{232}
(\byear{2009})
\end{barticle}
\endbibitem

\bibitem[\protect\citeauthoryear{{Guillot}}{1995}]{95guillot}
\begin{barticle}
\bauthor{\binits{T.} \bsnm{{Guillot}}},
\batitle{{Condensation of Methane, Ammonia, and Water and the Inhibition of
  Convection in Giant Planets}}.
\bjtitle{Science}
\bvolume{269},
\bfpage{1697}--\blpage{1699}
(\byear{1995}).
doi:\doiurl{10.1126/science.7569896}
\end{barticle}
\endbibitem

\bibitem[\protect\citeauthoryear{{Gulkis} et~al.}{1978}]{78gulkis}
\begin{barticle}
\bauthor{\binits{S.} \bsnm{{Gulkis}}},
\bauthor{\binits{M.A.} \bsnm{{Janssen}}},
\bauthor{\binits{E.T.} \bsnm{{Olsen}}},
\batitle{{Evidence for the depletion of ammonia in the Uranus atmosphere}}.
\bjtitle{Icarus}
\bvolume{34},
\bfpage{10}--\blpage{19}
(\byear{1978}).
doi:\doiurl{10.1016/0019-1035(78)90120-3}
\end{barticle}
\endbibitem

\bibitem[\protect\citeauthoryear{{Gurnett} et~al.}{1990}]{90gurnett}
\begin{barticle}
\bauthor{\binits{D.A.} \bsnm{{Gurnett}}},
\bauthor{\binits{W.S.} \bsnm{{Kurth}}},
\bauthor{\binits{I.H.} \bsnm{{Cairns}}},
\bauthor{\binits{L.J.} \bsnm{{Granroth}}},
\batitle{{Whistlers in Neptune's magnetosphere - Evidence of atmospheric
  lightning}}.
\bjtitle{Journal of Geophysical Research}
\bvolume{95},
\bfpage{20967}--\blpage{20976}
(\byear{1990}).
doi:\doiurl{10.1029/JA095iA12p20967}
\end{barticle}
\endbibitem

\bibitem[\protect\citeauthoryear{{Hammel} and {Lockwood}}{1997}]{97hammel}
\begin{barticle}
\bauthor{\binits{H.B.} \bsnm{{Hammel}}},
\bauthor{\binits{G.W.} \bsnm{{Lockwood}}},
\batitle{{Atmospheric Structure of Neptune in 1994, 1995, and 1996: HST Imaging
  at Multiple Wavelengths}}.
\bjtitle{Icarus}
\bvolume{129}(\bissue{2}),
\bfpage{466}--\blpage{481}
(\byear{1997}).
doi:\doiurl{10.1006/icar.1997.5764}
\end{barticle}
\endbibitem

\bibitem[\protect\citeauthoryear{{Hammel} and {Lockwood}}{2007}]{07hammel_var}
\begin{barticle}
\bauthor{\binits{H.B.} \bsnm{{Hammel}}},
\bauthor{\binits{G.W.} \bsnm{{Lockwood}}},
\batitle{{Long-term atmospheric variability on Uranus and Neptune}}.
\bjtitle{Icarus}
\bvolume{186},
\bfpage{291}--\blpage{301}
(\byear{2007}).
doi:\doiurl{10.1016/j.icarus.2006.08.027}
\end{barticle}
\endbibitem

\bibitem[\protect\citeauthoryear{{Hammel} et~al.}{1989}]{89hammel}
\begin{barticle}
\bauthor{\binits{H.B.} \bsnm{{Hammel}}},
\bauthor{\binits{K.H.} \bsnm{{Baines}}},
\bauthor{\binits{J.T.} \bsnm{{Bergstralh}}},
\batitle{{Vertical aerosol structure of Neptune - Constraints from
  center-to-limb profiles}}.
\bjtitle{Icarus}
\bvolume{80},
\bfpage{416}--\blpage{438}
(\byear{1989}).
doi:\doiurl{10.1016/0019-1035(89)90149-8}
\end{barticle}
\endbibitem

\bibitem[\protect\citeauthoryear{{Hammel} et~al.}{1995}]{95hammel}
\begin{barticle}
\bauthor{\binits{H.B.} \bsnm{{Hammel}}},
\bauthor{\binits{R.F.} \bsnm{{Beebe}}},
\bauthor{\binits{A.P.} \bsnm{{Ingersoll}}},
\bauthor{\binits{G.S.} \bsnm{{Orton}}},
\bauthor{\binits{J.R.} \bsnm{{Mills}}},
\bauthor{\binits{A.A.} \bsnm{{Simon}}},
\bauthor{\binits{P.} \bsnm{{Chodas}}},
\bauthor{\binits{J.T.} \bsnm{{Clarke}}},
\bauthor{\binits{E.} \bsnm{{de Jong}}},
\bauthor{\binits{T.E.} \bsnm{{Dowling}}},
\bauthor{\binits{J.} \bsnm{{Harrington}}},
\bauthor{\binits{L.F.} \bsnm{{Huber}}},
\bauthor{\binits{E.} \bsnm{{Karkoschka}}},
\bauthor{\binits{C.M.} \bsnm{{Santori}}},
\bauthor{\binits{A.} \bsnm{{Toigo}}},
\bauthor{\binits{D.} \bsnm{{Yeomans}}},
\bauthor{\binits{R.A.} \bsnm{{West}}},
\batitle{{HST Imaging of Atmospheric Phenomena Created by the Impact of Comet
  Shoemaker-Levy 9}}.
\bjtitle{Science}
\bvolume{267},
\bfpage{1288}--\blpage{1296}
(\byear{1995}).
doi:\doiurl{10.1126/science.7871425}
\end{barticle}
\endbibitem

\bibitem[\protect\citeauthoryear{{Hammel} et~al.}{2001}]{01hammel}
\begin{barticle}
\bauthor{\binits{H.B.} \bsnm{{Hammel}}},
\bauthor{\binits{K.} \bsnm{{Rages}}},
\bauthor{\binits{G.W.} \bsnm{{Lockwood}}},
\bauthor{\binits{E.} \bsnm{{Karkoschka}}},
\bauthor{\binits{I.} \bsnm{{de Pater}}},
\batitle{{New Measurements of the Winds of Uranus}}.
\bjtitle{Icarus}
\bvolume{153},
\bfpage{229}--\blpage{235}
(\byear{2001}).
doi:\doiurl{10.1006/icar.2001.6689}
\end{barticle}
\endbibitem

\bibitem[\protect\citeauthoryear{{Hammel} et~al.}{2005}]{05hammel}
\begin{barticle}
\bauthor{\binits{H.B.} \bsnm{{Hammel}}},
\bauthor{\binits{I.} \bsnm{{de Pater}}},
\bauthor{\binits{S.} \bsnm{{Gibbard}}},
\bauthor{\binits{G.W.} \bsnm{{Lockwood}}},
\bauthor{\binits{K.} \bsnm{{Rages}}},
\batitle{{Uranus in 2003: Zonal winds, banded structure, and discrete
  features}}.
\bjtitle{Icarus}
\bvolume{175},
\bfpage{534}--\blpage{545}
(\byear{2005}).
doi:\doiurl{10.1016/j.icarus.2004.11.012}
\end{barticle}
\endbibitem

\bibitem[\protect\citeauthoryear{{Hammel} et~al.}{2006}]{06hammel}
\begin{barticle}
\bauthor{\binits{H.B.} \bsnm{{Hammel}}},
\bauthor{\binits{D.K.} \bsnm{{Lynch}}},
\bauthor{\binits{R.W.} \bsnm{{Russell}}},
\bauthor{\binits{M.L.} \bsnm{{Sitko}}},
\bauthor{\binits{L.S.} \bsnm{{Bernstein}}},
\bauthor{\binits{T.} \bsnm{{Hewagama}}},
\batitle{{Mid-Infrared Ethane Emission on Neptune and Uranus}}.
\bjtitle{Astrophys. J. Letters}
\bvolume{644},
\bfpage{1326}--\blpage{1333}
(\byear{2006}).
doi:\doiurl{10.1086/503599}
\end{barticle}
\endbibitem

\bibitem[\protect\citeauthoryear{{Hammel} et~al.}{2007}]{07hammel}
\begin{barticle}
\bauthor{\binits{H.B.} \bsnm{{Hammel}}},
\bauthor{\binits{M.L.} \bsnm{{Sitko}}},
\bauthor{\binits{D.K.} \bsnm{{Lynch}}},
\bauthor{\binits{G.S.} \bsnm{{Orton}}},
\bauthor{\binits{R.W.} \bsnm{{Russell}}},
\bauthor{\binits{T.R.} \bsnm{{Geballe}}},
\bauthor{\binits{I.} \bsnm{{de Pater}}},
\batitle{{Distribution of Ethane and Methane Emission on Neptune}}.
\bjtitle{Astron. J.}
\bvolume{134},
\bfpage{637}--\blpage{641}
(\byear{2007}).
doi:\doiurl{10.1086/519382}
\end{barticle}
\endbibitem

\bibitem[\protect\citeauthoryear{{Hammel} et~al.}{2009}]{09hammel}
\begin{barticle}
\bauthor{\binits{H.B.} \bsnm{{Hammel}}},
\bauthor{\binits{L.A.} \bsnm{{Sromovsky}}},
\bauthor{\binits{P.M.} \bsnm{{Fry}}},
\bauthor{\binits{K.} \bsnm{{Rages}}},
\bauthor{\binits{M.} \bsnm{{Showalter}}},
\bauthor{\binits{I.} \bsnm{{de Pater}}},
\bauthor{\binits{M.A.} \bsnm{{van Dam}}},
\bauthor{\binits{R.P.} \bsnm{{LeBeau}}},
\bauthor{\binits{X.} \bsnm{{Deng}}},
\batitle{{The Dark Spot in the atmosphere of Uranus in 2006: Discovery,
  description, and dynamical simulations}}.
\bjtitle{Icarus}
\bvolume{201},
\bfpage{257}--\blpage{271}
(\byear{2009}).
doi:\doiurl{10.1016/j.icarus.2008.08.019}
\end{barticle}
\endbibitem

\bibitem[\protect\citeauthoryear{{Held} and {Hou}}{1980}]{80held}
\begin{barticle}
\bauthor{\binits{I.M.} \bsnm{{Held}}},
\bauthor{\binits{A.Y.} \bsnm{{Hou}}},
\batitle{{Nonlinear Axially Symmetric Circulations in a Nearly Inviscid
  Atmosphere.}}
\bjtitle{Journal of Atmospheric Sciences}
\bvolume{37},
\bfpage{515}--\blpage{533}
(\byear{1980}).
doi:\doiurl{10.1175/1520-0469(1980)037<0515:NASCIA>2.0.CO;2}
\end{barticle}
\endbibitem

\bibitem[\protect\citeauthoryear{{Herbert} et~al.}{1987}]{87herbert}
\begin{barticle}
\bauthor{\binits{F.} \bsnm{{Herbert}}},
\bauthor{\binits{B.R.} \bsnm{{Sandel}}},
\bauthor{\binits{R.V.} \bsnm{{Yelle}}},
\bauthor{\binits{J.B.} \bsnm{{Holberg}}},
\bauthor{\binits{A.L.} \bsnm{{Broadfoot}}},
\bauthor{\binits{D.E.} \bsnm{{Shemansky}}},
\bauthor{\binits{S.K.} \bsnm{{Atreya}}},
\bauthor{\binits{P.N.} \bsnm{{Romani}}},
\batitle{{The upper atmosphere of Uranus - EUV occultations observed by Voyager
  2}}.
\bjtitle{Journal of Geophysical Research}
\bvolume{92},
\bfpage{15093}--\blpage{15109}
(\byear{1987}).
doi:\doiurl{10.1029/JA092iA13p15093}
\end{barticle}
\endbibitem

\bibitem[\protect\citeauthoryear{{Hersant} et~al.}{2004}]{04hersant}
\begin{barticle}
\bauthor{\binits{F.} \bsnm{{Hersant}}},
\bauthor{\binits{D.} \bsnm{{Gautier}}},
\bauthor{\binits{J.I.} \bsnm{{Lunine}}},
\batitle{{Enrichment in volatiles in the giant planets of the Solar System}}.
\bjtitle{Plan. \& Space Sci.}
\bvolume{52},
\bfpage{623}--\blpage{641}
(\byear{2004}).
doi:\doiurl{10.1016/j.pss.2003.12.011}
\end{barticle}
\endbibitem

\bibitem[\protect\citeauthoryear{{Hofstadter} and
  {Butler}}{2003}]{03hofstadter}
\begin{barticle}
\bauthor{\binits{M.D.} \bsnm{{Hofstadter}}},
\bauthor{\binits{B.J.} \bsnm{{Butler}}},
\batitle{{Seasonal change in the deep atmosphere of Uranus}}.
\bjtitle{Icarus}
\bvolume{165},
\bfpage{168}--\blpage{180}
(\byear{2003}).
doi:\doiurl{10.1016/S0019-1035(03)00174-X}
\end{barticle}
\endbibitem

\bibitem[\protect\citeauthoryear{{Hofstadter} et~al.}{2004}]{04hofstadter_dps}
\begin{bchapter}
\bauthor{\binits{M.D.} \bsnm{{Hofstadter}}},
\bauthor{\binits{B.J.} \bsnm{{Butler}}},
\bauthor{\binits{H.B.} \bsnm{{Hammel}}},
\bauthor{\binits{M.J.} \bsnm{{Klein}}},
\bctitle{{The Discovery of Radio-Bright Northern Latitudes on Uranus:
  Implications for Weather and Climate}},
in \bbtitle{AAS/Division for Planetary Sciences Meeting Abstracts \#36}.
\bsertitle{AAS/Division for Planetary Sciences Meeting Abstracts},
\byear{2004},
pp. \bfpage{05}--\blpage{09}
\end{bchapter}
\endbibitem

\bibitem[\protect\citeauthoryear{Holton}{2004}]{04holton}
\begin{bbook}
\bauthor{\binits{J.R.} \bsnm{Holton}},
\bbtitle{{An Introduction to Dynamic Meteorology}}
(\bpublisher{Academic press}, \blocation{???}, \byear{2004}).
\bisbn{0123540151}
\end{bbook}
\endbibitem

\bibitem[\protect\citeauthoryear{{Hubbard} et~al.}{1991}]{91hubbard}
\begin{barticle}
\bauthor{\binits{W.B.} \bsnm{{Hubbard}}},
\bauthor{\binits{W.J.} \bsnm{{Nellis}}},
\bauthor{\binits{A.C.} \bsnm{{Mitchell}}},
\bauthor{\binits{N.C.} \bsnm{{Holmes}}},
\bauthor{\binits{S.S.} \bsnm{{Limaye}}},
\bauthor{\binits{P.C.} \bsnm{{McCandless}}},
\batitle{{Interior Structure of Neptune: Comparison with Uranus}}.
\bjtitle{Science}
\bvolume{253}(\bissue{5020}),
\bfpage{648}--\blpage{651}
(\byear{1991}).
doi:\doiurl{10.1126/science.253.5020.648}
\end{barticle}
\endbibitem

\bibitem[\protect\citeauthoryear{{Hueso} et~al.}{2017}]{17hueso}
\begin{barticle}
\bauthor{\binits{R.} \bsnm{{Hueso}}},
\bauthor{\binits{I.} \bsnm{{de Pater}}},
\bauthor{\binits{A.} \bsnm{{Simon}}},
\bauthor{\binits{A.} \bsnm{{S{\'a}nchez-Lavega}}},
\bauthor{\binits{M.} \bsnm{{Delcroix}}},
\bauthor{\binits{M.H.} \bsnm{{Wong}}},
\bauthor{\binits{J.W.} \bsnm{{Tollefson}}},
\bauthor{\binits{C.} \bsnm{{Baranec}}},
\bauthor{\binits{K.} \bsnm{{de Kleer}}},
\bauthor{\binits{S.H.} \bsnm{{Luszcz-Cook}}},
\batitle{{Neptune long-lived atmospheric features in 2013-2015 from small
  (28-cm) to large (10-m) telescopes}}.
\bjtitle{Icarus}
\bvolume{295},
\bfpage{89}--\blpage{109}
(\byear{2017}).
doi:\doiurl{10.1016/j.icarus.2017.06.009}
\end{barticle}
\endbibitem

\bibitem[\protect\citeauthoryear{{Ingersoll} et~al.}{2000}]{00ingersoll}
\begin{barticle}
\bauthor{\binits{A.P.} \bsnm{{Ingersoll}}},
\bauthor{\binits{G.P.} \bsnm{J.}},
\bauthor{\binits{B.} \bsnm{D.}},
\bauthor{\binits{A.R.} \bsnm{{Vasavada}}},
\bauthor{\bsnm{{Galileo Imaging Team}}},
\batitle{{Moist convection as an energy source for the large-scale motions in
  Jupiter's atmosphere}}.
\bjtitle{Nature}
\bvolume{403},
\bfpage{630}--\blpage{632}
(\byear{2000}).
doi:\doiurl{10.1038/35001021}
\end{barticle}
\endbibitem

\bibitem[\protect\citeauthoryear{{Ingersoll} et~al.}{2004}]{04ingersoll}
\begin{botherref}
\oauthor{\binits{A.P.} \bsnm{{Ingersoll}}},
\oauthor{\binits{T.E.} \bsnm{{Dowling}}},
\oauthor{\binits{P.J.} \bsnm{{Gierasch}}},
\oauthor{\binits{G.S.} \bsnm{{Orton}}},
\oauthor{\binits{P.L.} \bsnm{{Read}}},
\oauthor{\binits{A.} \bsnm{{S{\'a}nchez-Lavega}}},
\oauthor{\binits{A.P.} \bsnm{{Showman}}},
\oauthor{\binits{A.A.} \bsnm{{Simon-Miller}}},
\oauthor{\binits{A.R.} \bsnm{{Vasavada}}},
{Dynamics of Jupiter's atmosphere}
(Jupiter.~The Planet, Satellites and Magnetosphere, 2004),
pp. 105--128
\end{botherref}
\endbibitem

\bibitem[\protect\citeauthoryear{{Irwin} et~al.}{2010}]{10irwin}
\begin{barticle}
\bauthor{\binits{P.G.J.} \bsnm{{Irwin}}},
\bauthor{\binits{N.A.} \bsnm{{Teanby}}},
\bauthor{\binits{G.R.} \bsnm{{Davis}}},
\batitle{{Revised vertical cloud structure of Uranus from UKIRT/UIST
  observations and changes seen during Uranus{\textquoteright} Northern Spring
  Equinox from 2006 to 2008: Application of new methane absorption data and
  comparison with Neptune}}.
\bjtitle{Icarus}
\bvolume{208}(\bissue{2}),
\bfpage{913}--\blpage{926}
(\byear{2010}).
doi:\doiurl{10.1016/j.icarus.2010.03.017}
\end{barticle}
\endbibitem

\bibitem[\protect\citeauthoryear{{Irwin} et~al.}{2012}]{12irwin}
\begin{barticle}
\bauthor{\binits{P.G.J.} \bsnm{{Irwin}}},
\bauthor{\binits{N.A.} \bsnm{{Teanby}}},
\bauthor{\binits{G.R.} \bsnm{{Davis}}},
\bauthor{\binits{L.N.} \bsnm{{Fletcher}}},
\bauthor{\binits{G.S.} \bsnm{{Orton}}},
\bauthor{\binits{S.B.} \bsnm{{Calcutt}}},
\bauthor{\binits{D.S.} \bsnm{{Tice}}},
\bauthor{\binits{J.} \bsnm{{Hurley}}},
\batitle{{Further seasonal changes in Uranus{\textquoteright} cloud structure
  observed by Gemini-North and UKIRT}}.
\bjtitle{Icarus}
\bvolume{218}(\bissue{1}),
\bfpage{47}--\blpage{55}
(\byear{2012}).
doi:\doiurl{10.1016/j.icarus.2011.12.001}
\end{barticle}
\endbibitem

\bibitem[\protect\citeauthoryear{{Irwin} et~al.}{2016}]{16irwin_nep}
\begin{barticle}
\bauthor{\binits{P.G.J.} \bsnm{{Irwin}}},
\bauthor{\binits{L.N.} \bsnm{{Fletcher}}},
\bauthor{\binits{D.} \bsnm{{Tice}}},
\bauthor{\binits{S.J.} \bsnm{{Owen}}},
\bauthor{\binits{G.S.} \bsnm{{Orton}}},
\bauthor{\binits{N.A.} \bsnm{{Teanby}}},
\bauthor{\binits{G.R.} \bsnm{{Davis}}},
\batitle{{Time variability of Neptune's horizontal and vertical cloud structure
  revealed by VLT/SINFONI and Gemini/NIFS from 2009 to 2013}}.
\bjtitle{Icarus}
\bvolume{271},
\bfpage{418}--\blpage{437}
(\byear{2016}).
doi:\doiurl{10.1016/j.icarus.2016.01.015}
\end{barticle}
\endbibitem

\bibitem[\protect\citeauthoryear{{Irwin} et~al.}{2017}]{17irwin}
\begin{barticle}
\bauthor{\binits{P.G.J.} \bsnm{{Irwin}}},
\bauthor{\binits{M.H.} \bsnm{{Wong}}},
\bauthor{\binits{A.A.} \bsnm{{Simon}}},
\bauthor{\binits{G.S.} \bsnm{{Orton}}},
\bauthor{\binits{D.} \bsnm{{Toledo}}},
\batitle{{HST/WFC3 observations of Uranus' 2014 storm clouds and comparison
  with VLT/SINFONI and IRTF/Spex observations}}.
\bjtitle{Icarus}
\bvolume{288},
\bfpage{99}--\blpage{119}
(\byear{2017}).
doi:\doiurl{10.1016/j.icarus.2017.01.031}
\end{barticle}
\endbibitem

\bibitem[\protect\citeauthoryear{{Irwin} et~al.}{2018}]{18irwin_h2s}
\begin{barticle}
\bauthor{\binits{P.G.J.} \bsnm{{Irwin}}},
\bauthor{\binits{D.} \bsnm{{Toledo}}},
\bauthor{\binits{R.} \bsnm{{Garland}}},
\bauthor{\binits{N.A.} \bsnm{{Teanby}}},
\bauthor{\binits{L.N.} \bsnm{{Fletcher}}},
\bauthor{\binits{G.A.} \bsnm{{Orton}}},
\bauthor{\binits{B.} \bsnm{{B{\'e}zard}}},
\batitle{{Detection of hydrogen sulfide above the clouds in Uranus's
  atmosphere}}.
\bjtitle{Nature Astronomy}
\bvolume{2},
\bfpage{420}--\blpage{427}
(\byear{2018}).
doi:\doiurl{10.1038/s41550-018-0432-1}
\end{barticle}
\endbibitem

\bibitem[\protect\citeauthoryear{{Irwin} et~al.}{2019a}]{19irwin_muse}
\begin{barticle}
\bauthor{\binits{P.G.J.} \bsnm{{Irwin}}},
\bauthor{\binits{D.} \bsnm{{Toledo}}},
\bauthor{\binits{A.S.} \bsnm{{Braude}}},
\bauthor{\binits{R.} \bsnm{{Bacon}}},
\bauthor{\binits{P.M.} \bsnm{{Weilbacher}}},
\bauthor{\binits{N.A.} \bsnm{{Teanby}}},
\bauthor{\binits{L.N.} \bsnm{{Fletcher}}},
\bauthor{\binits{G.S.} \bsnm{{Orton}}},
\batitle{{Latitudinal variation in the abundance of methane (CH$_{4}$) above
  the clouds in Neptune's atmosphere from VLT/MUSE Narrow Field Mode
  Observations}}.
\bjtitle{Icarus}
\bvolume{331},
\bfpage{69}--\blpage{82}
(\byear{2019}a).
doi:\doiurl{10.1016/j.icarus.2019.05.011}
\end{barticle}
\endbibitem

\bibitem[\protect\citeauthoryear{{Irwin} et~al.}{2019b}]{19irwin_h2s}
\begin{barticle}
\bauthor{\binits{P.G.J.} \bsnm{{Irwin}}},
\bauthor{\binits{D.} \bsnm{{Toledo}}},
\bauthor{\binits{R.} \bsnm{{Garland}}},
\bauthor{\binits{N.A.} \bsnm{{Teanby}}},
\bauthor{\binits{L.N.} \bsnm{{Fletcher}}},
\bauthor{\binits{G.S.} \bsnm{{Orton}}},
\bauthor{\binits{B.} \bsnm{{B{\'e}zard}}},
\batitle{{Probable detection of hydrogen sulphide (H$_{2}$S) in Neptune's
  atmosphere}}.
\bjtitle{Icarus}
\bvolume{321},
\bfpage{550}--\blpage{563}
(\byear{2019}b).
doi:\doiurl{10.1016/j.icarus.2018.12.014}
\end{barticle}
\endbibitem

\bibitem[\protect\citeauthoryear{{Karkoschka}}{1998}]{98karkoschka}
\begin{barticle}
\bauthor{\binits{E.} \bsnm{{Karkoschka}}},
\batitle{{Clouds of High Contrast on Uranus}}.
\bjtitle{Science}
\bvolume{280},
\bfpage{570}
(\byear{1998}).
doi:\doiurl{10.1126/science.280.5363.570}
\end{barticle}
\endbibitem

\bibitem[\protect\citeauthoryear{{Karkoschka}}{2011}]{11karkoschka}
\begin{barticle}
\bauthor{\binits{E.} \bsnm{{Karkoschka}}},
\batitle{{Neptune's rotational period suggested by the extraordinary stability
  of two features}}.
\bjtitle{Icarus}
\bvolume{215},
\bfpage{439}--\blpage{448}
(\byear{2011}).
doi:\doiurl{10.1016/j.icarus.2011.05.013}
\end{barticle}
\endbibitem

\bibitem[\protect\citeauthoryear{{Karkoschka}}{2015}]{15karkoschka}
\begin{barticle}
\bauthor{\binits{E.} \bsnm{{Karkoschka}}},
\batitle{{Uranus' southern circulation revealed by Voyager 2: Unique
  characteristics}}.
\bjtitle{Icarus}
\bvolume{250},
\bfpage{294}--\blpage{307}
(\byear{2015}).
doi:\doiurl{10.1016/j.icarus.2014.12.003}
\end{barticle}
\endbibitem

\bibitem[\protect\citeauthoryear{{Karkoschka} and
  {Tomasko}}{2009}]{09karkoschka}
\begin{barticle}
\bauthor{\binits{E.} \bsnm{{Karkoschka}}},
\bauthor{\binits{M.} \bsnm{{Tomasko}}},
\batitle{{The haze and methane distributions on Uranus from HST-STIS
  spectroscopy}}.
\bjtitle{Icarus}
\bvolume{202},
\bfpage{287}--\blpage{309}
(\byear{2009}).
doi:\doiurl{10.1016/j.icarus.2009.02.010}
\end{barticle}
\endbibitem

\bibitem[\protect\citeauthoryear{{Karkoschka} and
  {Tomasko}}{2011}]{11karkoschka_ch4}
\begin{barticle}
\bauthor{\binits{E.} \bsnm{{Karkoschka}}},
\bauthor{\binits{M.G.} \bsnm{{Tomasko}}},
\batitle{{The haze and methane distributions on Neptune from HST-STIS
  spectroscopy}}.
\bjtitle{Icarus}
\bvolume{211},
\bfpage{780}--\blpage{797}
(\byear{2011}).
doi:\doiurl{10.1016/j.icarus.2010.08.013}
\end{barticle}
\endbibitem

\bibitem[\protect\citeauthoryear{{Kaspi} et~al.}{2013}]{13kaspi}
\begin{barticle}
\bauthor{\binits{Y.} \bsnm{{Kaspi}}},
\bauthor{\binits{A.P.} \bsnm{{Showman}}},
\bauthor{\binits{W.B.} \bsnm{{Hubbard}}},
\bauthor{\binits{O.} \bsnm{{Aharonson}}},
\bauthor{\binits{R.} \bsnm{{Helled}}},
\batitle{{Atmospheric confinement of jet streams on Uranus and Neptune}}.
\bjtitle{Nature}
\bvolume{497},
\bfpage{344}--\blpage{347}
(\byear{2013}).
doi:\doiurl{10.1038/nature12131}
\end{barticle}
\endbibitem

\bibitem[\protect\citeauthoryear{{LeBeau} and {Dowling}}{1998}]{98lebeau}
\begin{barticle}
\bauthor{\binits{R.P.} \bsnm{{LeBeau}}},
\bauthor{\binits{T.E.} \bsnm{{Dowling}}},
\batitle{{EPIC Simulations of Time-Dependent, Three-Dimensional Vortices with
  Application to Neptune's Great Dark SPOT}}.
\bjtitle{Icarus}
\bvolume{132},
\bfpage{239}--\blpage{265}
(\byear{1998}).
doi:\doiurl{10.1006/icar.1998.5918}
\end{barticle}
\endbibitem

\bibitem[\protect\citeauthoryear{{Leconte} et~al.}{2017}]{17leconte}
\begin{barticle}
\bauthor{\binits{J.} \bsnm{{Leconte}}},
\bauthor{\binits{F.} \bsnm{{Selsis}}},
\bauthor{\binits{F.} \bsnm{{Hersant}}},
\bauthor{\binits{T.} \bsnm{{Guillot}}},
\batitle{{Condensation-inhibited convection in hydrogen-rich atmospheres .
  Stability against double-diffusive processes and thermal profiles for
  Jupiter, Saturn, Uranus, and Neptune}}.
\bjtitle{\aap}
\bvolume{598},
\bfpage{98}
(\byear{2017}).
doi:\doiurl{10.1051/0004-6361/201629140}
\end{barticle}
\endbibitem

\bibitem[\protect\citeauthoryear{{Lellouch} et~al.}{2005}]{05lellouch}
\begin{barticle}
\bauthor{\binits{E.} \bsnm{{Lellouch}}},
\bauthor{\binits{R.} \bsnm{{Moreno}}},
\bauthor{\binits{G.} \bsnm{{Paubert}}},
\batitle{{A dual origin for Neptune's carbon monoxide?}}
\bjtitle{Astron. Astrophys}
\bvolume{430},
\bfpage{37}--\blpage{40}
(\byear{2005}).
doi:\doiurl{10.1051/0004-6361:200400127}
\end{barticle}
\endbibitem

\bibitem[\protect\citeauthoryear{{Lellouch} et~al.}{2010}]{10lellouch}
\begin{barticle}
\bauthor{\binits{E.} \bsnm{{Lellouch}}},
\bauthor{\binits{P.} \bsnm{{Hartogh}}},
\bauthor{\binits{H.} \bsnm{{Feuchtgruber}}},
\bauthor{\binits{B.} \bsnm{{Vandenbussche}}},
\bauthor{\binits{T.} \bsnm{{de Graauw}}},
\bauthor{\binits{R.} \bsnm{{Moreno}}},
\bauthor{\binits{C.} \bsnm{{Jarchow}}},
\bauthor{\binits{T.} \bsnm{{Cavali{\'e}}}},
\bauthor{\binits{G.} \bsnm{{Orton}}},
\bauthor{\binits{M.} \bsnm{{Banaszkiewicz}}},
\bauthor{\binits{M.I.} \bsnm{{Blecka}}},
\bauthor{\binits{D.} \bsnm{{Bockel{\'e}e-Morvan}}},
\bauthor{\binits{J.} \bsnm{{Crovisier}}},
\bauthor{\binits{T.} \bsnm{{Encrenaz}}},
\bauthor{\binits{T.} \bsnm{{Fulton}}},
\bauthor{\binits{M.} \bsnm{{K{\"u}ppers}}},
\bauthor{\binits{L.M.} \bsnm{{Lara}}},
\bauthor{\binits{D.C.} \bsnm{{Lis}}},
\bauthor{\binits{A.S.} \bsnm{{Medvedev}}},
\bauthor{\binits{M.} \bsnm{{Rengel}}},
\bauthor{\binits{H.} \bsnm{{Sagawa}}},
\bauthor{\binits{B.} \bsnm{{Swinyard}}},
\bauthor{\binits{S.} \bsnm{{Szutowicz}}},
\bauthor{\binits{F.} \bsnm{{Bensch}}},
\bauthor{\binits{E.} \bsnm{{Bergin}}},
\bauthor{\binits{F.} \bsnm{{Billebaud}}},
\bauthor{\binits{N.} \bsnm{{Biver}}},
\bauthor{\binits{G.A.} \bsnm{{Blake}}},
\bauthor{\binits{J.A.D.L.} \bsnm{{Blommaert}}},
\bauthor{\binits{J.} \bsnm{{Cernicharo}}},
\bauthor{\binits{R.} \bsnm{{Courtin}}},
\bauthor{\binits{G.R.} \bsnm{{Davis}}},
\bauthor{\binits{L.} \bsnm{{Decin}}},
\bauthor{\binits{P.} \bsnm{{Encrenaz}}},
\bauthor{\binits{A.} \bsnm{{Gonzalez}}},
\bauthor{\binits{E.} \bsnm{{Jehin}}},
\bauthor{\binits{M.} \bsnm{{Kidger}}},
\bauthor{\binits{D.} \bsnm{{Naylor}}},
\bauthor{\binits{G.} \bsnm{{Portyankina}}},
\bauthor{\binits{R.} \bsnm{{Schieder}}},
\bauthor{\binits{S.} \bsnm{{Sidher}}},
\bauthor{\binits{N.} \bsnm{{Thomas}}},
\bauthor{\binits{M.} \bsnm{{de Val-Borro}}},
\bauthor{\binits{E.} \bsnm{{Verdugo}}},
\bauthor{\binits{C.} \bsnm{{Waelkens}}},
\bauthor{\binits{H.} \bsnm{{Walker}}},
\bauthor{\binits{H.} \bsnm{{Aarts}}},
\bauthor{\binits{C.} \bsnm{{Comito}}},
\bauthor{\binits{J.H.} \bsnm{{Kawamura}}},
\bauthor{\binits{A.} \bsnm{{Maestrini}}},
\bauthor{\binits{T.} \bsnm{{Peacocke}}},
\bauthor{\binits{R.} \bsnm{{Teipen}}},
\bauthor{\binits{T.} \bsnm{{Tils}}},
\bauthor{\binits{K.} \bsnm{{Wildeman}}},
\batitle{{First results of Herschel-PACS observations of Neptune}}.
\bjtitle{Astron. Astrophys}
\bvolume{518},
\bfpage{152}
(\byear{2010}).
doi:\doiurl{10.1051/0004-6361/201014600}
\end{barticle}
\endbibitem

\bibitem[\protect\citeauthoryear{{Lewis}}{1969}]{69lewis}
\begin{barticle}
\bauthor{\binits{J.S.} \bsnm{{Lewis}}},
\batitle{{The clouds of Jupiter and the NH$_3$-H$_2$O and NH$_3$-H$_2$S
  systems}}.
\bjtitle{Icarus}
\bvolume{10},
\bfpage{365}--\blpage{378}
(\byear{1969}).
doi:\doiurl{10.1016/0019-1035(69)90091-8}
\end{barticle}
\endbibitem

\bibitem[\protect\citeauthoryear{{Li} and {Ingersoll}}{2015}]{15li}
\begin{barticle}
\bauthor{\binits{C.} \bsnm{{Li}}},
\bauthor{\binits{A.P.} \bsnm{{Ingersoll}}},
\batitle{{Moist convection in hydrogen atmospheres and the frequency of
  Saturn's giant storms}}.
\bjtitle{Nature Geoscience}
\bvolume{8},
\bfpage{398}--\blpage{403}
(\byear{2015}).
doi:\doiurl{10.1038/ngeo2405}
\end{barticle}
\endbibitem

\bibitem[\protect\citeauthoryear{{Li} et~al.}{2017}]{17li}
\begin{barticle}
\bauthor{\binits{C.} \bsnm{{Li}}},
\bauthor{\binits{A.} \bsnm{{Ingersoll}}},
\bauthor{\binits{M.} \bsnm{{Janssen}}},
\bauthor{\binits{S.} \bsnm{{Levin}}},
\bauthor{\binits{S.} \bsnm{{Bolton}}},
\bauthor{\binits{V.} \bsnm{{Adumitroaie}}},
\bauthor{\binits{M.} \bsnm{{Allison}}},
\bauthor{\binits{J.} \bsnm{{Arballo}}},
\bauthor{\binits{A.} \bsnm{{Bellotti}}},
\bauthor{\binits{S.} \bsnm{{Brown}}},
\bauthor{\binits{S.} \bsnm{{Ewald}}},
\bauthor{\binits{L.} \bsnm{{Jewell}}},
\bauthor{\binits{S.} \bsnm{{Misra}}},
\bauthor{\binits{G.} \bsnm{{Orton}}},
\bauthor{\binits{F.} \bsnm{{Oyafuso}}},
\bauthor{\binits{P.} \bsnm{{Steffes}}},
\bauthor{\binits{R.} \bsnm{{Williamson}}},
\batitle{{The distribution of ammonia on Jupiter from a preliminary inversion
  of Juno microwave radiometer data}}.
\bjtitle{Geophys. Res. Lett.}
\bvolume{44},
\bfpage{5317}--\blpage{5325}
(\byear{2017}).
doi:\doiurl{10.1002/2017GL073159}
\end{barticle}
\endbibitem

\bibitem[\protect\citeauthoryear{{Li} et~al.}{2018}]{18li}
\begin{barticle}
\bauthor{\binits{C.} \bsnm{{Li}}},
\bauthor{\binits{T.} \bsnm{{Le}}},
\bauthor{\binits{X.} \bsnm{{Zhang}}},
\bauthor{\binits{Y.L.} \bsnm{{Yung}}},
\batitle{{A high-performance atmospheric radiation package: With applications
  to the radiative energy budgets of giant planets}}.
\bjtitle{\jqsrt}
\bvolume{217},
\bfpage{353}--\blpage{362}
(\byear{2018}).
doi:\doiurl{10.1016/j.jqsrt.2018.06.002}
\end{barticle}
\endbibitem

\bibitem[\protect\citeauthoryear{{Limaye} and {Sromovsky}}{1991}]{91limaye}
\begin{barticle}
\bauthor{\binits{S.S.} \bsnm{{Limaye}}},
\bauthor{\binits{L.A.} \bsnm{{Sromovsky}}},
\batitle{{Winds of Neptune - Voyager observations of cloud motions}}.
\bjtitle{Journal of Geophysical Research}
\bvolume{96},
\bfpage{18}
(\byear{1991}).
doi:\doiurl{10.1029/91JA01701}
\end{barticle}
\endbibitem

\bibitem[\protect\citeauthoryear{{Lindal} et~al.}{1987}]{87lindal}
\begin{barticle}
\bauthor{\binits{G.F.} \bsnm{{Lindal}}},
\bauthor{\binits{J.R.} \bsnm{{Lyons}}},
\bauthor{\binits{D.N.} \bsnm{{Sweetnam}}},
\bauthor{\binits{V.R.} \bsnm{{Eshleman}}},
\bauthor{\binits{D.P.} \bsnm{{Hinson}}},
\batitle{{The atmosphere of Uranus - Results of radio occultation measurements
  with Voyager 2}}.
\bjtitle{Journal of Geophysical Research}
\bvolume{92},
\bfpage{14987}--\blpage{15001}
(\byear{1987}).
doi:\doiurl{10.1029/JA092iA13p14987}
\end{barticle}
\endbibitem

\bibitem[\protect\citeauthoryear{Lindal}{1992}]{92lindal_nep}
\begin{barticle}
\bauthor{\binits{G.F.} \bsnm{Lindal}},
\batitle{The atmosphere of neptune- an analysis of radio occultation data
  acquired with voyager 2}.
\bjtitle{Astronomical Journal}
\bvolume{103},
\bfpage{967}--\blpage{982}
(\byear{1992})
\end{barticle}
\endbibitem

\bibitem[\protect\citeauthoryear{{Little} et~al.}{1999}]{99little}
\begin{barticle}
\bauthor{\binits{B.} \bsnm{{Little}}},
\bauthor{\binits{C.D.} \bsnm{{Anger}}},
\bauthor{\binits{A.P.} \bsnm{{Ingersoll}}},
\bauthor{\binits{A.R.} \bsnm{{Vasavada}}},
\bauthor{\binits{D.A.} \bsnm{{Senske}}},
\bauthor{\binits{H.H.} \bsnm{{Breneman}}},
\bauthor{\binits{W.J.} \bsnm{{Borucki}}},
\bauthor{\bsnm{{The Galileo SSI Team}}},
\batitle{{Galileo Images of Lightning on Jupiter}}.
\bjtitle{Icarus}
\bvolume{142},
\bfpage{306}--\blpage{323}
(\byear{1999}).
doi:\doiurl{10.1006/icar.1999.6195}
\end{barticle}
\endbibitem

\bibitem[\protect\citeauthoryear{{Lockwood}}{2019}]{19lockwood}
\begin{barticle}
\bauthor{\binits{G.W.} \bsnm{{Lockwood}}},
\batitle{{Final compilation of photometry of Uranus and Neptune, 1972-2016.}}
\bjtitle{Icarus}
\bvolume{324},
\bfpage{77}--\blpage{85}
(\byear{2019}).
doi:\doiurl{10.1016/j.icarus.2019.01.024}
\end{barticle}
\endbibitem

\bibitem[\protect\citeauthoryear{{Lockwood} and
  {Jerzykiewicz}}{2006}]{06lockwood}
\begin{barticle}
\bauthor{\binits{G.W.} \bsnm{{Lockwood}}},
\bauthor{\binits{M.} \bsnm{{Jerzykiewicz}}},
\batitle{{Photometric variability of Uranus and Neptune, 1950-2004}}.
\bjtitle{Icarus}
\bvolume{180},
\bfpage{442}--\blpage{452}
(\byear{2006}).
doi:\doiurl{10.1016/j.icarus.2005.09.009}
\end{barticle}
\endbibitem

\bibitem[\protect\citeauthoryear{{Lodders} and {Fegley}}{1994}]{94lodders}
\begin{barticle}
\bauthor{\binits{K.} \bsnm{{Lodders}}},
\bauthor{\binits{B.} \bsnm{{Fegley}} \bsuffix{Jr.}},
\batitle{{The origin of carbon monoxide in Neptunes's atmosphere}}.
\bjtitle{Icarus}
\bvolume{112},
\bfpage{368}--\blpage{375}
(\byear{1994}).
doi:\doiurl{10.1006/icar.1994.1190}
\end{barticle}
\endbibitem

\bibitem[\protect\citeauthoryear{{Luszcz-Cook} et~al.}{2013}]{13luszczcook}
\begin{barticle}
\bauthor{\binits{S.H.} \bsnm{{Luszcz-Cook}}},
\bauthor{\binits{I.} \bsnm{{de Pater}}},
\bauthor{\binits{M.} \bsnm{{Wright}}},
\batitle{{Spatially-resolved millimeter-wavelength maps of Neptune}}.
\bjtitle{Icarus}
\bvolume{226},
\bfpage{437}--\blpage{454}
(\byear{2013})
\end{barticle}
\endbibitem

\bibitem[\protect\citeauthoryear{{Marten} et~al.}{1993}]{93marten}
\begin{barticle}
\bauthor{\binits{A.} \bsnm{{Marten}}},
\bauthor{\binits{D.} \bsnm{{Gautier}}},
\bauthor{\binits{T.} \bsnm{{Owen}}},
\bauthor{\binits{D.B.} \bsnm{{Sanders}}},
\bauthor{\binits{H.E.} \bsnm{{Matthews}}},
\bauthor{\binits{S.K.} \bsnm{{Atreya}}},
\bauthor{\binits{R.P.J.} \bsnm{{Tilanus}}},
\bauthor{\binits{J.R.} \bsnm{{Deane}}},
\batitle{{First observations of CO and HCN on Neptune and Uranus at millimeter
  wavelengths and the implications for atmospheric chemistry}}.
\bjtitle{ApJ}
\bvolume{406},
\bfpage{285}--\blpage{297}
(\byear{1993}).
doi:\doiurl{10.1086/172440}
\end{barticle}
\endbibitem

\bibitem[\protect\citeauthoryear{{Marten} et~al.}{2005}]{05marten}
\begin{barticle}
\bauthor{\binits{A.} \bsnm{{Marten}}},
\bauthor{\binits{H.E.} \bsnm{{Matthews}}},
\bauthor{\binits{T.} \bsnm{{Owen}}},
\bauthor{\binits{R.} \bsnm{{Moreno}}},
\bauthor{\binits{T.} \bsnm{{Hidayat}}},
\bauthor{\binits{Y.} \bsnm{{Biraud}}},
\batitle{{Improved constraints on Neptune's atmosphere from
  submillimetre-wavelength observations}}.
\bjtitle{\aap}
\bvolume{429},
\bfpage{1097}--\blpage{1105}
(\byear{2005}).
doi:\doiurl{10.1051/0004-6361:20041695}
\end{barticle}
\endbibitem

\bibitem[\protect\citeauthoryear{{Martin} et~al.}{2012}]{12martin}
\begin{barticle}
\bauthor{\binits{S.C.} \bsnm{{Martin}}},
\bauthor{\binits{I.} \bsnm{{de Pater}}},
\bauthor{\binits{P.} \bsnm{{Marcus}}},
\batitle{{Neptune's zonal winds from near-IR Keck adaptive optics imaging in
  August 2001}}.
\bjtitle{Astrophysics and Space Science}
\bvolume{337},
\bfpage{65}--\blpage{78}
(\byear{2012}).
doi:\doiurl{10.1007/s10509-011-0847-y}
\end{barticle}
\endbibitem

\bibitem[\protect\citeauthoryear{{Massie} and {Hunten}}{1982}]{82massie}
\begin{barticle}
\bauthor{\binits{S.T.} \bsnm{{Massie}}},
\bauthor{\binits{D.M.} \bsnm{{Hunten}}},
\batitle{{Conversion of para and ortho hydrogen in the Jovian planets}}.
\bjtitle{Icarus}
\bvolume{49},
\bfpage{213}--\blpage{226}
(\byear{1982}).
doi:\doiurl{10.1016/0019-1035(82)90073-2}
\end{barticle}
\endbibitem

\bibitem[\protect\citeauthoryear{{McMillan}}{1992}]{92mcmillan}
\begin{botherref}
\oauthor{\binits{W.W.} \bsnm{{McMillan}}},
{Revelations of a stratospheric circulation: The dynamical transport of
  hydrocarbons in the stratosphere of Uranus},
PhD thesis,
Johns Hopkins Univ., Baltimore, MD.,
1992
\end{botherref}
\endbibitem

\bibitem[\protect\citeauthoryear{{Melin} et~al.}{2018}]{18melin}
\begin{barticle}
\bauthor{\binits{H.} \bsnm{{Melin}}},
\bauthor{\binits{L.N.} \bsnm{{Fletcher}}},
\bauthor{\binits{P.T.} \bsnm{{Donnelly}}},
\bauthor{\binits{T.K.} \bsnm{{Greathouse}}},
\bauthor{\binits{J.H.} \bsnm{{Lacy}}},
\bauthor{\binits{G.S.} \bsnm{{Orton}}},
\bauthor{\binits{R.S.} \bsnm{{Giles}}},
\bauthor{\binits{J.A.} \bsnm{{Sinclair}}},
\bauthor{\binits{P.G.J.} \bsnm{{Irwin}}},
\batitle{{Assessing the long-term variability of acetylene and ethane in the
  stratosphere of Jupiter}}.
\bjtitle{Icarus}
\bvolume{305},
\bfpage{301}--\blpage{313}
(\byear{2018}).
doi:\doiurl{10.1016/j.icarus.2017.12.041}
\end{barticle}
\endbibitem

\bibitem[\protect\citeauthoryear{{Melin} et~al.}{2019}]{19melin}
\begin{botherref}
\oauthor{\binits{H.} \bsnm{{Melin}}},
\oauthor{\binits{L.N.} \bsnm{{Fletcher}}},
\oauthor{\binits{T.S.} \bsnm{{Stallard}}},
\oauthor{\binits{S.} \bsnm{{Miller}}},
\oauthor{\binits{L.M.} \bsnm{{Trafton}}},
\oauthor{\binits{R.} \bsnm{{Vervack}}},
\oauthor{\binits{L.} \bsnm{{Moore}}},
\oauthor{\binits{J.} \bsnm{{O'Donoghue}}},
\oauthor{\binits{T.R.} \bsnm{{Geballe}}},
\oauthor{\binits{L.} \bsnm{{Lamy}}},
\oauthor{\binits{C.} \bsnm{{Tao}}},
\oauthor{\binits{N.} \bsnm{{Chowdhury}}},
{Ground-based H3+ observations of Uranus: the long-term and the short-term}.
Royal Society of London Philosophical Transactions Series A
(2019)
\end{botherref}
\endbibitem

\bibitem[\protect\citeauthoryear{{Millot} et~al.}{2019}]{19millot}
\begin{barticle}
\bauthor{\binits{M.} \bsnm{{Millot}}},
\bauthor{\binits{F.} \bsnm{{Coppari}}},
\bauthor{\binits{J.R.} \bsnm{{Rygg}}},
\bauthor{\binits{A.} \bsnm{{Correa Barrios}}},
\bauthor{\binits{S.} \bsnm{{Hamel}}},
\bauthor{\binits{D.C.} \bsnm{{Swift}}},
\bauthor{\binits{J.H.} \bsnm{{Eggert}}},
\batitle{{Nanosecond X-ray diffraction of shock-compressed superionic water
  ice}}.
\bjtitle{\nat}
\bvolume{569}(\bissue{7755}),
\bfpage{251}--\blpage{255}
(\byear{2019}).
doi:\doiurl{10.1038/s41586-019-1114-6}
\end{barticle}
\endbibitem

\bibitem[\protect\citeauthoryear{{Molter} et~al.}{2019}]{18molter}
\begin{barticle}
\bauthor{\binits{E.} \bsnm{{Molter}}},
\bauthor{\binits{I.} \bsnm{{de Pater}}},
\bauthor{\binits{S.} \bsnm{{Luszcz-Cook}}},
\bauthor{\binits{R.} \bsnm{{Hueso}}},
\bauthor{\binits{J.} \bsnm{{Tollefson}}},
\bauthor{\binits{C.} \bsnm{{Alvarez}}},
\bauthor{\binits{A.} \bsnm{{S{\'a}nchez-Lavega}}},
\bauthor{\binits{M.H.} \bsnm{{Wong}}},
\bauthor{\binits{A.I.} \bsnm{{Hsu}}},
\bauthor{\binits{L.A.} \bsnm{{Sromovsky}}},
\batitle{{Analysis of Neptune's 2017 bright equatorial storm}}.
\bjtitle{Icarus}
\bvolume{321},
\bfpage{324}--\blpage{345}
(\byear{2019}).
doi:\doiurl{10.1016/j.icarus.2018.11.018}
\end{barticle}
\endbibitem

\bibitem[\protect\citeauthoryear{{Molter} et~al.}{2020}]{20molter}
\begin{botherref}
\oauthor{\binits{E.M.} \bsnm{{Molter}}},
\oauthor{\binits{I.} \bsnm{{de Pater}}},
\oauthor{\binits{S.H.} \bsnm{{Luszcz-Cook}}},
\oauthor{\binits{J.W.} \bsnm{{Tollefson}}},
\oauthor{\binits{R.J.} \bsnm{{Sault}}},
\oauthor{\binits{B.} \bsnm{{Butler}}},
\oauthor{\binits{D.} \bsnm{{de Boer}}},
{Tropospheric Composition and Circulation of Uranus with ALMA and the VLA}.
submitted
(2020)
\end{botherref}
\endbibitem

\bibitem[\protect\citeauthoryear{{Moreno} et~al.}{2009}]{09moreno_dps}
\begin{bchapter}
\bauthor{\binits{R.} \bsnm{{Moreno}}},
\bauthor{\binits{A.} \bsnm{{Marten}}},
\bauthor{\binits{E.} \bsnm{{Lellouch}}},
\bctitle{{Search for PH3 in the Atmospheres of Uranus and Neptune at Millimeter
  Wavelength}},
in \bbtitle{AAS/Division for Planetary Sciences Meeting Abstracts \#41}.
\bsertitle{AAS/Division for Planetary Sciences Meeting Abstracts},
\byear{2009},
pp. \bfpage{28}--\blpage{02}
\end{bchapter}
\endbibitem

\bibitem[\protect\citeauthoryear{{Moses} and {Poppe}}{2017}]{17moses}
\begin{botherref}
\oauthor{\binits{J.I.} \bsnm{{Moses}}},
\oauthor{\binits{A.R.} \bsnm{{Poppe}}},
{Dust Ablation on the Giant Planets: Consequences for Stratospheric
  Photochemistry}.
Icarus, submitted
(2017)
\end{botherref}
\endbibitem

\bibitem[\protect\citeauthoryear{{Moses} et~al.}{1992}]{92moses}
\begin{barticle}
\bauthor{\binits{J.I.} \bsnm{{Moses}}},
\bauthor{\binits{M.} \bsnm{{Allen}}},
\bauthor{\binits{Y.L.} \bsnm{{Yung}}},
\batitle{{Hydrocarbon nucleation and aerosol formation in Neptune's
  atmosphere}}.
\bjtitle{Icarus}
\bvolume{99},
\bfpage{318}--\blpage{346}
(\byear{1992}).
doi:\doiurl{10.1016/0019-1035(92)90149-2}
\end{barticle}
\endbibitem

\bibitem[\protect\citeauthoryear{{Moses} et~al.}{2005}]{05moses}
\begin{barticle}
\bauthor{\binits{J.I.} \bsnm{{Moses}}},
\bauthor{\binits{T.} \bsnm{{Fouchet}}},
\bauthor{\binits{B.} \bsnm{{B{\'e}zard}}},
\bauthor{\binits{G.R.} \bsnm{{Gladstone}}},
\bauthor{\binits{E.} \bsnm{{Lellouch}}},
\bauthor{\binits{H.} \bsnm{{Feuchtgruber}}},
\batitle{{Photochemistry and diffusion in Jupiter's stratosphere: Constraints
  from ISO observations and comparisons with other giant planets}}.
\bjtitle{Journal of Geophysical Research (Planets)}
\bvolume{110},
\bfpage{08001}
(\byear{2005}).
doi:\doiurl{10.1029/2005JE002411}
\end{barticle}
\endbibitem

\bibitem[\protect\citeauthoryear{{Moses} et~al.}{2018}]{18moses}
\begin{barticle}
\bauthor{\binits{J.I.} \bsnm{{Moses}}},
\bauthor{\binits{L.N.} \bsnm{{Fletcher}}},
\bauthor{\binits{T.K.} \bsnm{{Greathouse}}},
\bauthor{\binits{G.S.} \bsnm{{Orton}}},
\bauthor{\binits{V.} \bsnm{{Hue}}},
\batitle{{Seasonal stratospheric photochemistry on Uranus and Neptune}}.
\bjtitle{Icarus}
\bvolume{307},
\bfpage{124}--\blpage{145}
(\byear{2018}).
doi:\doiurl{10.1016/j.icarus.2018.02.004}
\end{barticle}
\endbibitem

\bibitem[\protect\citeauthoryear{{Mousis} et~al.}{2018}]{18mousis}
\begin{barticle}
\bauthor{\binits{O.} \bsnm{{Mousis}}},
\bauthor{\binits{D.H.} \bsnm{{Atkinson}}},
\bauthor{\binits{T.} \bsnm{{Cavali{\'e}}}},
\bauthor{\binits{L.N.} \bsnm{{Fletcher}}},
\bauthor{\binits{M.J.} \bsnm{{Amato}}},
\bauthor{\binits{S.} \bsnm{{Aslam}}},
\bauthor{\binits{F.} \bsnm{{Ferri}}},
\bauthor{\binits{J.-B.} \bsnm{{Renard}}},
\bauthor{\binits{T.} \bsnm{{Spilker}}},
\bauthor{\binits{E.} \bsnm{{Venkatapathy}}},
\bauthor{\binits{P.} \bsnm{{Wurz}}},
\bauthor{\binits{K.} \bsnm{{Aplin}}},
\bauthor{\binits{A.} \bsnm{{Coustenis}}},
\bauthor{\binits{M.} \bsnm{{Deleuil}}},
\bauthor{\binits{M.} \bsnm{{Dobrijevic}}},
\bauthor{\binits{T.} \bsnm{{Fouchet}}},
\bauthor{\binits{T.} \bsnm{{Guillot}}},
\bauthor{\binits{P.} \bsnm{{Hartogh}}},
\bauthor{\binits{T.} \bsnm{{Hewagama}}},
\bauthor{\binits{M.D.} \bsnm{{Hofstadter}}},
\bauthor{\binits{V.} \bsnm{{Hue}}},
\bauthor{\binits{R.} \bsnm{{Hueso}}},
\bauthor{\binits{J.-P.} \bsnm{{Lebreton}}},
\bauthor{\binits{E.} \bsnm{{Lellouch}}},
\bauthor{\binits{J.} \bsnm{{Moses}}},
\bauthor{\binits{G.S.} \bsnm{{Orton}}},
\bauthor{\binits{J.C.} \bsnm{{Pearl}}},
\bauthor{\binits{A.} \bsnm{{S{\'a}nchez-Lavega}}},
\bauthor{\binits{A.} \bsnm{{Simon}}},
\bauthor{\binits{O.} \bsnm{{Venot}}},
\bauthor{\binits{J.H.} \bsnm{{Waite}}},
\bauthor{\binits{R.K.} \bsnm{{Achterberg}}},
\bauthor{\binits{S.} \bsnm{{Atreya}}},
\bauthor{\binits{F.} \bsnm{{Billebaud}}},
\bauthor{\binits{M.} \bsnm{{Blanc}}},
\bauthor{\binits{F.} \bsnm{{Borget}}},
\bauthor{\binits{B.} \bsnm{{Brugger}}},
\bauthor{\binits{S.} \bsnm{{Charnoz}}},
\bauthor{\binits{T.} \bsnm{{Chiavassa}}},
\bauthor{\binits{V.} \bsnm{{Cottini}}},
\bauthor{\binits{L.} \bsnm{{d'Hendecourt}}},
\bauthor{\binits{G.} \bsnm{{Danger}}},
\bauthor{\binits{T.} \bsnm{{Encrenaz}}},
\bauthor{\binits{N.J.P.} \bsnm{{Gorius}}},
\bauthor{\binits{L.} \bsnm{{Jorda}}},
\bauthor{\binits{B.} \bsnm{{Marty}}},
\bauthor{\binits{R.} \bsnm{{Moreno}}},
\bauthor{\binits{A.} \bsnm{{Morse}}},
\bauthor{\binits{C.} \bsnm{{Nixon}}},
\bauthor{\binits{K.} \bsnm{{Reh}}},
\bauthor{\binits{T.} \bsnm{{Ronnet}}},
\bauthor{\binits{F.-X.} \bsnm{{Schmider}}},
\bauthor{\binits{S.} \bsnm{{Sheridan}}},
\bauthor{\binits{C.} \bsnm{{Sotin}}},
\bauthor{\binits{P.} \bsnm{{Vernazza}}},
\bauthor{\binits{G.L.} \bsnm{{Villanueva}}},
\batitle{{Scientific rationale for Uranus and Neptune in situ explorations}}.
\bjtitle{Plan. \& Space Sci.}
\bvolume{155},
\bfpage{12}--\blpage{40}
(\byear{2018}).
doi:\doiurl{10.1016/j.pss.2017.10.005}
\end{barticle}
\endbibitem

\bibitem[\protect\citeauthoryear{{Ness} et~al.}{1986}]{86ness}
\begin{barticle}
\bauthor{\binits{N.F.} \bsnm{{Ness}}},
\bauthor{\binits{M.H.} \bsnm{{Acuna}}},
\bauthor{\binits{K.W.} \bsnm{{Behannon}}},
\bauthor{\binits{L.F.} \bsnm{{Burlaga}}},
\bauthor{\binits{J.E.P.} \bsnm{{Connerney}}},
\bauthor{\binits{R.P.} \bsnm{{Lepping}}},
\bauthor{\binits{F.M.} \bsnm{{Neubauer}}},
\batitle{{Magnetic Fields at Uranus}}.
\bjtitle{Science}
\bvolume{233}(\bissue{4759}),
\bfpage{85}--\blpage{89}
(\byear{1986}).
doi:\doiurl{10.1126/science.233.4759.85}
\end{barticle}
\endbibitem

\bibitem[\protect\citeauthoryear{{Ness} et~al.}{1989}]{89ness}
\begin{barticle}
\bauthor{\binits{N.F.} \bsnm{{Ness}}},
\bauthor{\binits{M.H.} \bsnm{{Acuna}}},
\bauthor{\binits{L.F.} \bsnm{{Burlaga}}},
\bauthor{\binits{J.E.P.} \bsnm{{Connerney}}},
\bauthor{\binits{R.P.} \bsnm{{Lepping}}},
\bauthor{\binits{F.M.} \bsnm{{Neubauer}}},
\batitle{{Magnetic Fields at Neptune}}.
\bjtitle{Science}
\bvolume{246}(\bissue{4936}),
\bfpage{1473}--\blpage{1478}
(\byear{1989}).
doi:\doiurl{10.1126/science.246.4936.1473}
\end{barticle}
\endbibitem

\bibitem[\protect\citeauthoryear{{Nixon} et~al.}{2007}]{07nixon}
\begin{barticle}
\bauthor{\binits{C.A.} \bsnm{{Nixon}}},
\bauthor{\binits{R.K.} \bsnm{{Achterberg}}},
\bauthor{\binits{B.J.} \bsnm{{Conrath}}},
\bauthor{\binits{P.G.J.} \bsnm{{Irwin}}},
\bauthor{\binits{N.A.} \bsnm{{Teanby}}},
\bauthor{\binits{T.} \bsnm{{Fouchet}}},
\bauthor{\binits{P.D.} \bsnm{{Parrish}}},
\bauthor{\binits{P.N.} \bsnm{{Romani}}},
\bauthor{\binits{M.} \bsnm{{Abbas}}},
\bauthor{\binits{A.} \bsnm{{Leclair}}},
\bauthor{\binits{D.} \bsnm{{Strobel}}},
\bauthor{\binits{A.A.} \bsnm{{Simon-Miller}}},
\bauthor{\binits{D.J.} \bsnm{{Jennings}}},
\bauthor{\binits{F.M.} \bsnm{{Flasar}}},
\bauthor{\binits{V.G.} \bsnm{{Kunde}}},
\batitle{{Meridional variations of C$_{2}$H$_{2}$ and C$_{2}$H$_{6}$ in
  Jupiter's atmosphere from Cassini CIRS infrared spectra}}.
\bjtitle{Icarus}
\bvolume{188},
\bfpage{47}--\blpage{71}
(\byear{2007}).
doi:\doiurl{10.1016/j.icarus.2006.11.016}
\end{barticle}
\endbibitem

\bibitem[\protect\citeauthoryear{{Orton} et~al.}{1987}]{87orton}
\begin{barticle}
\bauthor{\binits{G.S.} \bsnm{{Orton}}},
\bauthor{\binits{D.K.} \bsnm{{Aitken}}},
\bauthor{\binits{C.} \bsnm{{Smith}}},
\bauthor{\binits{P.F.} \bsnm{{Roche}}},
\bauthor{\binits{J.} \bsnm{{Caldwell}}},
\bauthor{\binits{R.} \bsnm{{Snyder}}},
\batitle{{The spectra of Uranus and Neptune at 8-14 and 17-23 microns}}.
\bjtitle{Icarus}
\bvolume{70},
\bfpage{1}--\blpage{12}
(\byear{1987}).
doi:\doiurl{10.1016/0019-1035(87)90070-4}
\end{barticle}
\endbibitem

\bibitem[\protect\citeauthoryear{{Orton} et~al.}{1998}]{98orton}
\begin{barticle}
\bauthor{\binits{G.S.} \bsnm{{Orton}}},
\bauthor{\binits{B.M.} \bsnm{{Fisher}}},
\bauthor{\binits{K.H.} \bsnm{{Baines}}},
\bauthor{\binits{S.T.} \bsnm{{Stewart}}},
\bauthor{\binits{A.J.} \bsnm{{Friedson}}},
\bauthor{\binits{J.L.} \bsnm{{Ortiz}}},
\bauthor{\binits{M.} \bsnm{{Marinova}}},
\bauthor{\binits{M.} \bsnm{{Ressler}}},
\bauthor{\binits{A.} \bsnm{{Dayal}}},
\bauthor{\binits{W.} \bsnm{{Hoffmann}}},
\bauthor{\binits{J.} \bsnm{{Hora}}},
\bauthor{\binits{S.} \bsnm{{Hinkley}}},
\bauthor{\binits{V.} \bsnm{{Krishnan}}},
\bauthor{\binits{M.} \bsnm{{Masanovic}}},
\bauthor{\binits{J.} \bsnm{{Tesic}}},
\bauthor{\binits{A.} \bsnm{{Tziolas}}},
\bauthor{\binits{K.C.} \bsnm{{Parija}}},
\batitle{{Characteristics of the Galileo probe entry site from Earth-based
  remote sensing observations}}.
\bjtitle{J. Geophys. Res.}
\bvolume{103},
\bfpage{22791}--\blpage{22814}
(\byear{1998}).
doi:\doiurl{10.1029/98JE02380}
\end{barticle}
\endbibitem

\bibitem[\protect\citeauthoryear{{Orton} et~al.}{2007}]{07orton}
\begin{barticle}
\bauthor{\binits{G.S.} \bsnm{{Orton}}},
\bauthor{\binits{M.} \bsnm{{Gustafsson}}},
\bauthor{\binits{M.} \bsnm{{Burgdorf}}},
\bauthor{\binits{V.} \bsnm{{Meadows}}},
\batitle{{Revised Ab Initio Models for H$_2$-H$_2$ Collision Induced Absorption
  at Low Temperatures}}.
\bjtitle{Icarus}
\bvolume{189},
\bfpage{544}--\blpage{549}
(\byear{2007})
\end{barticle}
\endbibitem

\bibitem[\protect\citeauthoryear{{Orton} et~al.}{2012}]{12orton}
\begin{barticle}
\bauthor{\binits{G.S.} \bsnm{{Orton}}},
\bauthor{\binits{L.N.} \bsnm{{Fletcher}}},
\bauthor{\binits{J.} \bsnm{{Liu}}},
\bauthor{\binits{T.} \bsnm{{Schneider}}},
\bauthor{\binits{P.A.} \bsnm{{Yanamandra-Fisher}}},
\bauthor{\binits{I.} \bsnm{{de Pater}}},
\bauthor{\binits{M.} \bsnm{{Edwards}}},
\bauthor{\binits{T.R.} \bsnm{{Geballe}}},
\bauthor{\binits{H.B.} \bsnm{{Hammel}}},
\bauthor{\binits{T.} \bsnm{{Fujiyoshi}}},
\bauthor{\binits{T.} \bsnm{{Encrenaz}}},
\bauthor{\binits{E.} \bsnm{{Pantin}}},
\bauthor{\binits{O.} \bsnm{{Mousis}}},
\bauthor{\binits{T.} \bsnm{{Fuse}}},
\batitle{{Recovery and characterization of Neptune's near-polar stratospheric
  hot spot}}.
\bjtitle{Plan. \& Space Sci.}
\bvolume{61},
\bfpage{161}--\blpage{167}
(\byear{2012}).
doi:\doiurl{10.1016/j.pss.2011.06.013}
\end{barticle}
\endbibitem

\bibitem[\protect\citeauthoryear{{Orton} et~al.}{2014}]{14orton}
\begin{barticle}
\bauthor{\binits{G.S.} \bsnm{{Orton}}},
\bauthor{\binits{L.N.} \bsnm{{Fletcher}}},
\bauthor{\binits{J.I.} \bsnm{{Moses}}},
\bauthor{\binits{A.K.} \bsnm{{Mainzer}}},
\bauthor{\binits{D.} \bsnm{{Hines}}},
\bauthor{\binits{H.B.} \bsnm{{Hammel}}},
\bauthor{\binits{F.J.} \bsnm{{Martin-Torres}}},
\bauthor{\binits{M.} \bsnm{{Burgdorf}}},
\bauthor{\binits{C.} \bsnm{{Merlet}}},
\bauthor{\binits{M.R.} \bsnm{{Line}}},
\batitle{{Mid-infrared spectroscopy of Uranus from the Spitzer Infrared
  Spectrometer: 1. Determination of the mean temperature structure of the upper
  troposphere and stratosphere}}.
\bjtitle{Icarus}
\bvolume{243},
\bfpage{494}--\blpage{513}
(\byear{2014}).
doi:\doiurl{10.1016/j.icarus.2014.07.010}
\end{barticle}
\endbibitem

\bibitem[\protect\citeauthoryear{{Orton} et~al.}{2015}]{15orton}
\begin{barticle}
\bauthor{\binits{G.S.} \bsnm{{Orton}}},
\bauthor{\binits{L.N.} \bsnm{{Fletcher}}},
\bauthor{\binits{T.} \bsnm{{Encrenaz}}},
\bauthor{\binits{C.} \bsnm{{Leyrat}}},
\bauthor{\binits{H.G.} \bsnm{{Roe}}},
\bauthor{\binits{T.} \bsnm{{Fujiyoshi}}},
\bauthor{\binits{E.} \bsnm{{Pantin}}},
\batitle{{Thermal imaging of Uranus: Upper-tropospheric temperatures one season
  after Voyager}}.
\bjtitle{Icarus}
\bvolume{260},
\bfpage{94}--\blpage{102}
(\byear{2015}).
doi:\doiurl{10.1016/j.icarus.2015.07.004}
\end{barticle}
\endbibitem

\bibitem[\protect\citeauthoryear{{Orton} et~al.}{2018}]{18orton_cospar}
\begin{bchapter}
\bauthor{\binits{G.} \bsnm{{Orton}}},
\bauthor{\binits{J.} \bsnm{{Moses}}},
\bauthor{\binits{T.} \bsnm{{Encrenaz}}},
\bauthor{\binits{L.} \bsnm{{Fletcher}}},
\bauthor{\binits{T.} \bsnm{{Greathouse}}},
\bauthor{\binits{C.} \bsnm{{Leyrat}}},
\bauthor{\binits{J.} \bsnm{{Sinclair}}},
\bauthor{\binits{L.} \bsnm{{Trafton}}},
\bauthor{\binits{J.} \bsnm{{Lacy}}},
\bauthor{\binits{E.} \bsnm{{Pantin}}},
\bctitle{{Spatial Variability in the Stratosphere of Uranus}},
in \bbtitle{42nd COSPAR Scientific Assembly},
vol. \bseriesno{42},
\byear{2018},
pp. \bfpage{5}--\blpage{4318}
\end{bchapter}
\endbibitem

\bibitem[\protect\citeauthoryear{Orton et~al.}{2007}]{07orton_nep}
\begin{barticle}
\bauthor{\binits{G.S.} \bsnm{Orton}},
\bauthor{\binits{T.} \bsnm{Encrenaz}},
\bauthor{\binits{C.} \bsnm{Leyrat}},
\bauthor{\binits{R.} \bsnm{Puetter}},
\bauthor{\binits{A.J.} \bsnm{Friedson}},
\batitle{{Evidence for methane escape and strong seasonal and dynamical
  perturbations of Neptune's atmospheric temperatures}}.
\bjtitle{A\&A}
\bvolume{473},
\bfpage{5}--\blpage{8}
(\byear{2007})
\end{barticle}
\endbibitem

\bibitem[\protect\citeauthoryear{{Pearl} and {Conrath}}{1991}]{91pearl}
\begin{barticle}
\bauthor{\binits{J.C.} \bsnm{{Pearl}}},
\bauthor{\binits{B.J.} \bsnm{{Conrath}}},
\batitle{{The albedo, effective temperature, and energy balance of Neptune, as
  determined from Voyager data}}.
\bjtitle{J. Geophys. Res.}
\bvolume{96},
\bfpage{18921}
(\byear{1991})
\end{barticle}
\endbibitem

\bibitem[\protect\citeauthoryear{{Pollack} et~al.}{1987}]{87pollack}
\begin{barticle}
\bauthor{\binits{J.B.} \bsnm{{Pollack}}},
\bauthor{\binits{K.} \bsnm{{Rages}}},
\bauthor{\binits{S.K.} \bsnm{{Pope}}},
\bauthor{\binits{M.G.} \bsnm{{Tomasko}}},
\bauthor{\binits{P.N.} \bsnm{{Romani}}},
\bauthor{\binits{S.K.} \bsnm{{Atreya}}},
\batitle{{Nature of the stratospheric haze on Uranus: Evidence for condensed
  hydrocarbons}}.
\bjtitle{\jgr}
\bvolume{92}(\bissue{A13}),
\bfpage{15037}--\blpage{15065}
(\byear{1987}).
doi:\doiurl{10.1029/JA092iA13p15037}
\end{barticle}
\endbibitem

\bibitem[\protect\citeauthoryear{{Poppe}}{2016}]{16poppe}
\begin{barticle}
\bauthor{\binits{A.R.} \bsnm{{Poppe}}},
\batitle{{An improved model for interplanetary dust fluxes in the outer Solar
  System}}.
\bjtitle{Icarus}
\bvolume{264},
\bfpage{369}--\blpage{386}
(\byear{2016}).
doi:\doiurl{10.1016/j.icarus.2015.10.001}
\end{barticle}
\endbibitem

\bibitem[\protect\citeauthoryear{{Rages} et~al.}{2004}]{04rages}
\begin{barticle}
\bauthor{\binits{K.A.} \bsnm{{Rages}}},
\bauthor{\binits{H.B.} \bsnm{{Hammel}}},
\bauthor{\binits{A.J.} \bsnm{{Friedson}}},
\batitle{{Evidence for temporal change at Uranus' south pole}}.
\bjtitle{Icarus}
\bvolume{172}(\bissue{2}),
\bfpage{548}--\blpage{554}
(\byear{2004}).
doi:\doiurl{10.1016/j.icarus.2004.07.009}
\end{barticle}
\endbibitem

\bibitem[\protect\citeauthoryear{{Rages} et~al.}{1991}]{91rages}
\begin{barticle}
\bauthor{\binits{K.} \bsnm{{Rages}}},
\bauthor{\binits{J.B.} \bsnm{{Pollack}}},
\bauthor{\binits{M.G.} \bsnm{{Tomasko}}},
\bauthor{\binits{L.R.} \bsnm{{Doose}}},
\batitle{{Properties of scatterers in the troposphere and lower stratosphere of
  Uranus based on Voyager imaging data}}.
\bjtitle{Icarus}
\bvolume{89}(\bissue{2}),
\bfpage{359}--\blpage{376}
(\byear{1991}).
doi:\doiurl{10.1016/0019-1035(91)90183-T}
\end{barticle}
\endbibitem

\bibitem[\protect\citeauthoryear{{Roman} et~al.}{2018}]{18roman}
\begin{barticle}
\bauthor{\binits{M.T.} \bsnm{{Roman}}},
\bauthor{\binits{D.} \bsnm{{Banfield}}},
\bauthor{\binits{P.J.} \bsnm{{Gierasch}}},
\batitle{{Aerosols and methane in the ice giant atmospheres inferred from
  spatially resolved, near-infrared spectra: I. Uranus, 2001-2007}}.
\bjtitle{Icarus}
\bvolume{310},
\bfpage{54}--\blpage{76}
(\byear{2018}).
doi:\doiurl{10.1016/j.icarus.2017.10.036}
\end{barticle}
\endbibitem

\bibitem[\protect\citeauthoryear{{Roman} et~al.}{2020}]{19roman}
\begin{barticle}
\bauthor{\binits{M.T.} \bsnm{{Roman}}},
\bauthor{\binits{L.N.} \bsnm{{Fletcher}}},
\bauthor{\binits{G.S.} \bsnm{{Orton}}},
\bauthor{\binits{N.} \bsnm{{Rowe-Gurney}}},
\bauthor{\binits{P.G.J.} \bsnm{{Irwin}}},
\batitle{{Uranus in Northern Midspring: Persistent Atmospheric Temperatures and
  Circulations Inferred from Thermal Imaging}}.
\bjtitle{Astronomical Journal}
\bvolume{159}(\bissue{2}),
\bfpage{45}
(\byear{2020}).
doi:\doiurl{10.3847/1538-3881/ab5dc7}
\end{barticle}
\endbibitem

\bibitem[\protect\citeauthoryear{{Romani} et~al.}{1993}]{93romani}
\begin{barticle}
\bauthor{\binits{P.N.} \bsnm{{Romani}}},
\bauthor{\binits{J.} \bsnm{{Bishop}}},
\bauthor{\binits{B.} \bsnm{{Bezard}}},
\bauthor{\binits{S.} \bsnm{{Atreya}}},
\batitle{{Methane photochemistry on Neptune - Ethane and acetylene mixing
  ratios and haze production}}.
\bjtitle{Icarus}
\bvolume{106},
\bfpage{442}
(\byear{1993}).
doi:\doiurl{10.1006/icar.1993.1184}
\end{barticle}
\endbibitem

\bibitem[\protect\citeauthoryear{Salyk et~al.}{2006}]{06salyk}
\begin{barticle}
\bauthor{\binits{C.} \bsnm{Salyk}},
\bauthor{\binits{A.P.} \bsnm{Ingersoll}},
\bauthor{\binits{J.} \bsnm{Lorre}},
\bauthor{\binits{A.} \bsnm{Vasavada}},
\bauthor{\binits{A.D.} \bsnm{Del~Genio}},
\batitle{{Interaction between eddies and mean flow in Jupiter's atmosphere:
  Analysis of Cassini imaging data}}.
\bjtitle{Icarus}
\bvolume{185}(\bissue{2}),
\bfpage{430}--\blpage{442}
(\byear{2006})
\end{barticle}
\endbibitem

\bibitem[\protect\citeauthoryear{{Sanchez-Lavega}
  et~al.}{2019}]{18sanchez_jets}
\begin{bchapter}
\bauthor{\binits{A.} \bsnm{{Sanchez-Lavega}}},
\bauthor{\binits{L.A.} \bsnm{{Sromovsky}}},
\bauthor{\binits{A.P.} \bsnm{{Showman}}},
\bauthor{\binits{A.D.} \bsnm{{Del Genio}}},
\bauthor{\binits{R.M.B.} \bsnm{{Young}}},
\bauthor{\binits{R.} \bsnm{{Hueso}}},
\bauthor{\binits{E.} \bsnm{{Garc{\'{\i}}a-Melendo}}},
\bauthor{\binits{Y.} \bsnm{{Kaspi}}},
\bauthor{\binits{G.S.} \bsnm{{Orton}}},
\bauthor{\binits{N.} \bsnm{{Barrado-Izagirre}}},
\bauthor{\binits{D.S.} \bsnm{{Choi}}},
\bauthor{\binits{J.M.} \bsnm{{Barbara}}},
\bctitle{{Gas Giants}},
ed. by \beditor{\binits{B.} \bsnm{{Galperin}}},
\beditor{\binits{P.L.} \bsnm{{Read}}}
(\bpublisher{Cambridge University Press}, \blocation{???}, \byear{2019}).
\bcomment{Chap. 4}
\end{bchapter}
\endbibitem

\bibitem[\protect\citeauthoryear{{Showman} and {de Pater}}{2005}]{05showman}
\begin{barticle}
\bauthor{\binits{A.P.} \bsnm{{Showman}}},
\bauthor{\binits{I.} \bsnm{{de Pater}}},
\batitle{{Dynamical implications of Jupiter's tropospheric ammonia abundance}}.
\bjtitle{Icarus}
\bvolume{174},
\bfpage{192}--\blpage{204}
(\byear{2005}).
doi:\doiurl{10.1016/j.icarus.2004.10.004}
\end{barticle}
\endbibitem

\bibitem[\protect\citeauthoryear{{Showman} et~al.}{2013}]{13showman}
\begin{botherref}
\oauthor{\binits{A.P.} \bsnm{{Showman}}},
\oauthor{\binits{R.D.} \bsnm{{Wordsworth}}},
\oauthor{\binits{T.M.} \bsnm{{Merlis}}},
\oauthor{\binits{Y.} \bsnm{{Kaspi}}},
{Atmospheric Circulation of Terrestrial Exoplanets},
ed. by S.J. {Mackwell}, A.A. {Simon-Miller}, J.W. {Harder}, M.A. {Bullock}
2013,
pp. 277--326.
doi:\doiurl{10.2458/azu\_uapress\_9780816530595-ch12}
\end{botherref}
\endbibitem

\bibitem[\protect\citeauthoryear{{Showman} et~al.}{2018}]{18showman}
\begin{bchapter}
\bauthor{\binits{A.P.} \bsnm{{Showman}}},
\bauthor{\binits{A.P.} \bsnm{{Ingersoll}}},
\bauthor{\binits{R.K.} \bsnm{{Achterberg}}},
\bauthor{\binits{Y.} \bsnm{{Kaspi}}},
\bctitle{{The Global Atmospheric Circulation of Saturn}},
ed. by \beditor{\binits{K.H.} \bsnm{{Baines}}},
\beditor{\binits{F.M.} \bsnm{{Flasar}}},
\beditor{\binits{N.} \bsnm{{Krupp}}},
\beditor{\binits{T.S.} \bsnm{{Stallard}}}
(\bpublisher{Cambridge Planetary Science. Cambridge Univ. Press, New York},
  \blocation{???}, \byear{2018}).
\bcomment{Chap. 11}
\end{bchapter}
\endbibitem

\bibitem[\protect\citeauthoryear{{Showman} and {Polvani}}{2011}]{11showman}
\begin{barticle}
\bauthor{\binits{A.P.} \bsnm{{Showman}}},
\bauthor{\binits{L.M.} \bsnm{{Polvani}}},
\batitle{{Equatorial Superrotation on Tidally Locked Exoplanets}}.
\bjtitle{\apj}
\bvolume{738}(\bissue{1}),
\bfpage{71}
(\byear{2011}).
doi:\doiurl{10.1088/0004-637X/738/1/71}
\end{barticle}
\endbibitem

\bibitem[\protect\citeauthoryear{{Simon} et~al.}{2019}]{19simon}
\begin{barticle}
\bauthor{\binits{A.A.} \bsnm{{Simon}}},
\bauthor{\binits{M.H.} \bsnm{{Wong}}},
\bauthor{\binits{A.I.} \bsnm{{Hsu}}},
\batitle{{Formation of a New Great Dark Spot on Neptune in 2018}}.
\bjtitle{\grl}
\bvolume{46}(\bissue{6}),
\bfpage{3108}--\blpage{3113}
(\byear{2019}).
doi:\doiurl{10.1029/2019GL081961}
\end{barticle}
\endbibitem

\bibitem[\protect\citeauthoryear{{Simon} et~al.}{2016}]{16simon}
\begin{barticle}
\bauthor{\binits{A.A.} \bsnm{{Simon}}},
\bauthor{\binits{J.F.} \bsnm{{Rowe}}},
\bauthor{\binits{P.} \bsnm{{Gaulme}}},
\bauthor{\binits{H.B.} \bsnm{{Hammel}}},
\bauthor{\binits{S.L.} \bsnm{{Casewell}}},
\bauthor{\binits{J.J.} \bsnm{{Fortney}}},
\bauthor{\binits{J.E.} \bsnm{{Gizis}}},
\bauthor{\binits{J.J.} \bsnm{{Lissauer}}},
\bauthor{\binits{R.} \bsnm{{Morales-Juberias}}},
\bauthor{\binits{G.S.} \bsnm{{Orton}}},
\bauthor{\binits{M.H.} \bsnm{{Wong}}},
\bauthor{\binits{M.S.} \bsnm{{Marley}}},
\batitle{{Neptune's Dynamic Atmosphere from Kepler K2 Observations:
  Implications for Brown Dwarf Light Curve Analyses}}.
\bjtitle{{Astrophys. J.}}
\bvolume{817},
\bfpage{162}
(\byear{2016}).
doi:\doiurl{10.3847/0004-637X/817/2/162}
\end{barticle}
\endbibitem

\bibitem[\protect\citeauthoryear{{Sinclair} et~al.}{2020}]{20sinclair}
\begin{botherref}
\oauthor{\binits{J.A.} \bsnm{{Sinclair}}},
\oauthor{\binits{G.S.} \bsnm{{Orton}}},
\oauthor{\binits{L.N.} \bsnm{{Fletcher}}},
\oauthor{\binits{M.T.} \bsnm{{Roman}}},
\oauthor{\binits{I.} \bsnm{{de Pater}}},
\oauthor{\binits{T.} \bsnm{{Encrenaz}}},
\oauthor{\binits{H.B.} \bsnm{{Hammel}}},
\oauthor{\binits{R.S.} \bsnm{{Giles}}},
\oauthor{\binits{T.} \bsnm{{Velusamy}}},
\oauthor{\binits{J.I.} \bsnm{{Moses}}},
\oauthor{\binits{P.G.J.} \bsnm{{Irwin}}},
\oauthor{\binits{T.W.} \bsnm{{Momary}}},
\oauthor{\binits{N.} \bsnm{{Rowe-Gurney}}},
\oauthor{\binits{F.} \bsnm{{Tabataba-Vakili}}},
{Spatial structure in Neptune's 7.90-$\mu$m stratospheric CH$_4$ emission as
  measured by VLT-VISIR}.
Icarus, submitted
(2020)
\end{botherref}
\endbibitem

\bibitem[\protect\citeauthoryear{{Smith} et~al.}{1986}]{86smith}
\begin{barticle}
\bauthor{\binits{B.A.} \bsnm{{Smith}}},
\bauthor{\binits{L.A.} \bsnm{{Soderblom}}},
\bauthor{\binits{R.} \bsnm{{Beebe}}},
\bauthor{\binits{D.} \bsnm{{Bliss}}},
\bauthor{\binits{R.H.} \bsnm{{Brown}}},
\bauthor{\binits{S.A.} \bsnm{{Collins}}},
\bauthor{\binits{J.M.} \bsnm{{Boyce}}},
\bauthor{\binits{G.A.} \bsnm{{Briggs}}},
\bauthor{\binits{A.} \bsnm{{Brahic}}},
\bauthor{\binits{J.N.} \bsnm{{Cuzzi}}},
\bauthor{\binits{D.} \bsnm{{Morrison}}},
\batitle{{Voyager 2 in the Uranian system - Imaging science results}}.
\bjtitle{Science}
\bvolume{233},
\bfpage{43}--\blpage{64}
(\byear{1986}).
doi:\doiurl{10.1126/science.233.4759.43}
\end{barticle}
\endbibitem

\bibitem[\protect\citeauthoryear{{Smith} et~al.}{1989}]{89smith}
\begin{barticle}
\bauthor{\binits{B.A.} \bsnm{{Smith}}},
\bauthor{\binits{L.A.} \bsnm{{Soderblom}}},
\bauthor{\binits{D.} \bsnm{{Banfield}}},
\bauthor{\binits{C.} \bsnm{{Barnet}}},
\bauthor{\binits{A.T.} \bsnm{{Basilevsky}}},
\bauthor{\binits{R.F.} \bsnm{{Beebe}}},
\bauthor{\binits{K.} \bsnm{{Bollinger}}},
\bauthor{\binits{J.M.} \bsnm{{Boyce}}},
\bauthor{\binits{A.} \bsnm{{Brahic}}},
\bauthor{\binits{G.A.} \bsnm{{Briggs}}},
\bauthor{\binits{R.H.} \bsnm{{Brown}}},
\bauthor{\binits{C.} \bsnm{{Chyba}}},
\bauthor{\binits{S.A.} \bsnm{{Collins}}},
\bauthor{\binits{T.} \bsnm{{Colvin}}},
\bauthor{\binits{A.F.} \bsnm{{Cook}} \bsuffix{II}},
\bauthor{\binits{D.} \bsnm{{Crisp}}},
\bauthor{\binits{S.K.} \bsnm{{Croft}}},
\bauthor{\binits{D.} \bsnm{{Cruikshank}}},
\bauthor{\binits{J.N.} \bsnm{{Cuzzi}}},
\bauthor{\binits{G.E.} \bsnm{{Danielson}}},
\bauthor{\binits{M.E.} \bsnm{{Davies}}},
\bauthor{\binits{E.} \bsnm{{De Jong}}},
\bauthor{\binits{L.} \bsnm{{Dones}}},
\bauthor{\binits{D.} \bsnm{{Godfrey}}},
\bauthor{\binits{J.} \bsnm{{Goguen}}},
\bauthor{\binits{I.} \bsnm{{Grenier}}},
\bauthor{\binits{V.R.} \bsnm{{Haemmerle}}},
\bauthor{\binits{H.} \bsnm{{Hammel}}},
\bauthor{\binits{C.J.} \bsnm{{Hansen}}},
\bauthor{\binits{C.P.} \bsnm{{Helfenstein}}},
\bauthor{\binits{C.} \bsnm{{Howell}}},
\bauthor{\binits{G.E.} \bsnm{{Hunt}}},
\bauthor{\binits{A.P.} \bsnm{{Ingersoll}}},
\bauthor{\binits{T.V.} \bsnm{{Johnson}}},
\bauthor{\binits{J.} \bsnm{{Kargel}}},
\bauthor{\binits{R.} \bsnm{{Kirk}}},
\bauthor{\binits{D.I.} \bsnm{{Kuehn}}},
\bauthor{\binits{S.} \bsnm{{Limaye}}},
\bauthor{\binits{H.} \bsnm{{Masursky}}},
\bauthor{\binits{A.} \bsnm{{McEwen}}},
\bauthor{\binits{D.} \bsnm{{Morrison}}},
\bauthor{\binits{T.} \bsnm{{Owen}}},
\bauthor{\binits{W.} \bsnm{{Owen}}},
\bauthor{\binits{J.B.} \bsnm{{Pollack}}},
\bauthor{\binits{C.C.} \bsnm{{Porco}}},
\bauthor{\binits{K.} \bsnm{{Rages}}},
\bauthor{\binits{P.} \bsnm{{Rogers}}},
\bauthor{\binits{D.} \bsnm{{Rudy}}},
\bauthor{\binits{C.} \bsnm{{Sagan}}},
\bauthor{\binits{J.} \bsnm{{Schwartz}}},
\bauthor{\binits{E.M.} \bsnm{{Shoemaker}}},
\bauthor{\binits{M.} \bsnm{{Showalter}}},
\bauthor{\binits{B.} \bsnm{{Sicardy}}},
\bauthor{\binits{D.} \bsnm{{Simonelli}}},
\bauthor{\binits{J.} \bsnm{{Spencer}}},
\bauthor{\binits{L.A.} \bsnm{{Sromovsky}}},
\bauthor{\binits{C.} \bsnm{{Stoker}}},
\bauthor{\binits{R.G.} \bsnm{{Strom}}},
\bauthor{\binits{V.E.} \bsnm{{Suomi}}},
\bauthor{\binits{S.P.} \bsnm{{Synott}}},
\bauthor{\binits{R.J.} \bsnm{{Terrile}}},
\bauthor{\binits{P.} \bsnm{{Thomas}}},
\bauthor{\binits{W.R.} \bsnm{{Thompson}}},
\bauthor{\binits{A.} \bsnm{{Verbiscer}}},
\bauthor{\binits{J.} \bsnm{{Veverka}}},
\batitle{{Voyager 2 at Neptune - Imaging science results}}.
\bjtitle{Science}
\bvolume{246},
\bfpage{1422}--\blpage{1449}
(\byear{1989}).
doi:\doiurl{10.1126/science.246.4936.1422}
\end{barticle}
\endbibitem

\bibitem[\protect\citeauthoryear{{Soderlund} et~al.}{2013}]{13soderlund}
\begin{barticle}
\bauthor{\binits{K.M.} \bsnm{{Soderlund}}},
\bauthor{\binits{M.H.} \bsnm{{Heimpel}}},
\bauthor{\binits{E.M.} \bsnm{{King}}},
\bauthor{\binits{J.M.} \bsnm{{Aurnou}}},
\batitle{{Turbulent models of ice giant internal dynamics: Dynamos, heat
  transfer, and zonal flows}}.
\bjtitle{Icarus}
\bvolume{224}(\bissue{1}),
\bfpage{97}--\blpage{113}
(\byear{2013}).
doi:\doiurl{10.1016/j.icarus.2013.02.014}
\end{barticle}
\endbibitem

\bibitem[\protect\citeauthoryear{{Sromovsky} and {Fry}}{2005}]{05sromovsky}
\begin{barticle}
\bauthor{\binits{L.A.} \bsnm{{Sromovsky}}},
\bauthor{\binits{P.M.} \bsnm{{Fry}}},
\batitle{{Dynamics of cloud features on Uranus}}.
\bjtitle{Icarus}
\bvolume{179},
\bfpage{459}--\blpage{484}
(\byear{2005}).
doi:\doiurl{10.1016/j.icarus.2005.07.022}
\end{barticle}
\endbibitem

\bibitem[\protect\citeauthoryear{{Sromovsky} et~al.}{2002}]{02sromovsky}
\begin{barticle}
\bauthor{\binits{L.A.} \bsnm{{Sromovsky}}},
\bauthor{\binits{P.M.} \bsnm{{Fry}}},
\bauthor{\binits{K.H.} \bsnm{{Baines}}},
\batitle{{The Unusual Dynamics of Northern Dark Spots on Neptune}}.
\bjtitle{Icarus}
\bvolume{156}(\bissue{1}),
\bfpage{16}--\blpage{36}
(\byear{2002}).
doi:\doiurl{10.1006/icar.2001.6761}
\end{barticle}
\endbibitem

\bibitem[\protect\citeauthoryear{{Sromovsky} et~al.}{1993}]{93sromovsky}
\begin{barticle}
\bauthor{\binits{L.A.} \bsnm{{Sromovsky}}},
\bauthor{\binits{S.S.} \bsnm{{Limaye}}},
\bauthor{\binits{P.M.} \bsnm{{Fry}}},
\batitle{{Dynamics of Neptune's Major Cloud Features}}.
\bjtitle{Icarus}
\bvolume{105},
\bfpage{110}--\blpage{141}
(\byear{1993}).
doi:\doiurl{10.1006/icar.1993.1114}
\end{barticle}
\endbibitem

\bibitem[\protect\citeauthoryear{{Sromovsky} et~al.}{1995}]{95sromovsky}
\begin{barticle}
\bauthor{\binits{L.A.} \bsnm{{Sromovsky}}},
\bauthor{\binits{S.S.} \bsnm{{Limaye}}},
\bauthor{\binits{P.M.} \bsnm{{Fry}}},
\batitle{{Clouds and circulation on Neptune: Implications of 1991 HST
  observations.}}
\bjtitle{Icarus}
\bvolume{118},
\bfpage{25}--\blpage{38}
(\byear{1995}).
doi:\doiurl{10.1006/icar.1995.1175}
\end{barticle}
\endbibitem

\bibitem[\protect\citeauthoryear{{Sromovsky} et~al.}{2001}]{01sromovsky}
\begin{barticle}
\bauthor{\binits{L.A.} \bsnm{{Sromovsky}}},
\bauthor{\binits{P.M.} \bsnm{{Fry}}},
\bauthor{\binits{T.E.} \bsnm{{Dowling}}},
\bauthor{\binits{K.H.} \bsnm{{Baines}}},
\bauthor{\binits{S.S.} \bsnm{{Limaye}}},
\batitle{{Coordinated 1996 HST and IRTF Imaging of Neptune and Triton. III.
  Neptune's Atmospheric Circulation and Cloud Structure}}.
\bjtitle{Icarus}
\bvolume{149},
\bfpage{459}--\blpage{488}
(\byear{2001}).
doi:\doiurl{10.1006/icar.2000.6564}
\end{barticle}
\endbibitem

\bibitem[\protect\citeauthoryear{{Sromovsky} et~al.}{2007}]{07sromovsky}
\begin{barticle}
\bauthor{\binits{L.A.} \bsnm{{Sromovsky}}},
\bauthor{\binits{P.M.} \bsnm{{Fry}}},
\bauthor{\binits{H.B.} \bsnm{{Hammel}}},
\bauthor{\binits{I.} \bsnm{{de Pater}}},
\bauthor{\binits{K.A.} \bsnm{{Rages}}},
\bauthor{\binits{M.R.} \bsnm{{Showalter}}},
\batitle{{Dynamics, evolution, and structure of Uranus' brightest cloud
  feature}}.
\bjtitle{Icarus}
\bvolume{192}(\bissue{2}),
\bfpage{558}--\blpage{575}
(\byear{2007}).
doi:\doiurl{10.1016/j.icarus.2007.05.015}
\end{barticle}
\endbibitem

\bibitem[\protect\citeauthoryear{{Sromovsky} et~al.}{2009}]{09sromovsky}
\begin{barticle}
\bauthor{\binits{L.A.} \bsnm{{Sromovsky}}},
\bauthor{\binits{P.M.} \bsnm{{Fry}}},
\bauthor{\binits{H.B.} \bsnm{{Hammel}}},
\bauthor{\binits{W.M.} \bsnm{{Ahue}}},
\bauthor{\binits{I.} \bsnm{{de Pater}}},
\bauthor{\binits{K.A.} \bsnm{{Rages}}},
\bauthor{\binits{M.R.} \bsnm{{Showalter}}},
\bauthor{\binits{M.A.} \bsnm{{van Dam}}},
\batitle{{Uranus at equinox: Cloud morphology and dynamics}}.
\bjtitle{Icarus}
\bvolume{203},
\bfpage{265}--\blpage{286}
(\byear{2009}).
doi:\doiurl{10.1016/j.icarus.2009.04.015}
\end{barticle}
\endbibitem

\bibitem[\protect\citeauthoryear{{Sromovsky} et~al.}{2012}]{12sromovsky}
\begin{barticle}
\bauthor{\binits{L.A.} \bsnm{{Sromovsky}}},
\bauthor{\binits{P.M.} \bsnm{{Fry}}},
\bauthor{\binits{H.B.} \bsnm{{Hammel}}},
\bauthor{\binits{I.} \bsnm{{de Pater}}},
\bauthor{\binits{K.A.} \bsnm{{Rages}}},
\batitle{{Post-equinox dynamics and polar cloud structure on Uranus}}.
\bjtitle{Icarus}
\bvolume{220}(\bissue{2}),
\bfpage{694}--\blpage{712}
(\byear{2012}).
doi:\doiurl{10.1016/j.icarus.2012.05.029}
\end{barticle}
\endbibitem

\bibitem[\protect\citeauthoryear{{Sromovsky} et~al.}{2014}]{14sromovsky}
\begin{barticle}
\bauthor{\binits{L.A.} \bsnm{{Sromovsky}}},
\bauthor{\binits{E.} \bsnm{{Karkoschka}}},
\bauthor{\binits{P.M.} \bsnm{{Fry}}},
\bauthor{\binits{H.B.} \bsnm{{Hammel}}},
\bauthor{\binits{I.} \bsnm{{de Pater}}},
\bauthor{\binits{K.} \bsnm{{Rages}}},
\batitle{{Methane depletion in both polar regions of Uranus inferred from
  HST/STIS and Keck/NIRC2 observations}}.
\bjtitle{Icarus}
\bvolume{238},
\bfpage{137}--\blpage{155}
(\byear{2014}).
doi:\doiurl{10.1016/j.icarus.2014.05.016}
\end{barticle}
\endbibitem

\bibitem[\protect\citeauthoryear{{Sromovsky} et~al.}{2015}]{15sromovsky}
\begin{barticle}
\bauthor{\binits{L.A.} \bsnm{{Sromovsky}}},
\bauthor{\binits{I.} \bsnm{{de Pater}}},
\bauthor{\binits{P.M.} \bsnm{{Fry}}},
\bauthor{\binits{H.B.} \bsnm{{Hammel}}},
\bauthor{\binits{P.} \bsnm{{Marcus}}},
\batitle{{High S/N Keck and Gemini AO imaging of Uranus during 2012-2014: New
  cloud patterns, increasing activity, and improved wind measurements}}.
\bjtitle{Icarus}
\bvolume{258},
\bfpage{192}--\blpage{223}
(\byear{2015}).
doi:\doiurl{10.1016/j.icarus.2015.05.029}
\end{barticle}
\endbibitem

\bibitem[\protect\citeauthoryear{{Sromovsky} et~al.}{2019}]{18sromovsky}
\begin{barticle}
\bauthor{\binits{L.A.} \bsnm{{Sromovsky}}},
\bauthor{\binits{E.} \bsnm{{Karkoschka}}},
\bauthor{\binits{P.M.} \bsnm{{Fry}}},
\bauthor{\binits{I.} \bsnm{{de Pater}}},
\bauthor{\binits{H.B.} \bsnm{{Hammel}}},
\batitle{{The methane distribution and polar brightening on Uranus based on
  HST/STIS, Keck/NIRC2, and IRTF/SpeX observations through 2015}}.
\bjtitle{Icarus}
\bvolume{317},
\bfpage{266}--\blpage{306}
(\byear{2019}).
doi:\doiurl{10.1016/j.icarus.2018.06.026}
\end{barticle}
\endbibitem

\bibitem[\protect\citeauthoryear{{Stauffer} et~al.}{2016}]{16stauffer}
\begin{barticle}
\bauthor{\binits{J.} \bsnm{{Stauffer}}},
\bauthor{\binits{M.S.} \bsnm{{Marley}}},
\bauthor{\binits{J.E.} \bsnm{{Gizis}}},
\bauthor{\binits{L.} \bsnm{{Rebull}}},
\bauthor{\binits{S.J.} \bsnm{{Carey}}},
\bauthor{\binits{J.} \bsnm{{Krick}}},
\bauthor{\binits{J.G.} \bsnm{{Ingalls}}},
\bauthor{\binits{P.} \bsnm{{Lowrance}}},
\bauthor{\binits{W.} \bsnm{{Glaccum}}},
\bauthor{\binits{J.D.} \bsnm{{Kirkpatrick}}},
\bauthor{\binits{A.A.} \bsnm{{Simon}}},
\bauthor{\binits{M.H.} \bsnm{{Wong}}},
\batitle{{Spitzer Space Telescope Mid-IR Light Curves of Neptune}}.
\bjtitle{Astronomical Journal}
\bvolume{152},
\bfpage{142}
(\byear{2016}).
doi:\doiurl{10.3847/0004-6256/152/5/142}
\end{barticle}
\endbibitem

\bibitem[\protect\citeauthoryear{{Stevens} et~al.}{1993}]{93stevens}
\begin{barticle}
\bauthor{\binits{M.H.} \bsnm{{Stevens}}},
\bauthor{\binits{D.F.} \bsnm{{Strobel}}},
\bauthor{\binits{F.} \bsnm{{Herbert}}},
\batitle{{An Analysis of the Voyager 2 Ultraviolet Spectrometer Occultation
  Data at Uranus: Inferring Heat Sources and Model Atmospheres}}.
\bjtitle{Icarus}
\bvolume{101}(\bissue{1}),
\bfpage{45}--\blpage{63}
(\byear{1993}).
doi:\doiurl{10.1006/icar.1993.1005}
\end{barticle}
\endbibitem

\bibitem[\protect\citeauthoryear{{Stratman} et~al.}{2001}]{01stratman}
\begin{barticle}
\bauthor{\binits{P.W.} \bsnm{{Stratman}}},
\bauthor{\binits{A.P.} \bsnm{{Showman}}},
\bauthor{\binits{T.E.} \bsnm{{Dowling}}},
\bauthor{\binits{L.A.} \bsnm{{Sromovsky}}},
\batitle{{EPIC Simulations of Bright Companions to Neptune's Great Dark
  Spots}}.
\bjtitle{Icarus}
\bvolume{151},
\bfpage{275}--\blpage{285}
(\byear{2001}).
doi:\doiurl{10.1006/icar.2001.6603}
\end{barticle}
\endbibitem

\bibitem[\protect\citeauthoryear{{Teanby} et~al.}{2019}]{19teanby}
\begin{barticle}
\bauthor{\binits{N.A.} \bsnm{{Teanby}}},
\bauthor{\binits{P.G.J.} \bsnm{{Irwin}}},
\bauthor{\binits{J.I.} \bsnm{{Moses}}},
\batitle{{Neptune's carbon monoxide profile and phosphine upper limits from
  Herschel/SPIRE: Implications for interior structure and formation}}.
\bjtitle{Icarus}
\bvolume{319},
\bfpage{86}--\blpage{98}
(\byear{2019}).
doi:\doiurl{10.1016/j.icarus.2018.09.014}
\end{barticle}
\endbibitem

\bibitem[\protect\citeauthoryear{{Tice} et~al.}{2013}]{13tice}
\begin{barticle}
\bauthor{\binits{D.S.} \bsnm{{Tice}}},
\bauthor{\binits{P.G.J.} \bsnm{{Irwin}}},
\bauthor{\binits{L.N.} \bsnm{{Fletcher}}},
\bauthor{\binits{N.A.} \bsnm{{Teanby}}},
\bauthor{\binits{J.} \bsnm{{Hurley}}},
\bauthor{\binits{G.S.} \bsnm{{Orton}}},
\bauthor{\binits{G.R.} \bsnm{{Davis}}},
\batitle{{Uranus' cloud particle properties and latitudinal methane variation
  from IRTF SpeX observations}}.
\bjtitle{Icarus}
\bvolume{223},
\bfpage{684}--\blpage{698}
(\byear{2013}).
doi:\doiurl{10.1016/j.icarus.2013.01.006}
\end{barticle}
\endbibitem

\bibitem[\protect\citeauthoryear{{Toledo} et~al.}{2018}]{18toledo}
\begin{barticle}
\bauthor{\binits{D.} \bsnm{{Toledo}}},
\bauthor{\binits{P.G.J.} \bsnm{{Irwin}}},
\bauthor{\binits{N.A.} \bsnm{{Teanby}}},
\bauthor{\binits{A.A.} \bsnm{{Simon}}},
\bauthor{\binits{M.H.} \bsnm{{Wong}}},
\bauthor{\binits{G.S.} \bsnm{{Orton}}},
\batitle{{Uranus's Northern Polar Cap in 2014}}.
\bjtitle{\grl}
\bvolume{45}(\bissue{11}),
\bfpage{5329}--\blpage{5335}
(\byear{2018}).
doi:\doiurl{10.1029/2018GL077654}
\end{barticle}
\endbibitem

\bibitem[\protect\citeauthoryear{{Toledo} et~al.}{2019}]{19toledo_ura}
\begin{barticle}
\bauthor{\binits{D.} \bsnm{{Toledo}}},
\bauthor{\binits{P.G.J.} \bsnm{{Irwin}}},
\bauthor{\binits{P.} \bsnm{{Rannou}}},
\bauthor{\binits{N.A.} \bsnm{{Teanby}}},
\bauthor{\binits{A.A.} \bsnm{{Simon}}},
\bauthor{\binits{M.H.} \bsnm{{Wong}}},
\bauthor{\binits{G.S.} \bsnm{{Orton}}},
\batitle{{Constraints on Uranus's haze structure, formation and transport}}.
\bjtitle{Icarus}
\bvolume{333},
\bfpage{1}--\blpage{11}
(\byear{2019}).
doi:\doiurl{10.1016/j.icarus.2019.05.018}
\end{barticle}
\endbibitem

\bibitem[\protect\citeauthoryear{{Tollefson} et~al.}{2018}]{18tollefson}
\begin{barticle}
\bauthor{\binits{J.} \bsnm{{Tollefson}}},
\bauthor{\binits{I.} \bsnm{{de Pater}}},
\bauthor{\binits{P.S.} \bsnm{{Marcus}}},
\bauthor{\binits{S.} \bsnm{{Luszcz-Cook}}},
\bauthor{\binits{L.A.} \bsnm{{Sromovsky}}},
\bauthor{\binits{P.M.} \bsnm{{Fry}}},
\bauthor{\binits{L.N.} \bsnm{{Fletcher}}},
\bauthor{\binits{M.H.} \bsnm{{Wong}}},
\batitle{{Vertical wind shear in Neptune's upper atmosphere explained with a
  modified thermal wind equation}}.
\bjtitle{Icarus}
\bvolume{311},
\bfpage{317}--\blpage{339}
(\byear{2018}).
doi:\doiurl{10.1016/j.icarus.2018.04.009}
\end{barticle}
\endbibitem

\bibitem[\protect\citeauthoryear{{Tollefson} et~al.}{2019}]{19tollefson}
\begin{botherref}
\oauthor{\binits{J.} \bsnm{{Tollefson}}},
\oauthor{\binits{I.} \bsnm{{de Pater}}},
\oauthor{\binits{S.} \bsnm{{Luszcz-Cook}}},
\oauthor{\binits{D.} \bsnm{{DeBoer}}},
{Neptune's Latitudinal Variations as Viewed with ALMA}.
arXiv e-prints,
1905--03384
(2019)
\end{botherref}
\endbibitem

\bibitem[\protect\citeauthoryear{{Tyler} et~al.}{1986}]{86tyler}
\begin{barticle}
\bauthor{\binits{G.L.} \bsnm{{Tyler}}},
\bauthor{\binits{D.N.} \bsnm{{Sweetnam}}},
\bauthor{\binits{J.D.} \bsnm{{Anderson}}},
\bauthor{\binits{J.K.} \bsnm{{Campbell}}},
\bauthor{\binits{V.R.} \bsnm{{Eshleman}}},
\bauthor{\binits{D.P.} \bsnm{{Hinson}}},
\bauthor{\binits{G.S.} \bsnm{{Levy}}},
\bauthor{\binits{G.F.} \bsnm{{Lindal}}},
\bauthor{\binits{E.A.} \bsnm{{Marouf}}},
\bauthor{\binits{R.A.} \bsnm{{Simpson}}},
\batitle{{Voyager 2 Radio Science Observations of the Uranian System:
  Atmosphere, Rings, and Satellites}}.
\bjtitle{Science}
\bvolume{233}(\bissue{4759}),
\bfpage{79}--\blpage{84}
(\byear{1986}).
doi:\doiurl{10.1126/science.233.4759.79}
\end{barticle}
\endbibitem

\bibitem[\protect\citeauthoryear{{Tyler} et~al.}{1989}]{89tyler}
\begin{barticle}
\bauthor{\binits{G.L.} \bsnm{{Tyler}}},
\bauthor{\binits{D.N.} \bsnm{{Sweetnam}}},
\bauthor{\binits{J.D.} \bsnm{{Anderson}}},
\bauthor{\binits{S.E.} \bsnm{{Borutzki}}},
\bauthor{\binits{J.K.} \bsnm{{Campbell}}},
\bauthor{\binits{V.R.} \bsnm{{Eshleman}}},
\bauthor{\binits{D.L.} \bsnm{{Gresh}}},
\bauthor{\binits{E.M.} \bsnm{{Gurrola}}},
\bauthor{\binits{D.P.} \bsnm{{Hinson}}},
\bauthor{\binits{N.} \bsnm{{Kawashima}}},
\batitle{{Voyager Radio Science Observations of Neptune and Triton}}.
\bjtitle{Science}
\bvolume{246}(\bissue{4936}),
\bfpage{1466}--\blpage{1473}
(\byear{1989}).
doi:\doiurl{10.1126/science.246.4936.1466}
\end{barticle}
\endbibitem

\bibitem[\protect\citeauthoryear{{Vallis}}{2006}]{06vallis}
\begin{bbook}
\bauthor{\binits{G.K.} \bsnm{{Vallis}}},
\bbtitle{{Atmospheric and Oceanic Fluid Dynamics}}
\byear{2006},
p. \bfpage{770}.
doi:\doiurl{10.2277/0521849691}
\end{bbook}
\endbibitem

\bibitem[\protect\citeauthoryear{{Visscher} and {Fegley}}{2005}]{05visscher}
\begin{barticle}
\bauthor{\binits{C.} \bsnm{{Visscher}}},
\bauthor{\binits{B.J.} \bsnm{{Fegley}}},
\batitle{{Chemical Constraints on the Water and Total Oxygen Abundances in the
  Deep Atmosphere of Saturn}}.
\bjtitle{Astrophys. J.}
\bvolume{623},
\bfpage{1221}--\blpage{1227}
(\byear{2005}).
doi:\doiurl{10.1086/428493}
\end{barticle}
\endbibitem

\bibitem[\protect\citeauthoryear{{Weidenschilling} and
  {Lewis}}{1973}]{73weidenschilling}
\begin{barticle}
\bauthor{\binits{S.J.} \bsnm{{Weidenschilling}}},
\bauthor{\binits{J.S.} \bsnm{{Lewis}}},
\batitle{{Atmospheric and cloud structures of the jovian planets}}.
\bjtitle{Icarus}
\bvolume{20},
\bfpage{465}--\blpage{476}
(\byear{1973})
\end{barticle}
\endbibitem

\bibitem[\protect\citeauthoryear{{Wong} et~al.}{2018}]{18wong}
\begin{barticle}
\bauthor{\binits{M.H.} \bsnm{{Wong}}},
\bauthor{\binits{J.} \bsnm{{Tollefson}}},
\bauthor{\binits{A.I.} \bsnm{{Hsu}}},
\bauthor{\binits{I.} \bsnm{{de Pater}}},
\bauthor{\binits{A.A.} \bsnm{{Simon}}},
\bauthor{\binits{R.} \bsnm{{Hueso}}},
\bauthor{\binits{A.} \bsnm{{S{\'a}nchez-Lavega}}},
\bauthor{\binits{L.} \bsnm{{Sromovsky}}},
\bauthor{\binits{P.} \bsnm{{Fry}}},
\bauthor{\binits{S.} \bsnm{{Luszcz-Cook}}},
\batitle{{A New Dark Vortex on Neptune}}.
\bjtitle{\aj}
\bvolume{155}(\bissue{3}),
\bfpage{117}
(\byear{2018}).
doi:\doiurl{10.3847/1538-3881/aaa6d6}
\end{barticle}
\endbibitem

\bibitem[\protect\citeauthoryear{{Yelle} et~al.}{1989}]{89yelle}
\begin{barticle}
\bauthor{\binits{R.V.} \bsnm{{Yelle}}},
\bauthor{\binits{J.C.} \bsnm{{McConnell}}},
\bauthor{\binits{D.F.} \bsnm{{Strobel}}},
\batitle{{The far ultraviolet reflection spectrum of Uranus - Results from the
  Voyager encounter}}.
\bjtitle{Icarus}
\bvolume{77},
\bfpage{439}--\blpage{456}
(\byear{1989}).
doi:\doiurl{10.1016/0019-1035(89)90098-5}
\end{barticle}
\endbibitem

\bibitem[\protect\citeauthoryear{{Zarka} and {Pedersen}}{1986}]{86zarka}
\begin{barticle}
\bauthor{\binits{P.} \bsnm{{Zarka}}},
\bauthor{\binits{B.M.} \bsnm{{Pedersen}}},
\batitle{{Radio detection of Uranian lightning by Voyager 2}}.
\bjtitle{Nature}
\bvolume{323},
\bfpage{605}--\blpage{608}
(\byear{1986}).
doi:\doiurl{10.1038/323605a0}
\end{barticle}
\endbibitem

\bibitem[\protect\citeauthoryear{{Zhang} et~al.}{2013}]{13zhang}
\begin{barticle}
\bauthor{\binits{X.} \bsnm{{Zhang}}},
\bauthor{\binits{C.A.} \bsnm{{Nixon}}},
\bauthor{\binits{R.L.} \bsnm{{Shia}}},
\bauthor{\binits{R.A.} \bsnm{{West}}},
\bauthor{\binits{P.G.J.} \bsnm{{Irwin}}},
\bauthor{\binits{R.V.} \bsnm{{Yelle}}},
\bauthor{\binits{M.A.} \bsnm{{Allen}}},
\bauthor{\binits{Y.L.} \bsnm{{Yung}}},
\batitle{{Radiative forcing of the stratosphere of Jupiter, Part I: Atmospheric
  cooling rates from Voyager to Cassini}}.
\bjtitle{Plan. \& Space Sci.}
\bvolume{88},
\bfpage{3}--\blpage{25}
(\byear{2013}).
doi:\doiurl{10.1016/j.pss.2013.07.005}
\end{barticle}
\endbibitem

\end{thebibliography}

%
%



\end{document}